\documentclass[aps,prl,nofootinbib,twocolumn,showpacs,superscriptaddress,letterpaper,amsmath,amsfonts,amssymb, preprintnumbers,longbibliography]{revtex4-1}
\usepackage{mathtools,amssymb,braket,mathrsfs,tikz,xspace,xcolor,float,color,siunitx,comment,afterpage,multirow,array,hhline,tensor,hyperref, textcomp,tikz,amsthm,amsmath}
\usepackage[utf8]{inputenc}
\usepackage{appendix}
\usepackage[caption=false]{subfig}
\usepackage{hyperref}

\newcommand{\CleverName}{\textsc{FORCE}\xspace}

\DeclareRobustCommand{\Tab}[1]{Table~\ref{#1}}

\DeclareRobustCommand{\Fig}[1]{Fig.~\ref{#1}}

\DeclareRobustCommand{\Eq}[1]{Eq.~(\ref{#1})}
\DeclareRobustCommand{\Eqs}[2]{Eqs.~(\ref{#1}) and (\ref{#2})}

\DeclareRobustCommand{\Reff}[1]{Ref.~\cite{#1}}
\DeclareRobustCommand{\Reffs}[1]{Refs.~\cite{#1}}

\definecolor{jdtcolor}{rgb}{0.8,0,0}
\definecolor{emmcolor}{rgb}{0,0.8,0}
\definecolor{rawcolor}{rgb}{0,0,0.8}

\begin{document}
\title{Anomaly Detection in Collider Physics via Factorized Observables}
\preprint{MIT-CTP/5644}

\author{Eric M. Metodiev}
\email{metodiev@mit.edu}
\affiliation{Center for Theoretical Physics, Massachusetts Institute of Technology, Cambridge, MA 02139, USA}
\affiliation{The NSF AI Institute for Artificial Intelligence and Fundamental Interactions}

\author{Jesse Thaler}
\email{jthaler@mit.edu}
\affiliation{Center for Theoretical Physics, Massachusetts Institute of Technology, Cambridge, MA 02139, USA}
\affiliation{The NSF AI Institute for Artificial Intelligence and Fundamental Interactions}

\author{Raymond Wynne}
\email{rwynne@caltech.edu}
\affiliation{The NSF AI Institute for Artificial Intelligence and Fundamental Interactions}
\affiliation{Department of Electrical Engineering and Computer Science,
Massachusetts Institute of Technology, Cambridge, MA 02139, USA}
\affiliation{Division of Physics, Mathematics and Astronomy, Caltech, Pasadena, CA, USA}

\flushbottom

\begin{abstract}
%
To maximize the discovery potential of high-energy colliders, experimental searches should be sensitive to unforeseen new physics scenarios.
This goal has motivated the use of machine learning for unsupervised anomaly detection.
In this paper, we introduce a new anomaly detection strategy called \CleverName:  factorized observables for regressing conditional expectations.
Our approach is based on the inductive bias of factorization, which is the idea that the physics governing different energy scales can be treated as approximately independent.
Assuming factorization holds separately for signal and background processes, the appearance of non-trivial correlations between low- and high-energy observables is a robust indicator of new physics.
Under the most restrictive form of factorization, a machine-learned model trained to identify such correlations will in fact converge to the optimal new physics classifier.
We test \CleverName on a benchmark anomaly detection task for the Large Hadron Collider involving collimated sprays of particles called jets.
By teasing out correlations between the kinematics and substructure of jets, our method can reliably extract percent-level signal fractions.
This strategy for uncovering new physics adds to the growing toolbox of anomaly detection methods for collider physics with a complementary set of assumptions.
\end{abstract}

\maketitle

Despite the excellent targeted search efforts of multiple experiments, no conclusive evidence for new physics has been seen at the Large Hadron Collider (LHC) since the Higgs boson discovery in 2012~\cite{Aad:2012tfa,Chatrchyan:2012ufa}.
It is difficult, however, to exclude the possibility that new physics might exist in a form that has yet to be theoretically predicted.
Although targeted searches for a specific scenario (or class of scenarios) might yield a serendipitous discovery, they could lack sensitivity to even sizeable amounts of unforeseen new physics in LHC data.
To enable the broadest coverage for collider searches, robust techniques are needed to probe generic deviations from the Standard Model.
This goal has inspired the development of several anomaly detection approaches for collider physics \cite{Aguilar-Saavedra:2017rzt, Collins:2018epr, DAgnolo:2018cun, DeSimone:2018efk, Hajer:2018kqm, Farina:2018fyg, Heimel:2018mkt, 1809.02977, Cerri:2018anq, Collins:2019jip, Roy:2019jae, Dillon:2019cqt, Blance:2019ibf, Romao:2019dvs, Mullin:2019mmh, DAgnolo:2019vbw, Nachman:2020lpy, Andreassen:2020nkr, Amram:2020ykb, Romao:2020ojy, Knapp:2020dde, ATLAS:2020iwa, Dillon:2020quc, CrispimRomao:2020ucc, Cheng:2020dal, Khosa:2020qrz, Thaprasop:2020mzp, Aguilar-Saavedra:2020uhm, Alexander:2020mbx, Benkendorfer:2020gek, Pol:2020weg, Mikuni:2020qds, vanBeekveld:2020txa, Park:2020pak, Faroughy:2020gas, Stein:2020rou, Kasieczka:2021xcg, Chakravarti:2021svb, Batson:2021agz, Blance:2021gcs, Bortolato:2021zic, Collins:2021nxn, Dillon:2021nxw, Finke:2021sdf, Shih:2021kbt, Atkinson:2021nlt, Kahn:2021drv, Aarrestad:2021oeb, Dorigo:2021iyy, Caron:2021wmq, Govorkova:2021hqu, Kasieczka:2021tew, Volkovich:2021txe, Govorkova:2021utb, Hallin:2021wme, Ostdiek:2021bem, Fraser:2021lxm, Jawahar:2021vyu, Herrero-Garcia:2021goa, Lester:2021aks, Aguilar-Saavedra:2021utu, Tombs:2021wae, Mikuni:2021nwn, Chekanov:2021pus, dAgnolo:2021aun, Canelli:2021aps, Ngairangbam:2021yma, Aguilar-Saavedra:2022ejy, Buss:2022lxw, Bradshaw:2022qev, Birman:2022xzu, Raine:2022hht, Letizia:2022xbe, Fanelli:2022xwl, Finke:2022lsu, Verheyen:2022tov, Dillon:2022tmm, Alvi:2022fkk, Dillon:2022mkq, Caron:2022wrw, Park:2022zov, Kasieczka:2022naq, Kamenik:2022qxs, Hallin:2022eoq, Araz:2022zxk, Mastandrea:2022vas, Schuhmacher:2023pro, Golling:2023juz, Roche:2023int, Sengupta:2023xqy, Aguilar-Saavedra:2023pde,Vaslin:2023lig, ATLAS:2023azi, Mikuni:2023tok, Golling:2023yjq, Chekanov:2023uot, CMSECAL:2023fvz, Bickendorf:2023nej, Finke:2023ltw, Buhmann:2023acn, Freytsis:2023cjr}, which have recently found experimental applications~\cite{ATLAS:2020iwa, ATLAS:2023azi}.
%

%
Any anomaly detection technique must make assumptions about what constitutes an anomaly, which then implies limitations on its sensitivity.
One class of techniques uses comparisons between data and simulation to detect anomalous events~\cite{Aguilar-Saavedra:2017rzt,DAgnolo:2018cun,DeSimone:2018efk}; this approach is susceptible to detector or generator mismodeling and may confuse poorly modeled regions of phase space for new physics.
A more data-driven approach assumes that new physics will appear as a localized cluster in phase space~\cite{Collins:2018epr,Collins:2019jip, Benkendorfer:2020gek}; 
this is an excellent inductive bias to detect mass resonances, but limits the types of models that can be probed.
The most unstructured techniques, such as autoencoder reconstruction losses, operationally define the notion of anomalous events via the choice of machine learning architecture~\cite{Farina:2018fyg, Heimel:2018mkt, Roy:2019jae};
since they lack controlled assumptions, it is challenging to determine the applicability of such methods to particular new physics scenarios.

\pagebreak[3]

\begin{figure}[t]
\centering
\includegraphics[width=0.8\columnwidth]{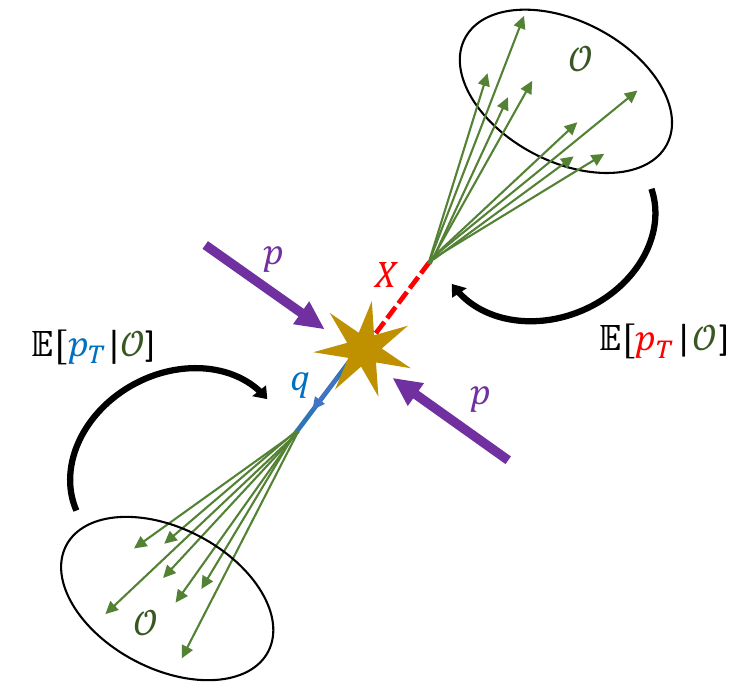}
\caption{\label{fig:illustration} Illustration of the \CleverName anomaly detection approach, applied to a dijet search.
A machine-learning model is trained to predict the kinematics of a jet from its substructure.
The model output converges to the optimal new physics jet classifier assuming factorization holds for both the signal and background processes.}
\end{figure}

In this paper, we introduce an anomaly detection strategy called \CleverName---factorized observables for regressing conditional expectations---based on the inductive bias of factorization.
Factorization occurs when the physics governing high-energy scales is approximately independent from those governing low-energy scales.
Jet production offers a canonical example of factorization at colliders, where the processes that determine the kinematics and flavors of high-energy partons are approximately independent of the dynamics that yield collimated sprays of low-energy hadrons.
As illustrated in \Fig{fig:illustration}, a machine learning model can be trained to predict the kinematics of a jet from its boost-invariant substructure.
Kinematics and substructure are approximately independent in the absence of new physics, so if the model learns non-trivial correlations, then this indicates a possible anomaly.
Our approach does not require simulated data, works even if the new physics is non-resonant, and provably converges to the optimal classifier assuming strict factorization.
\CleverName builds upon previous uses of factorized structures to estimate backgrounds~\cite{Cohen:2014epa, Lin:2019htn}, train data-driven collider classifiers~\cite{Dery:2017fap,Cohen:2017exh,Metodiev:2017vrx,Komiske:2018oaa,Sirunyan:2019jud,Amram:2020ykb}, and disentangle particle flavors using topic modeling~\cite{Metodiev:2018ftz,Komiske:2018vkc,Aad:2019onw,Komiske:2022vxg}.

To demonstrate the \CleverName approach, we perform a case study involving jets \cite{Salam:2010nqg,Larkoski:2017jix,Kogler:2018hem, Marzani:2019hun}.
Jets are proxies for the partons or resonances produced in high-energy collisions, with the kinematics of a jet reflecting the kinematics of its initiating particle.
Jets then acquire substructure through lower-energy processes, such as decays of intermediate-scale resonances or showering/hadronization in quantum chromodynamics (QCD).
Many new physics scenarios involve jet production, making jets a key target for anomaly detection.


In the soft-collinear limit of QCD, the substructure of a jet factorizes from its kinematics \cite{Bauer:2000yr,Bauer:2001ct,Bauer:2001yt,Bauer:2002nz,Bauer:2002nz,Beneke:2002ph} (see also \cite{Collins:1989gx,Bauer:2008jx,Feige:2014wja,Sterman:2022gyf}).
Factorization also holds for the decay of an intermediate-scale resonance in the narrow width approximation \cite{Berdine:2007uv,Uhlemann:2008pm}, such as for a Lorentz-boosted $W$/$Z$ boson, Higgs boson, or top quark.
Therefore, at leading order in the high-$p_T$ limit, the kinematics of a jet is determined by its transverse momentum $p_T$ and rapidity $y$.
Let $\mathcal O$ be a list of jet substructure observables, possibly high dimensional.
Then, assuming factorization holds, the distribution of jet kinematics and substructure obey:
\begin{equation}\label{eq:factorized}
P(p_T, y, \mathcal O) \approx \sum_i f_i\,P_i(p_T, y)\,P_i(\mathcal O|p_T),
\end{equation}
where $P_i$ refers to the probability density function, $i\in \{q,g,W,t,\cdots\}$ labels the types of initiating particle, and $f_i$ is the fraction of jets initiated by that particle type.
Factorization imposes a non-trivial constraint that $P_i(\mathcal O|p_T)$ is independent of $y$ for each $i$ \textit{and} that a finite sum over $i$ is sufficient to model the distribution.

If we make an even more restrictive assumption that $\mathcal{O}$ consists of scale- and boost-invariant observables, with no conditional $p_T$ dependence, then we can write:
\begin{equation}\label{eq:independent}
    P_i(\mathcal O|p_T) \approx P_i(\mathcal O).
\end{equation}
Examples of such quasi-invariant observables are 
$N$-subjettiness ratios~\cite{Thaler:2010tr,Thaler:2011gf}, $D_2$~\cite{Larkoski:2014gra}, $D_3$~\cite{Larkoski:2014zma}, and $N_i$~\cite{Moult:2016cvt}.
Here, we take the jet substructure to be dominated by the initiating particle's flavor and independent of the remainder of the event, up to subleading corrections.
The factorized structure of \Eqs{eq:factorized}{eq:independent} is what we will exploit for anomaly detection using \CleverName.

Consider the case of only two jet types: background ($B$) QCD jets from high-energy quarks and gluons, and signal ($S$) jets from the hadronic decay of a new particle.
To simplify the algebra, we marginalize over $y$.
Via \Eqs{eq:factorized}{eq:independent}, i.e.\ assuming $p_T$ and $\mathcal{O}$ are independent in both the signal and background processes, the joint distribution of jet kinematics and substructure is:
\begin{align}\label{eq:factorized2}
P(p_T,\mathcal O) = f_S\ P_S(p_T)\,P_S(\mathcal O) + f_B\, P_B(p_T)\,P_B(\mathcal O),
\end{align}
where $f_S$ is the fraction of new physics events and $f_B$ is the fraction of QCD events, with $f_S + f_B = 1$.
Our goal is to discover and characterize the new physics signal in a data-driven manner.

The key insight behind \CleverName is that a machine-learned model trained to predict a jet's $p_T$ from its substructure observable $\mathcal O$ yields the optimal $S$ versus $B$ classifier (assuming factorization and with sufficient training and statistics).
By ``predict'', we mean learning the conditional expectation value:
\begin{equation}
\hat{p}_T(\mathcal O) \equiv \mathbb{E}[p_T | \mathcal O],
\end{equation}
which is a function of $\mathcal O$ that can be learned from minimizing the mean-squared error; see the Supplemental Material.
With a single factorized process, $\hat{p}_T(\mathcal O)$ would be independent of $\mathcal{O}$, but the sum of two factorized processes yields non-trivial $\mathcal{O}$ dependence.
To see this, recall from the Neyman-Pearson lemma~\cite{nplemma} that the signal-to-background likelihood ratio is the optimal new physics classifier derivable from $\mathcal{O}$:
\begin{equation}
L_{S / B}(\mathcal O) = \frac{P_S(\mathcal O)}{P_B(\mathcal O)}.
\end{equation}
(A stronger classifier might exist if one includes $p_T$ information, but that requires a priori knowledge of ${P_S(p_T)}/{P_B(p_T)}$.)
From \Eq{eq:factorized2}, the conditional distribution can be written as 
\begin{align}\label{eq:conddist}
P(p_T | \mathcal O)
& = \frac{(1- f_S)\,P_B(p_T) + f_S\, L_{S / B}(\mathcal O) \, P_S(p_T)}{ 1-f_S+ f_S\, L_{S / B}(\mathcal O)}.
\end{align}
Taking the expectation value with respect to $p_T$ yields:
\begin{align}
\label{eq:cond_exp}
\hat{p}_T(\mathcal O) = \langle p_T \rangle_B + f_S \frac{(\langle p_T\rangle_S-\langle p_T\rangle_B)L_{S / B}(\mathcal O)}{1-f_S + f_S \,L_{S / B}(\mathcal O)}.
\end{align}
Remarkably, $\hat{p}_T(\mathcal O)$ is monotonically related to $ L_{S / B}(\mathcal O)$, so it also defines optimal decision boundaries.
A similar observation underpins anomaly detection methods based on classification without labels~\cite{Metodiev:2017vrx,Collins:2018epr,Collins:2019jip}.
To our knowledge, the first proof that optimal classifiers can be defined through regression (as opposed to classification) appears in \Reff{Komiske:2022vxg}.
Note that the factorization assumption is crucial for learning a monotone of $ L_{S / B}(\mathcal O)$ without explicit knowledge of $P_S(\mathcal{O})$ or $P_B(\mathcal{O})$ individually.

Thus, assuming \textbf{F}actorized \textbf{O}bservables, \textbf{R}egressing the \textbf{C}onditional \textbf{E}xpectation furnishes a powerful probe of new physics, justifying the \CleverName acronym.
Interestingly, the same logic holds with more than one type of new particle, such as $pp \to XY$, as long as $\langle p_T\rangle_X = \langle p_T \rangle_Y$ as expected from momentum conservation.
If $\langle p_T \rangle_S > \langle p_T \rangle_B$, then \Eq{eq:cond_exp} defines an optimal tagger; otherwise, it defines an optimal anti-tagger.
In the absence of new physics ($f_S=0$) or if the signal and background have the same average kinematics ($\langle p_T \rangle_S = \langle p_T \rangle_B$), then \Eq{eq:cond_exp} simply returns the expectation value $\langle p_T \rangle$ with no observable dependence.
Deviations of the model output from $\langle p_T \rangle$ are therefore a
harbinger for a new type of factorized object in the data (or a violation of the factorization assumption).

\pagebreak[3]

In summary, \CleverName proceeds as follows:
\begin{enumerate}
\item {\bf Define} approximately factorized objects (e.g.~jets) with kinematics $p_T$ and scale-/boost-invariant substructure $\mathcal O$.
\item {\bf Train} a machine-learning model $\hat{p}_T(\mathcal O)$ to predict $p_T$ from $\mathcal O$ with the mean-squared error loss.
\item {\bf Classify} anomalous objects via the model output.
\end{enumerate}
%
%
Of course, real collider data is richer than the simple two-category case in \Eq{eq:factorized2}.
QCD jets themselves are admixtures of quark and gluon jets, each with slightly different kinematics and substructure.
Multiple effects can violate the strict version of factorization in \Eq{eq:independent}, such as partial containment of particle decay products in the jet cone or the logarithmic scale-dependence of QCD due to the running of the strong coupling constant.
Further, certain known Standard-Model processes, such as jets from hadronically decaying $W$/$Z$/Higgs bosons or top quarks, may be considered anomalous beyond the QCD dijet background by our formulation.
This behavior may in fact be desirable, and ``re-discovering'' these particles may be an interesting way to benchmark this technique in data.
More broadly, though, the general structure of factorization motivates \CleverName as a new physics search strategy.

We now showcase \CleverName for a new physics search involving dijets.
Our case study is based on the development dataset~\cite{gregor_kasieczka_2019_2629073} from the LHC Olympics 2020 Anomaly Detection Challenge~\cite{Kasieczka:2021xcg}.
This simulated dataset consists of 1 million QCD dijet events and up to 100 thousand $W'\to XY$ events, with the $X$ and $Y$ particles decaying to two quarks.
The masses of the three new particles are $m_{W'} = 3.5$~TeV, $m_X = 500$~GeV, and $m_Y = 100$~GeV.
The $X$ and $Y$ particles are boosted, giving rise to a dijet resonance with two-pronged jet substructure.
While the signal has a $W'$ mass peak, this feature is not used for \CleverName training.

The LHC Olympics dataset is generated with \textsc{Pythia} 8.219~\cite{Sjostrand:2007gs,Sjostrand:2014zea} and simulated with \textsc{Delphes}~3.4.1~\cite{deFavereau:2013fsa}, excluding pileup or multiple parton interactions.
Events are selected to have at least one $R=1.0$ anti-$k_T$~\cite{Cacciari:2008gp} jet with transverse momentum $p_T > 1.2$~TeV and pseudorapidity $|\eta|<2.5$.
Jets are clustered via the anti-$k_T$ algorithm with a radius of $R=1.0$ using \textsc{FastJet 3.3.3}~\cite{Cacciari:2008gp,Cacciari:2011ma}.
The leading two jets, i.e.\ those with highest transverse momenta, are recorded in each event as a proxy for the products of the high-energy scattering process.
Both jets are used in the analysis, so the anomalies are defined over jets (instead of over events).

%
For our substructure observables $\mathcal{O}$, we use energy flow polynomials (EFPs)~\cite{Komiske:2017aww,Komiske:2019asc}.
As reviewed in the Supplemental Material, EFPs arise from a systematic expansion in energies and angles, and they are sensitive to a broad range of jet features, including the two-prong substructure of the boosted $X$ and $Y$ particles.
We compute all 13 EFPs up to and including degree 3 using \textsc{EnergyFlow 1.0.3}~\cite{EnergyFlow}, using $z_i = p_{T,i}$ as the energy variable and $\theta_{ij} = (p_i^\mu p_{j\mu}/p_{T,i} p_{T,j})^{1/2}$ as the angular variable.
To satisfy \Eq{eq:independent}, the EFPs need to be made scale- and boost-invariant.
Quasi-scale-invariance can be achieved by normalizing the energies to sum to unity.
As boosts transverse to the beamline approximately scale energies by $\gamma$ and angles by $1/\gamma$, the EFPs can be made quasi-boost-invariant by rescaling them via:
\begin{equation}
\text{EFP} \to \frac{\text{EFP}}{\left(\displaystyle\sum_{i=1}^M p_{T,i}\right)^{N-2d} \left(\displaystyle\sum_{i=1}^M \sum_{j=1}^M p_{T,i} p_{T,j} \theta_{ij}\right)^{d}},
\end{equation}
where $N$ and $d$ are the energy and angular degrees of the polynomial.
This rescaling reduces our basis to seven independent elements.
We note that observables desired to be independent of $p_T$ have been employed in prior work on anomaly detection \cite{Collins:2019jip} and jet-tagging \cite{Aguilar-Saavedra:2017rzt}.
In the Supplemental Material, we show how FORCE performance degrades without this normalization. 
Interestingly, existing observables for multi-prong new physics searches, such as $D_2$~\cite{Larkoski:2014gra}, emerge naturally as elements of this quasi-invariant basis.

\begin{figure*}[t]
\centering
\subfloat[][]{\label{fig:dijet} \includegraphics[width=0.98\columnwidth]{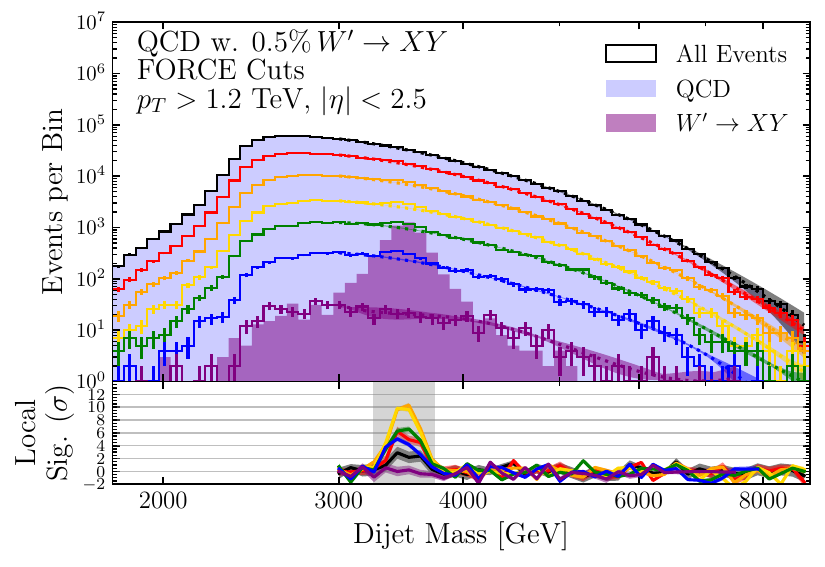}}
$\qquad$
\subfloat[][]
{\label{fig:jet} \includegraphics[width=0.98\columnwidth]{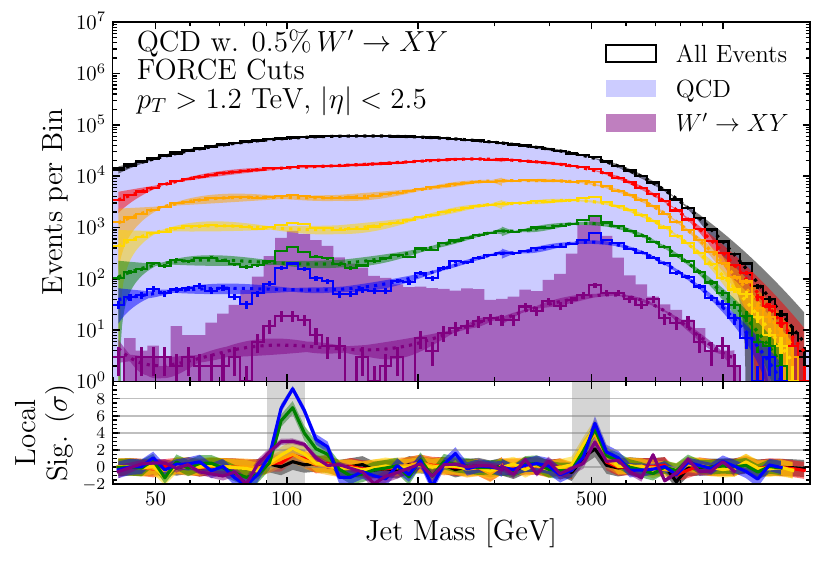}}%
\caption{\label{fig:dijetmassdists}
The \CleverName method applied to a dijet search with a 0.5\% new physics signal fraction ($f_S = 0.005$), where the same cut on $\hat{p}_T(\mathcal O)$ is imposed on both jets.
Shown are (a) dijet and (b) jet mass distributions in the top panels and local significance values in the lower panels.
The shaded regions ``QCD'' and ``$W' \to XY$'' refer to the truth distributions after baseline kinematic selections.
The solid lines indicate the data, whereas the dashed lines and shaded areas indicate the background predictions and uncertainties derived from the Legendre fits.
The black curve indicates no cuts on the trained model output $\hat{p}_T(\mathcal O)$, while the sweep from red to purple corresponds to cuts that increase the lower bound on $\hat{p}_T(\mathcal O)$, with specific cut values chosen manually to highlight the qualitative behavior. 
Tighter cuts, where the model predictions are further from plain QCD jets, clearly identify the new physics signal with a dijet mass peak at $m_{W'}=3.5$~TeV and individual jet mass peaks at $m_X = 500$~GeV and $m_Y = 100$~GeV.
See the Supplemental Material for different values of $f_S$.
}
\label{fig:mass_plots}
\end{figure*}

The \CleverName method works with any machine-learning algorithm whose output $\hat{p}_T(\mathcal O)$ converges to the conditional expectation $\mathbb{E}[p_T|\mathcal O]$.
We use a fully-connected neural network consisting of three dense layers with 50 nodes per layer, as well as L2 kernel and bias regularization of $10^{-5}$ in each layer.
Between each dense layer is a dropout \cite{dropout} layer with $p=0.1$.
Neural networks are implemented and trained with \textsc{Keras}~\cite{keras} using \textsc{TensorFlow}~\cite{tensorflow}, optimized with \textsc{ADAM}~\cite{kingma2017adam} with a patience parameter of 10.
Since our method is fully unsupervised, seeing no signal/background labels, we utilize the full dataset in training.
(In a full analysis, it might be preferable to use statistically independent samples for training and testing.)
Our code implementing \CleverName is publicly available on GitHub~\cite{repocode}.

%
The dijet and jet mass distributions are shown in \Fig{fig:mass_plots} after applying \CleverName for a signal fraction of $f_S=0.005$.
Here, we impose a cut on both jets that enforces their model output $\hat{p}_T(\mathcal O)$ to be above the same threshold.
With a strict enough cut, the signal clearly manifests as a peak at $m_{W'}=3.5$~TeV in the dijet mass distribution, and peaks at $m_X = 500$~GeV and $m_Y=100$~GeV in the individual jet mass distribution.
To estimate the local significance, a background fit is performed using Legendre polynomials outside of the shaded signal region, using fifth order as the central value and between second and seventh orders for the uncertainty band.
For the dijet mass background fit, we use data above $3$~TeV, and for the jet mass background fits, below $300$~GeV for $m_Y$ and above $300$~GeV for $m_X$.
We find a boost in significance, where a pre-cut excess of $2\sigma$ for the $W'$ and $X$ are increased to $> 5\sigma$, while a pre-cut excess of $1 \sigma$ for the $Y$ is increased to $> 5\sigma$.
Note that although the new physics in this case study appears as a resonance in the jet and dijet mass distributions, a resonance is not a requirement of the \CleverName method.
(Without a bump-like feature, though, one would have to leverage some other method for background estimation.)
Further, by imposing quasi-boost/scale invariance, the model output is largely decorrelated from jet mass (see Fig.~4 in the Supplemental Material and related discussion in \Reffs{Dolen:2016kst,Shimmin:2017mfk,Moult:2017okx}).

\begin{figure}
\centering
\includegraphics[width=0.96\columnwidth]{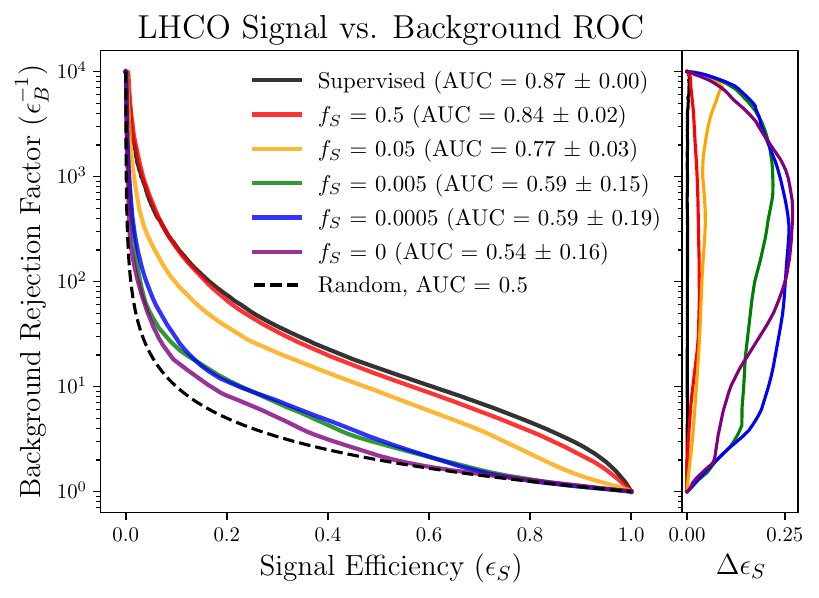}
\caption{\label{fig:aucs}
Per-jet classification performance of \CleverName in a dijet search with different signal fractions $f_S$.
The background rejection factor is shown as a function of (left panel) the signal efficiency and (right panel) the standard deviation of the signal efficiency over 10 trainings.
See the Supplemental Material for a discussion of the $f_S = 0$ limit, where non-trivial performance can arise from a breakdown of factorization.
}
\end{figure}

%
To test the robustness of \CleverName, we apply our method on a range of signal fractions $f_S$.
For stability in this analysis, we train with an equally mixed dataset of 100,000 signal and 100,000 background events, using sample weights in Keras to mimic a signal fraction $f_S$.
We then test the model on the full dataset.
To account for variability and obtain error bars, we train an ensemble of 10 different models.
The average per-jet classification performance is shown in \Fig{fig:aucs}, where we plot the background rejection factor as a function of signal efficiency.
Here, the per-jet performance is evaluated only on the learned $\hat{p}_T(\mathcal O)$, not including any additional features such as jet mass or assumptions about the $W' \to XY$ event topology.
(For these reasons, one cannot directly compare our results to those of previous LHC Olympians.)
In the large signal limit, \CleverName approaches the optimal supervised classifier, as predicted by \Eq{eq:cond_exp}.
As the signal fraction decreases, the performance degrades and the variability increases, but there are still substantial gains in sensitivity.
Note that the $f_S=0$ limit still yields reasonable classification performance; this is possibly due to deviations from strict factorization, as discussed further in the Supplemental Material.
(Alternatively, since we are applying a no-signal model to a dataset with signal, proper convergence might not be achievable off the data manifold.)

%
Having established the desired behavior of \CleverName on a benchmark collider search, it is worth remarking on several important points.
First, our method is based on the inductive bias of factorization, so the performance we saw in the dijet analysis may not translate to other scenarios.
This reflects a universal challenge for all approaches to anomaly detection, where the performance of the method depends on the applicability of the assumptions.
Nevertheless, limits can be set on the parameters of specific new physics scenarios (even post hoc) by performing pseudo-experiments, injecting various amounts of signal, and repeating the procedure to establish confidence intervals.
Second, as the signal fraction decreases, the performance of the learned model becomes highly sensitive to parameter initialization and statistical fluctuations.
To ensure robust behavior in this regime, we recommend \CleverName be paired with a regularization method like ensemble learning ~\cite{DBLP:journals/corr/abs-2104-02395}.
Third, detector effects can introduce factorization-violating effects, so it may be beneficial to apply \CleverName after multi-dimensional unfolding is applied to the data~\cite{Andreassen:2019cjw,Bellagente:2020piv,Andreassen:2021zzk,Arratia:2021otl}.
Jet grooming techniques~\cite{Dasgupta:2013ihk,Larkoski:2014wba,Frye:2016aiz} might also improve the factorized behavior of jets at the theoretical level.
Finally, we emphasize that no strategy can outperform a targeted search (i.e.~hypothesis test) for a specific model, and that the power of data-driven approaches such as \CleverName is in broadening the space of new physics scenarios that can be probed.

In summary, we introduced \CleverName: an anomaly detection strategy for factorized new physics.
By training a machine-learning model to predict the kinematics of factorized objects from their scale- and boost-invariant substructure, we obtain a powerful classifier directly from observed data.
We showcased \CleverName on a benchmark search for new physics in the dijet final state, where it successfully identified a new physics signal.
This work contributes to a growing body of work where powerful computational tools from machine learning are combined with deep theoretical principles to unlock novel collider data analysis strategies.
Furthermore, the \CleverName method can be easily integrated into these prior methods when viewing the model output as an observable with high discrimination power.

The \CleverName framework shifts the discussion of new physics searches from specific models to their general factorized structure, with machine-learning techniques performing detailed observable-level analyses.
It would be interesting to generalize \CleverName to handle more than one kinematic feature and more than two event categories, which would be important to handle multiple background components.
It would also be interesting to combine our reasoning with the factorization of the full event energy flow~\cite{Bauer:2008jx}, which may help reframe anomaly detection in the language of ``theory space''~\cite{Komiske:2019fks,Komiske:2020qhg}.
Though we focused on jets and jet substructure here, this method applies more broadly to any factorized probability distributions, in collider physics and beyond.
Data-driven searches hold the potential to fundamentally surprise us, not only by discovering new physics, but by uncovering it in forms that we have failed to imagine.

\acknowledgments
The authors are grateful to the organizers of the LHC Olympics 2020 Anomaly Detection Challenge for stimulating this research direction and producing excellent public datasets.
We are grateful to Samuel Alipour-fard, Rikab Gambhir, Patrick Komiske, Benjamin Nachman, and Nilai Sarda for helpful comments and discussions.
JT is supported by the National Science Foundation under Cooperative Agreement PHY-2019786 (\href{https://iaifi.org}{The NSF AI Institute for Artificial Intelligence and Fundamental Interactions}) and by the Simons Foundation through Investigator grant 929241.
This work was supported by the Office of Nuclear Physics of the U.S. Department of Energy (DOE) under Grant No.\ DE-SC0011090 and by the DOE Office of High Energy Physics under grants DE-SC0012567 and DE-SC0019128.

\bibliography{anomalydetection_arxiv_v 2, other_v2}

\begin{thebibliography}{162}%
\makeatletter
\providecommand \@ifxundefined [1]{%
 \@ifx{#1\undefined}
}%
\providecommand \@ifnum [1]{%
 \ifnum #1\expandafter \@firstoftwo
 \else \expandafter \@secondoftwo
 \fi
}%
\providecommand \@ifx [1]{%
 \ifx #1\expandafter \@firstoftwo
 \else \expandafter \@secondoftwo
 \fi
}%
\providecommand \natexlab [1]{#1}%
\providecommand \enquote  [1]{``#1''}%
\providecommand \bibnamefont  [1]{#1}%
\providecommand \bibfnamefont [1]{#1}%
\providecommand \citenamefont [1]{#1}%
\providecommand \href@noop [0]{\@secondoftwo}%
\providecommand \href [0]{\begingroup \@sanitize@url \@href}%
\providecommand \@href[1]{\@@startlink{#1}\@@href}%
\providecommand \@@href[1]{\endgroup#1\@@endlink}%
\providecommand \@sanitize@url [0]{\catcode `\\12\catcode `\$12\catcode `\&12\catcode `\#12\catcode `\^12\catcode `\_12\catcode `\%12\relax}%
\providecommand \@@startlink[1]{}%
\providecommand \@@endlink[0]{}%
\providecommand \url  [0]{\begingroup\@sanitize@url \@url }%
\providecommand \@url [1]{\endgroup\@href {#1}{\urlprefix }}%
\providecommand \urlprefix  [0]{URL }%
\providecommand \Eprint [0]{\href }%
\providecommand \doibase [0]{http://dx.doi.org/}%
\providecommand \selectlanguage [0]{\@gobble}%
\providecommand \bibinfo  [0]{\@secondoftwo}%
\providecommand \bibfield  [0]{\@secondoftwo}%
\providecommand \translation [1]{[#1]}%
\providecommand \BibitemOpen [0]{}%
\providecommand \bibitemStop [0]{}%
\providecommand \bibitemNoStop [0]{.\EOS\space}%
\providecommand \EOS [0]{\spacefactor3000\relax}%
\providecommand \BibitemShut  [1]{\csname bibitem#1\endcsname}%
\let\auto@bib@innerbib\@empty
\bibitem [{\citenamefont {Aad}\ \emph {et~al.}(2012)\citenamefont {Aad} \emph {et~al.}}]{Aad:2012tfa}%
  \BibitemOpen
  \bibfield  {author} {\bibinfo {author} {\bibfnamefont {Georges}\ \bibnamefont {Aad}} \emph {et~al.} (\bibinfo {collaboration} {ATLAS}),\ }\bibfield  {title} {\enquote {\bibinfo {title} {{Observation of a new particle in the search for the Standard Model Higgs boson with the ATLAS detector at the LHC}},}\ }\href {\doibase 10.1016/j.physletb.2012.08.020} {\bibfield  {journal} {\bibinfo  {journal} {Phys. Lett. B}\ }\textbf {\bibinfo {volume} {716}},\ \bibinfo {pages} {1--29} (\bibinfo {year} {2012})},\ \Eprint {http://arxiv.org/abs/1207.7214} {arXiv:1207.7214 [hep-ex]} \BibitemShut {NoStop}%
\bibitem [{\citenamefont {Chatrchyan}\ \emph {et~al.}(2012)\citenamefont {Chatrchyan} \emph {et~al.}}]{Chatrchyan:2012ufa}%
  \BibitemOpen
  \bibfield  {author} {\bibinfo {author} {\bibfnamefont {Serguei}\ \bibnamefont {Chatrchyan}} \emph {et~al.} (\bibinfo {collaboration} {CMS}),\ }\bibfield  {title} {\enquote {\bibinfo {title} {{Observation of a New Boson at a Mass of 125 GeV with the CMS Experiment at the LHC}},}\ }\href {\doibase 10.1016/j.physletb.2012.08.021} {\bibfield  {journal} {\bibinfo  {journal} {Phys. Lett. B}\ }\textbf {\bibinfo {volume} {716}},\ \bibinfo {pages} {30--61} (\bibinfo {year} {2012})},\ \Eprint {http://arxiv.org/abs/1207.7235} {arXiv:1207.7235 [hep-ex]} \BibitemShut {NoStop}%
\bibitem [{\citenamefont {Aguilar-Saavedra}\ \emph {et~al.}(2017)\citenamefont {Aguilar-Saavedra}, \citenamefont {Collins},\ and\ \citenamefont {Mishra}}]{Aguilar-Saavedra:2017rzt}%
  \BibitemOpen
  \bibfield  {author} {\bibinfo {author} {\bibfnamefont {J.~A.}\ \bibnamefont {Aguilar-Saavedra}}, \bibinfo {author} {\bibfnamefont {Jack~H.}\ \bibnamefont {Collins}}, \ and\ \bibinfo {author} {\bibfnamefont {Rashmish~K.}\ \bibnamefont {Mishra}},\ }\bibfield  {title} {\enquote {\bibinfo {title} {{A generic anti-QCD jet tagger}},}\ }\href {\doibase 10.1007/JHEP11(2017)163} {\bibfield  {journal} {\bibinfo  {journal} {JHEP}\ }\textbf {\bibinfo {volume} {11}},\ \bibinfo {pages} {163} (\bibinfo {year} {2017})},\ \Eprint {http://arxiv.org/abs/1709.01087} {arXiv:1709.01087 [hep-ph]} \BibitemShut {NoStop}%
\bibitem [{\citenamefont {Collins}\ \emph {et~al.}(2018)\citenamefont {Collins}, \citenamefont {Howe},\ and\ \citenamefont {Nachman}}]{Collins:2018epr}%
  \BibitemOpen
  \bibfield  {author} {\bibinfo {author} {\bibfnamefont {Jack~H.}\ \bibnamefont {Collins}}, \bibinfo {author} {\bibfnamefont {Kiel}\ \bibnamefont {Howe}}, \ and\ \bibinfo {author} {\bibfnamefont {Benjamin}\ \bibnamefont {Nachman}},\ }\bibfield  {title} {\enquote {\bibinfo {title} {{Anomaly Detection for Resonant New Physics with Machine Learning}},}\ }\href {\doibase 10.1103/PhysRevLett.121.241803} {\bibfield  {journal} {\bibinfo  {journal} {Phys. Rev. Lett.}\ }\textbf {\bibinfo {volume} {121}},\ \bibinfo {pages} {241803} (\bibinfo {year} {2018})},\ \Eprint {http://arxiv.org/abs/1805.02664} {arXiv:1805.02664 [hep-ph]} \BibitemShut {NoStop}%
\bibitem [{\citenamefont {D'Agnolo}\ and\ \citenamefont {Wulzer}(2019)}]{DAgnolo:2018cun}%
  \BibitemOpen
  \bibfield  {author} {\bibinfo {author} {\bibfnamefont {Raffaele~Tito}\ \bibnamefont {D'Agnolo}}\ and\ \bibinfo {author} {\bibfnamefont {Andrea}\ \bibnamefont {Wulzer}},\ }\bibfield  {title} {\enquote {\bibinfo {title} {{Learning New Physics from a Machine}},}\ }\href {\doibase 10.1103/PhysRevD.99.015014} {\bibfield  {journal} {\bibinfo  {journal} {Phys. Rev. D}\ }\textbf {\bibinfo {volume} {99}},\ \bibinfo {pages} {015014} (\bibinfo {year} {2019})},\ \Eprint {http://arxiv.org/abs/1806.02350} {arXiv:1806.02350 [hep-ph]} \BibitemShut {NoStop}%
\bibitem [{\citenamefont {De~Simone}\ and\ \citenamefont {Jacques}(2019)}]{DeSimone:2018efk}%
  \BibitemOpen
  \bibfield  {author} {\bibinfo {author} {\bibfnamefont {Andrea}\ \bibnamefont {De~Simone}}\ and\ \bibinfo {author} {\bibfnamefont {Thomas}\ \bibnamefont {Jacques}},\ }\bibfield  {title} {\enquote {\bibinfo {title} {{Guiding New Physics Searches with Unsupervised Learning}},}\ }\href {\doibase 10.1140/epjc/s10052-019-6787-3} {\bibfield  {journal} {\bibinfo  {journal} {Eur. Phys. J. C}\ }\textbf {\bibinfo {volume} {79}},\ \bibinfo {pages} {289} (\bibinfo {year} {2019})},\ \Eprint {http://arxiv.org/abs/1807.06038} {arXiv:1807.06038 [hep-ph]} \BibitemShut {NoStop}%
\bibitem [{\citenamefont {Hajer}\ \emph {et~al.}(2020)\citenamefont {Hajer}, \citenamefont {Li}, \citenamefont {Liu},\ and\ \citenamefont {Wang}}]{Hajer:2018kqm}%
  \BibitemOpen
  \bibfield  {author} {\bibinfo {author} {\bibfnamefont {Jan}\ \bibnamefont {Hajer}}, \bibinfo {author} {\bibfnamefont {Ying-Ying}\ \bibnamefont {Li}}, \bibinfo {author} {\bibfnamefont {Tao}\ \bibnamefont {Liu}}, \ and\ \bibinfo {author} {\bibfnamefont {He}~\bibnamefont {Wang}},\ }\bibfield  {title} {\enquote {\bibinfo {title} {{Novelty Detection Meets Collider Physics}},}\ }\href {\doibase 10.1103/PhysRevD.101.076015} {\bibfield  {journal} {\bibinfo  {journal} {Phys. Rev. D}\ }\textbf {\bibinfo {volume} {101}},\ \bibinfo {pages} {076015} (\bibinfo {year} {2020})},\ \Eprint {http://arxiv.org/abs/1807.10261} {arXiv:1807.10261 [hep-ph]} \BibitemShut {NoStop}%
\bibitem [{\citenamefont {Farina}\ \emph {et~al.}(2020)\citenamefont {Farina}, \citenamefont {Nakai},\ and\ \citenamefont {Shih}}]{Farina:2018fyg}%
  \BibitemOpen
  \bibfield  {author} {\bibinfo {author} {\bibfnamefont {Marco}\ \bibnamefont {Farina}}, \bibinfo {author} {\bibfnamefont {Yuichiro}\ \bibnamefont {Nakai}}, \ and\ \bibinfo {author} {\bibfnamefont {David}\ \bibnamefont {Shih}},\ }\bibfield  {title} {\enquote {\bibinfo {title} {{Searching for New Physics with Deep Autoencoders}},}\ }\href {\doibase 10.1103/PhysRevD.101.075021} {\bibfield  {journal} {\bibinfo  {journal} {Phys. Rev. D}\ }\textbf {\bibinfo {volume} {101}},\ \bibinfo {pages} {075021} (\bibinfo {year} {2020})},\ \Eprint {http://arxiv.org/abs/1808.08992} {arXiv:1808.08992 [hep-ph]} \BibitemShut {NoStop}%
\bibitem [{\citenamefont {Heimel}\ \emph {et~al.}(2019)\citenamefont {Heimel}, \citenamefont {Kasieczka}, \citenamefont {Plehn},\ and\ \citenamefont {Thompson}}]{Heimel:2018mkt}%
  \BibitemOpen
  \bibfield  {author} {\bibinfo {author} {\bibfnamefont {Theo}\ \bibnamefont {Heimel}}, \bibinfo {author} {\bibfnamefont {Gregor}\ \bibnamefont {Kasieczka}}, \bibinfo {author} {\bibfnamefont {Tilman}\ \bibnamefont {Plehn}}, \ and\ \bibinfo {author} {\bibfnamefont {Jennifer~M.}\ \bibnamefont {Thompson}},\ }\bibfield  {title} {\enquote {\bibinfo {title} {{QCD or What?}}}\ }\href {\doibase 10.21468/SciPostPhys.6.3.030} {\bibfield  {journal} {\bibinfo  {journal} {SciPost Phys.}\ }\textbf {\bibinfo {volume} {6}},\ \bibinfo {pages} {030} (\bibinfo {year} {2019})},\ \Eprint {http://arxiv.org/abs/1808.08979} {arXiv:1808.08979 [hep-ph]} \BibitemShut {NoStop}%
\bibitem [{\citenamefont {Casa}\ and\ \citenamefont {Menardi}(2018)}]{1809.02977}%
  \BibitemOpen
  \bibfield  {author} {\bibinfo {author} {\bibfnamefont {Alessandro}\ \bibnamefont {Casa}}\ and\ \bibinfo {author} {\bibfnamefont {Giovanna}\ \bibnamefont {Menardi}},\ }\bibfield  {title} {\enquote {\bibinfo {title} {{Nonparametric semisupervised classification for signal detection in high energy physics}},}\ }\href@noop {} {\  (\bibinfo {year} {2018})},\ \Eprint {http://arxiv.org/abs/1809.02977} {arXiv:1809.02977 [stat.AP]} \BibitemShut {NoStop}%
\bibitem [{\citenamefont {Cerri}\ \emph {et~al.}(2019)\citenamefont {Cerri}, \citenamefont {Nguyen}, \citenamefont {Pierini}, \citenamefont {Spiropulu},\ and\ \citenamefont {Vlimant}}]{Cerri:2018anq}%
  \BibitemOpen
  \bibfield  {author} {\bibinfo {author} {\bibfnamefont {Olmo}\ \bibnamefont {Cerri}}, \bibinfo {author} {\bibfnamefont {Thong~Q.}\ \bibnamefont {Nguyen}}, \bibinfo {author} {\bibfnamefont {Maurizio}\ \bibnamefont {Pierini}}, \bibinfo {author} {\bibfnamefont {Maria}\ \bibnamefont {Spiropulu}}, \ and\ \bibinfo {author} {\bibfnamefont {Jean-Roch}\ \bibnamefont {Vlimant}},\ }\bibfield  {title} {\enquote {\bibinfo {title} {{Variational Autoencoders for New Physics Mining at the Large Hadron Collider}},}\ }\href {\doibase 10.1007/JHEP05(2019)036} {\bibfield  {journal} {\bibinfo  {journal} {JHEP}\ }\textbf {\bibinfo {volume} {05}},\ \bibinfo {pages} {036} (\bibinfo {year} {2019})},\ \Eprint {http://arxiv.org/abs/1811.10276} {arXiv:1811.10276 [hep-ex]} \BibitemShut {NoStop}%
\bibitem [{\citenamefont {Collins}\ \emph {et~al.}(2019)\citenamefont {Collins}, \citenamefont {Howe},\ and\ \citenamefont {Nachman}}]{Collins:2019jip}%
  \BibitemOpen
  \bibfield  {author} {\bibinfo {author} {\bibfnamefont {Jack~H.}\ \bibnamefont {Collins}}, \bibinfo {author} {\bibfnamefont {Kiel}\ \bibnamefont {Howe}}, \ and\ \bibinfo {author} {\bibfnamefont {Benjamin}\ \bibnamefont {Nachman}},\ }\bibfield  {title} {\enquote {\bibinfo {title} {{Extending the search for new resonances with machine learning}},}\ }\href {\doibase 10.1103/PhysRevD.99.014038} {\bibfield  {journal} {\bibinfo  {journal} {Phys. Rev. D}\ }\textbf {\bibinfo {volume} {99}},\ \bibinfo {pages} {014038} (\bibinfo {year} {2019})},\ \Eprint {http://arxiv.org/abs/1902.02634} {arXiv:1902.02634 [hep-ph]} \BibitemShut {NoStop}%
\bibitem [{\citenamefont {Roy}\ and\ \citenamefont {Vijay}(2019)}]{Roy:2019jae}%
  \BibitemOpen
  \bibfield  {author} {\bibinfo {author} {\bibfnamefont {Tuhin~S.}\ \bibnamefont {Roy}}\ and\ \bibinfo {author} {\bibfnamefont {Aravind~H.}\ \bibnamefont {Vijay}},\ }\bibfield  {title} {\enquote {\bibinfo {title} {{A robust anomaly finder based on autoencoders}},}\ }\href@noop {} {\  (\bibinfo {year} {2019})},\ \Eprint {http://arxiv.org/abs/1903.02032} {arXiv:1903.02032 [hep-ph]} \BibitemShut {NoStop}%
\bibitem [{\citenamefont {Dillon}\ \emph {et~al.}(2019)\citenamefont {Dillon}, \citenamefont {Faroughy},\ and\ \citenamefont {Kamenik}}]{Dillon:2019cqt}%
  \BibitemOpen
  \bibfield  {author} {\bibinfo {author} {\bibfnamefont {Barry~M.}\ \bibnamefont {Dillon}}, \bibinfo {author} {\bibfnamefont {Darius~A.}\ \bibnamefont {Faroughy}}, \ and\ \bibinfo {author} {\bibfnamefont {Jernej~F.}\ \bibnamefont {Kamenik}},\ }\bibfield  {title} {\enquote {\bibinfo {title} {{Uncovering latent jet substructure}},}\ }\href {\doibase 10.1103/PhysRevD.100.056002} {\bibfield  {journal} {\bibinfo  {journal} {Phys. Rev. D}\ }\textbf {\bibinfo {volume} {100}},\ \bibinfo {pages} {056002} (\bibinfo {year} {2019})},\ \Eprint {http://arxiv.org/abs/1904.04200} {arXiv:1904.04200 [hep-ph]} \BibitemShut {NoStop}%
\bibitem [{\citenamefont {Blance}\ \emph {et~al.}(2019)\citenamefont {Blance}, \citenamefont {Spannowsky},\ and\ \citenamefont {Waite}}]{Blance:2019ibf}%
  \BibitemOpen
  \bibfield  {author} {\bibinfo {author} {\bibfnamefont {Andrew}\ \bibnamefont {Blance}}, \bibinfo {author} {\bibfnamefont {Michael}\ \bibnamefont {Spannowsky}}, \ and\ \bibinfo {author} {\bibfnamefont {Philip}\ \bibnamefont {Waite}},\ }\bibfield  {title} {\enquote {\bibinfo {title} {{Adversarially-trained autoencoders for robust unsupervised new physics searches}},}\ }\href {\doibase 10.1007/JHEP10(2019)047} {\bibfield  {journal} {\bibinfo  {journal} {JHEP}\ }\textbf {\bibinfo {volume} {10}},\ \bibinfo {pages} {047} (\bibinfo {year} {2019})},\ \Eprint {http://arxiv.org/abs/1905.10384} {arXiv:1905.10384 [hep-ph]} \BibitemShut {NoStop}%
\bibitem [{\citenamefont {Rom\~ao Crispim}\ \emph {et~al.}(2020)\citenamefont {Rom\~ao Crispim}, \citenamefont {Castro}, \citenamefont {Pedro},\ and\ \citenamefont {Vale}}]{Romao:2019dvs}%
  \BibitemOpen
  \bibfield  {author} {\bibinfo {author} {\bibfnamefont {M.}~\bibnamefont {Rom\~ao Crispim}}, \bibinfo {author} {\bibfnamefont {N.~F.}\ \bibnamefont {Castro}}, \bibinfo {author} {\bibfnamefont {R.}~\bibnamefont {Pedro}}, \ and\ \bibinfo {author} {\bibfnamefont {T.}~\bibnamefont {Vale}},\ }\bibfield  {title} {\enquote {\bibinfo {title} {{Transferability of Deep Learning Models in Searches for New Physics at Colliders}},}\ }\href {\doibase 10.1103/PhysRevD.101.035042} {\bibfield  {journal} {\bibinfo  {journal} {Phys. Rev. D}\ }\textbf {\bibinfo {volume} {101}},\ \bibinfo {pages} {035042} (\bibinfo {year} {2020})},\ \Eprint {http://arxiv.org/abs/1912.04220} {arXiv:1912.04220 [hep-ph]} \BibitemShut {NoStop}%
\bibitem [{\citenamefont {Mullin}\ \emph {et~al.}(2021)\citenamefont {Mullin}, \citenamefont {Nicholls}, \citenamefont {Pacey}, \citenamefont {Parker}, \citenamefont {White},\ and\ \citenamefont {Williams}}]{Mullin:2019mmh}%
  \BibitemOpen
  \bibfield  {author} {\bibinfo {author} {\bibfnamefont {Anna}\ \bibnamefont {Mullin}}, \bibinfo {author} {\bibfnamefont {Stuart}\ \bibnamefont {Nicholls}}, \bibinfo {author} {\bibfnamefont {Holly}\ \bibnamefont {Pacey}}, \bibinfo {author} {\bibfnamefont {Michael}\ \bibnamefont {Parker}}, \bibinfo {author} {\bibfnamefont {Martin}\ \bibnamefont {White}}, \ and\ \bibinfo {author} {\bibfnamefont {Sarah}\ \bibnamefont {Williams}},\ }\bibfield  {title} {\enquote {\bibinfo {title} {{Does SUSY have friends? A new approach for LHC event analysis}},}\ }\href {\doibase 10.1007/JHEP02(2021)160} {\bibfield  {journal} {\bibinfo  {journal} {JHEP}\ }\textbf {\bibinfo {volume} {02}},\ \bibinfo {pages} {160} (\bibinfo {year} {2021})},\ \Eprint {http://arxiv.org/abs/1912.10625} {arXiv:1912.10625 [hep-ph]} \BibitemShut {NoStop}%
\bibitem [{\citenamefont {D'Agnolo}\ \emph {et~al.}(2021)\citenamefont {D'Agnolo}, \citenamefont {Grosso}, \citenamefont {Pierini}, \citenamefont {Wulzer},\ and\ \citenamefont {Zanetti}}]{DAgnolo:2019vbw}%
  \BibitemOpen
  \bibfield  {author} {\bibinfo {author} {\bibfnamefont {Raffaele~Tito}\ \bibnamefont {D'Agnolo}}, \bibinfo {author} {\bibfnamefont {Gaia}\ \bibnamefont {Grosso}}, \bibinfo {author} {\bibfnamefont {Maurizio}\ \bibnamefont {Pierini}}, \bibinfo {author} {\bibfnamefont {Andrea}\ \bibnamefont {Wulzer}}, \ and\ \bibinfo {author} {\bibfnamefont {Marco}\ \bibnamefont {Zanetti}},\ }\bibfield  {title} {\enquote {\bibinfo {title} {{Learning multivariate new physics}},}\ }\href {\doibase 10.1140/epjc/s10052-021-08853-y} {\bibfield  {journal} {\bibinfo  {journal} {Eur. Phys. J. C}\ }\textbf {\bibinfo {volume} {81}},\ \bibinfo {pages} {89} (\bibinfo {year} {2021})},\ \Eprint {http://arxiv.org/abs/1912.12155} {arXiv:1912.12155 [hep-ph]} \BibitemShut {NoStop}%
\bibitem [{\citenamefont {Nachman}\ and\ \citenamefont {Shih}(2020)}]{Nachman:2020lpy}%
  \BibitemOpen
  \bibfield  {author} {\bibinfo {author} {\bibfnamefont {Benjamin}\ \bibnamefont {Nachman}}\ and\ \bibinfo {author} {\bibfnamefont {David}\ \bibnamefont {Shih}},\ }\bibfield  {title} {\enquote {\bibinfo {title} {{Anomaly Detection with Density Estimation}},}\ }\href {\doibase 10.1103/PhysRevD.101.075042} {\bibfield  {journal} {\bibinfo  {journal} {Phys. Rev. D}\ }\textbf {\bibinfo {volume} {101}},\ \bibinfo {pages} {075042} (\bibinfo {year} {2020})},\ \Eprint {http://arxiv.org/abs/2001.04990} {arXiv:2001.04990 [hep-ph]} \BibitemShut {NoStop}%
\bibitem [{\citenamefont {Andreassen}\ \emph {et~al.}(2020{\natexlab{a}})\citenamefont {Andreassen}, \citenamefont {Nachman},\ and\ \citenamefont {Shih}}]{Andreassen:2020nkr}%
  \BibitemOpen
  \bibfield  {author} {\bibinfo {author} {\bibfnamefont {Anders}\ \bibnamefont {Andreassen}}, \bibinfo {author} {\bibfnamefont {Benjamin}\ \bibnamefont {Nachman}}, \ and\ \bibinfo {author} {\bibfnamefont {David}\ \bibnamefont {Shih}},\ }\bibfield  {title} {\enquote {\bibinfo {title} {{Simulation Assisted Likelihood-free Anomaly Detection}},}\ }\href {\doibase 10.1103/PhysRevD.101.095004} {\bibfield  {journal} {\bibinfo  {journal} {Phys. Rev. D}\ }\textbf {\bibinfo {volume} {101}},\ \bibinfo {pages} {095004} (\bibinfo {year} {2020}{\natexlab{a}})},\ \Eprint {http://arxiv.org/abs/2001.05001} {arXiv:2001.05001 [hep-ph]} \BibitemShut {NoStop}%
\bibitem [{\citenamefont {Amram}\ and\ \citenamefont {Suarez}(2021)}]{Amram:2020ykb}%
  \BibitemOpen
  \bibfield  {author} {\bibinfo {author} {\bibfnamefont {Oz}~\bibnamefont {Amram}}\ and\ \bibinfo {author} {\bibfnamefont {Cristina~Mantilla}\ \bibnamefont {Suarez}},\ }\bibfield  {title} {\enquote {\bibinfo {title} {{Tag N\textquoteright{} Train: a technique to train improved classifiers on unlabeled data}},}\ }\href {\doibase 10.1007/JHEP01(2021)153} {\bibfield  {journal} {\bibinfo  {journal} {JHEP}\ }\textbf {\bibinfo {volume} {01}},\ \bibinfo {pages} {153} (\bibinfo {year} {2021})},\ \Eprint {http://arxiv.org/abs/2002.12376} {arXiv:2002.12376 [hep-ph]} \BibitemShut {NoStop}%
\bibitem [{\citenamefont {Crispim Rom\~ao}\ \emph {et~al.}(2021{\natexlab{a}})\citenamefont {Crispim Rom\~ao}, \citenamefont {Castro}, \citenamefont {Milhano}, \citenamefont {Pedro},\ and\ \citenamefont {Vale}}]{Romao:2020ojy}%
  \BibitemOpen
  \bibfield  {author} {\bibinfo {author} {\bibfnamefont {M.}~\bibnamefont {Crispim Rom\~ao}}, \bibinfo {author} {\bibfnamefont {N.~F.}\ \bibnamefont {Castro}}, \bibinfo {author} {\bibfnamefont {J.~G.}\ \bibnamefont {Milhano}}, \bibinfo {author} {\bibfnamefont {R.}~\bibnamefont {Pedro}}, \ and\ \bibinfo {author} {\bibfnamefont {T.}~\bibnamefont {Vale}},\ }\bibfield  {title} {\enquote {\bibinfo {title} {{Use of a generalized energy Mover\textquoteright{}s distance in the search for rare phenomena at colliders}},}\ }\href {\doibase 10.1140/epjc/s10052-021-08891-6} {\bibfield  {journal} {\bibinfo  {journal} {Eur. Phys. J. C}\ }\textbf {\bibinfo {volume} {81}},\ \bibinfo {pages} {192} (\bibinfo {year} {2021}{\natexlab{a}})},\ \Eprint {http://arxiv.org/abs/2004.09360} {arXiv:2004.09360 [hep-ph]} \BibitemShut {NoStop}%
\bibitem [{\citenamefont {Knapp}\ \emph {et~al.}(2021)\citenamefont {Knapp}, \citenamefont {Cerri}, \citenamefont {Dissertori}, \citenamefont {Nguyen}, \citenamefont {Pierini},\ and\ \citenamefont {Vlimant}}]{Knapp:2020dde}%
  \BibitemOpen
  \bibfield  {author} {\bibinfo {author} {\bibfnamefont {Oliver}\ \bibnamefont {Knapp}}, \bibinfo {author} {\bibfnamefont {Olmo}\ \bibnamefont {Cerri}}, \bibinfo {author} {\bibfnamefont {Guenther}\ \bibnamefont {Dissertori}}, \bibinfo {author} {\bibfnamefont {Thong~Q.}\ \bibnamefont {Nguyen}}, \bibinfo {author} {\bibfnamefont {Maurizio}\ \bibnamefont {Pierini}}, \ and\ \bibinfo {author} {\bibfnamefont {Jean-Roch}\ \bibnamefont {Vlimant}},\ }\bibfield  {title} {\enquote {\bibinfo {title} {{Adversarially Learned Anomaly Detection on CMS Open Data: re-discovering the top quark}},}\ }\href {\doibase 10.1140/epjp/s13360-021-01109-4} {\bibfield  {journal} {\bibinfo  {journal} {Eur. Phys. J. Plus}\ }\textbf {\bibinfo {volume} {136}},\ \bibinfo {pages} {236} (\bibinfo {year} {2021})},\ \Eprint {http://arxiv.org/abs/2005.01598} {arXiv:2005.01598 [hep-ex]} \BibitemShut {NoStop}%
\bibitem [{\citenamefont {Aad}\ \emph {et~al.}(2020)\citenamefont {Aad} \emph {et~al.}}]{ATLAS:2020iwa}%
  \BibitemOpen
  \bibfield  {author} {\bibinfo {author} {\bibfnamefont {Georges}\ \bibnamefont {Aad}} \emph {et~al.} (\bibinfo {collaboration} {ATLAS}),\ }\bibfield  {title} {\enquote {\bibinfo {title} {{Dijet resonance search with weak supervision using $\sqrt{s}=13$ TeV $pp$ collisions in the ATLAS detector}},}\ }\href {\doibase 10.1103/PhysRevLett.125.131801} {\bibfield  {journal} {\bibinfo  {journal} {Phys. Rev. Lett.}\ }\textbf {\bibinfo {volume} {125}},\ \bibinfo {pages} {131801} (\bibinfo {year} {2020})},\ \Eprint {http://arxiv.org/abs/2005.02983} {arXiv:2005.02983 [hep-ex]} \BibitemShut {NoStop}%
\bibitem [{\citenamefont {Dillon}\ \emph {et~al.}(2020)\citenamefont {Dillon}, \citenamefont {Faroughy}, \citenamefont {Kamenik},\ and\ \citenamefont {Szewc}}]{Dillon:2020quc}%
  \BibitemOpen
  \bibfield  {author} {\bibinfo {author} {\bibfnamefont {B.~M.}\ \bibnamefont {Dillon}}, \bibinfo {author} {\bibfnamefont {D.~A.}\ \bibnamefont {Faroughy}}, \bibinfo {author} {\bibfnamefont {J.~F.}\ \bibnamefont {Kamenik}}, \ and\ \bibinfo {author} {\bibfnamefont {M.}~\bibnamefont {Szewc}},\ }\bibfield  {title} {\enquote {\bibinfo {title} {{Learning the latent structure of collider events}},}\ }\href {\doibase 10.1007/JHEP10(2020)206} {\bibfield  {journal} {\bibinfo  {journal} {JHEP}\ }\textbf {\bibinfo {volume} {10}},\ \bibinfo {pages} {206} (\bibinfo {year} {2020})},\ \Eprint {http://arxiv.org/abs/2005.12319} {arXiv:2005.12319 [hep-ph]} \BibitemShut {NoStop}%
\bibitem [{\citenamefont {Crispim Rom\~ao}\ \emph {et~al.}(2021{\natexlab{b}})\citenamefont {Crispim Rom\~ao}, \citenamefont {Castro},\ and\ \citenamefont {Pedro}}]{CrispimRomao:2020ucc}%
  \BibitemOpen
  \bibfield  {author} {\bibinfo {author} {\bibfnamefont {M.}~\bibnamefont {Crispim Rom\~ao}}, \bibinfo {author} {\bibfnamefont {N.~F.}\ \bibnamefont {Castro}}, \ and\ \bibinfo {author} {\bibfnamefont {R.}~\bibnamefont {Pedro}},\ }\bibfield  {title} {\enquote {\bibinfo {title} {{Finding New Physics without learning about it: Anomaly Detection as a tool for Searches at Colliders}},}\ }\href {\doibase 10.1140/epjc/s10052-021-09813-2} {\bibfield  {journal} {\bibinfo  {journal} {Eur. Phys. J. C}\ }\textbf {\bibinfo {volume} {81}},\ \bibinfo {pages} {27} (\bibinfo {year} {2021}{\natexlab{b}})},\ \bibinfo {note} {[Erratum: Eur.Phys.J.C 81, 1020 (2021)]},\ \Eprint {http://arxiv.org/abs/2006.05432} {arXiv:2006.05432 [hep-ph]} \BibitemShut {NoStop}%
\bibitem [{\citenamefont {Cheng}\ \emph {et~al.}(2023)\citenamefont {Cheng}, \citenamefont {Arguin}, \citenamefont {Leissner-Martin}, \citenamefont {Pilette},\ and\ \citenamefont {Golling}}]{Cheng:2020dal}%
  \BibitemOpen
  \bibfield  {author} {\bibinfo {author} {\bibfnamefont {Taoli}\ \bibnamefont {Cheng}}, \bibinfo {author} {\bibfnamefont {Jean-Fran\c{c}ois}\ \bibnamefont {Arguin}}, \bibinfo {author} {\bibfnamefont {Julien}\ \bibnamefont {Leissner-Martin}}, \bibinfo {author} {\bibfnamefont {Jacinthe}\ \bibnamefont {Pilette}}, \ and\ \bibinfo {author} {\bibfnamefont {Tobias}\ \bibnamefont {Golling}},\ }\bibfield  {title} {\enquote {\bibinfo {title} {{Variational autoencoders for anomalous jet tagging}},}\ }\href {\doibase 10.1103/PhysRevD.107.016002} {\bibfield  {journal} {\bibinfo  {journal} {Phys. Rev. D}\ }\textbf {\bibinfo {volume} {107}},\ \bibinfo {pages} {016002} (\bibinfo {year} {2023})},\ \Eprint {http://arxiv.org/abs/2007.01850} {arXiv:2007.01850 [hep-ph]} \BibitemShut {NoStop}%
\bibitem [{\citenamefont {Khosa}\ and\ \citenamefont {Sanz}(2023)}]{Khosa:2020qrz}%
  \BibitemOpen
  \bibfield  {author} {\bibinfo {author} {\bibfnamefont {Charanjit~Kaur}\ \bibnamefont {Khosa}}\ and\ \bibinfo {author} {\bibfnamefont {Veronica}\ \bibnamefont {Sanz}},\ }\bibfield  {title} {\enquote {\bibinfo {title} {{Anomaly Awareness}},}\ }\href {\doibase 10.21468/SciPostPhys.15.2.053} {\bibfield  {journal} {\bibinfo  {journal} {SciPost Phys.}\ }\textbf {\bibinfo {volume} {15}},\ \bibinfo {pages} {053} (\bibinfo {year} {2023})},\ \Eprint {http://arxiv.org/abs/2007.14462} {arXiv:2007.14462 [cs.LG]} \BibitemShut {NoStop}%
\bibitem [{\citenamefont {Thaprasop}\ \emph {et~al.}(2021)\citenamefont {Thaprasop}, \citenamefont {Zhou}, \citenamefont {Steinheimer},\ and\ \citenamefont {Herold}}]{Thaprasop:2020mzp}%
  \BibitemOpen
  \bibfield  {author} {\bibinfo {author} {\bibfnamefont {Punnathat}\ \bibnamefont {Thaprasop}}, \bibinfo {author} {\bibfnamefont {Kai}\ \bibnamefont {Zhou}}, \bibinfo {author} {\bibfnamefont {Jan}\ \bibnamefont {Steinheimer}}, \ and\ \bibinfo {author} {\bibfnamefont {Christoph}\ \bibnamefont {Herold}},\ }\bibfield  {title} {\enquote {\bibinfo {title} {{Unsupervised Outlier Detection in Heavy-Ion Collisions}},}\ }\href {\doibase 10.1088/1402-4896/abf214} {\bibfield  {journal} {\bibinfo  {journal} {Phys. Scripta}\ }\textbf {\bibinfo {volume} {96}},\ \bibinfo {pages} {064003} (\bibinfo {year} {2021})},\ \Eprint {http://arxiv.org/abs/2007.15830} {arXiv:2007.15830 [hep-ex]} \BibitemShut {NoStop}%
\bibitem [{\citenamefont {Aguilar-Saavedra}\ \emph {et~al.}(2021)\citenamefont {Aguilar-Saavedra}, \citenamefont {Joaquim},\ and\ \citenamefont {Seabra}}]{Aguilar-Saavedra:2020uhm}%
  \BibitemOpen
  \bibfield  {author} {\bibinfo {author} {\bibfnamefont {J.~A.}\ \bibnamefont {Aguilar-Saavedra}}, \bibinfo {author} {\bibfnamefont {F.~R.}\ \bibnamefont {Joaquim}}, \ and\ \bibinfo {author} {\bibfnamefont {J.~F.}\ \bibnamefont {Seabra}},\ }\bibfield  {title} {\enquote {\bibinfo {title} {{Mass Unspecific Supervised Tagging (MUST) for boosted jets}},}\ }\href {\doibase 10.1007/JHEP03(2021)012} {\bibfield  {journal} {\bibinfo  {journal} {JHEP}\ }\textbf {\bibinfo {volume} {03}},\ \bibinfo {pages} {012} (\bibinfo {year} {2021})},\ \bibinfo {note} {[Erratum: JHEP 04, 133 (2021)]},\ \Eprint {http://arxiv.org/abs/2008.12792} {arXiv:2008.12792 [hep-ph]} \BibitemShut {NoStop}%
\bibitem [{\citenamefont {Alexander}\ \emph {et~al.}(2020)\citenamefont {Alexander}, \citenamefont {Gleyzer}, \citenamefont {Parul}, \citenamefont {Reddy}, \citenamefont {Toomey}, \citenamefont {Usai},\ and\ \citenamefont {Von~Klar}}]{Alexander:2020mbx}%
  \BibitemOpen
  \bibfield  {author} {\bibinfo {author} {\bibfnamefont {Stephon}\ \bibnamefont {Alexander}}, \bibinfo {author} {\bibfnamefont {Sergei}\ \bibnamefont {Gleyzer}}, \bibinfo {author} {\bibfnamefont {Hanna}\ \bibnamefont {Parul}}, \bibinfo {author} {\bibfnamefont {Pranath}\ \bibnamefont {Reddy}}, \bibinfo {author} {\bibfnamefont {Michael~W.}\ \bibnamefont {Toomey}}, \bibinfo {author} {\bibfnamefont {Emanuele}\ \bibnamefont {Usai}}, \ and\ \bibinfo {author} {\bibfnamefont {Ryker}\ \bibnamefont {Von~Klar}},\ }\bibfield  {title} {\enquote {\bibinfo {title} {{Decoding Dark Matter Substructure without Supervision}},}\ }\href@noop {} {\  (\bibinfo {year} {2020})},\ \Eprint {http://arxiv.org/abs/2008.12731} {arXiv:2008.12731 [astro-ph.CO]} \BibitemShut {NoStop}%
\bibitem [{\citenamefont {Benkendorfer}\ \emph {et~al.}(2021)\citenamefont {Benkendorfer}, \citenamefont {Pottier},\ and\ \citenamefont {Nachman}}]{Benkendorfer:2020gek}%
  \BibitemOpen
  \bibfield  {author} {\bibinfo {author} {\bibfnamefont {Kees}\ \bibnamefont {Benkendorfer}}, \bibinfo {author} {\bibfnamefont {Luc~Le}\ \bibnamefont {Pottier}}, \ and\ \bibinfo {author} {\bibfnamefont {Benjamin}\ \bibnamefont {Nachman}},\ }\bibfield  {title} {\enquote {\bibinfo {title} {{Simulation-assisted decorrelation for resonant anomaly detection}},}\ }\href {\doibase 10.1103/PhysRevD.104.035003} {\bibfield  {journal} {\bibinfo  {journal} {Phys. Rev. D}\ }\textbf {\bibinfo {volume} {104}},\ \bibinfo {pages} {035003} (\bibinfo {year} {2021})},\ \Eprint {http://arxiv.org/abs/2009.02205} {arXiv:2009.02205 [hep-ph]} \BibitemShut {NoStop}%
\bibitem [{\citenamefont {Pol}\ \emph {et~al.}(2020)\citenamefont {Pol}, \citenamefont {Berger}, \citenamefont {Cerminara}, \citenamefont {Germain},\ and\ \citenamefont {Pierini}}]{Pol:2020weg}%
  \BibitemOpen
  \bibfield  {author} {\bibinfo {author} {\bibfnamefont {Adrian~Alan}\ \bibnamefont {Pol}}, \bibinfo {author} {\bibfnamefont {Victor}\ \bibnamefont {Berger}}, \bibinfo {author} {\bibfnamefont {Gianluca}\ \bibnamefont {Cerminara}}, \bibinfo {author} {\bibfnamefont {Cecile}\ \bibnamefont {Germain}}, \ and\ \bibinfo {author} {\bibfnamefont {Maurizio}\ \bibnamefont {Pierini}},\ }\bibfield  {title} {\enquote {\bibinfo {title} {{Anomaly Detection With Conditional Variational Autoencoders}},}\ }in\ \href@noop {} {\emph {\bibinfo {booktitle} {{Eighteenth International Conference on Machine Learning and Applications}}}}\ (\bibinfo {year} {2020})\ \Eprint {http://arxiv.org/abs/2010.05531} {arXiv:2010.05531 [cs.LG]} \BibitemShut {NoStop}%
\bibitem [{\citenamefont {Mikuni}\ and\ \citenamefont {Canelli}(2021)}]{Mikuni:2020qds}%
  \BibitemOpen
  \bibfield  {author} {\bibinfo {author} {\bibfnamefont {Vinicius}\ \bibnamefont {Mikuni}}\ and\ \bibinfo {author} {\bibfnamefont {Florencia}\ \bibnamefont {Canelli}},\ }\bibfield  {title} {\enquote {\bibinfo {title} {{Unsupervised clustering for collider physics}},}\ }\href {\doibase 10.1103/PhysRevD.103.092007} {\bibfield  {journal} {\bibinfo  {journal} {Phys. Rev. D}\ }\textbf {\bibinfo {volume} {103}},\ \bibinfo {pages} {092007} (\bibinfo {year} {2021})},\ \Eprint {http://arxiv.org/abs/2010.07106} {arXiv:2010.07106 [physics.data-an]} \BibitemShut {NoStop}%
\bibitem [{\citenamefont {van Beekveld}\ \emph {et~al.}(2021)\citenamefont {van Beekveld}, \citenamefont {Caron}, \citenamefont {Hendriks}, \citenamefont {Jackson}, \citenamefont {Leinweber}, \citenamefont {Otten}, \citenamefont {Patrick}, \citenamefont {Ruiz De~Austri}, \citenamefont {Santoni},\ and\ \citenamefont {White}}]{vanBeekveld:2020txa}%
  \BibitemOpen
  \bibfield  {author} {\bibinfo {author} {\bibfnamefont {Melissa}\ \bibnamefont {van Beekveld}}, \bibinfo {author} {\bibfnamefont {Sascha}\ \bibnamefont {Caron}}, \bibinfo {author} {\bibfnamefont {Luc}\ \bibnamefont {Hendriks}}, \bibinfo {author} {\bibfnamefont {Paul}\ \bibnamefont {Jackson}}, \bibinfo {author} {\bibfnamefont {Adam}\ \bibnamefont {Leinweber}}, \bibinfo {author} {\bibfnamefont {Sydney}\ \bibnamefont {Otten}}, \bibinfo {author} {\bibfnamefont {Riley}\ \bibnamefont {Patrick}}, \bibinfo {author} {\bibfnamefont {Roberto}\ \bibnamefont {Ruiz De~Austri}}, \bibinfo {author} {\bibfnamefont {Marco}\ \bibnamefont {Santoni}}, \ and\ \bibinfo {author} {\bibfnamefont {Martin}\ \bibnamefont {White}},\ }\bibfield  {title} {\enquote {\bibinfo {title} {{Combining outlier analysis algorithms to identify new physics at the LHC}},}\ }\href {\doibase 10.1007/JHEP09(2021)024} {\bibfield  {journal} {\bibinfo  {journal} {JHEP}\ }\textbf {\bibinfo {volume} {09}},\ \bibinfo {pages} {024} (\bibinfo {year} {2021})},\
  \Eprint {http://arxiv.org/abs/2010.07940} {arXiv:2010.07940 [hep-ph]} \BibitemShut {NoStop}%
\bibitem [{\citenamefont {Park}\ \emph {et~al.}(2020)\citenamefont {Park}, \citenamefont {Rankin}, \citenamefont {Udrescu}, \citenamefont {Yunus},\ and\ \citenamefont {Harris}}]{Park:2020pak}%
  \BibitemOpen
  \bibfield  {author} {\bibinfo {author} {\bibfnamefont {Sang~Eon}\ \bibnamefont {Park}}, \bibinfo {author} {\bibfnamefont {Dylan}\ \bibnamefont {Rankin}}, \bibinfo {author} {\bibfnamefont {Silviu-Marian}\ \bibnamefont {Udrescu}}, \bibinfo {author} {\bibfnamefont {Mikaeel}\ \bibnamefont {Yunus}}, \ and\ \bibinfo {author} {\bibfnamefont {Philip}\ \bibnamefont {Harris}},\ }\bibfield  {title} {\enquote {\bibinfo {title} {{Quasi Anomalous Knowledge: Searching for new physics with embedded knowledge}},}\ }\href {\doibase 10.1007/JHEP06(2021)030} {\bibfield  {journal} {\bibinfo  {journal} {JHEP}\ }\textbf {\bibinfo {volume} {21}},\ \bibinfo {pages} {030} (\bibinfo {year} {2020})},\ \Eprint {http://arxiv.org/abs/2011.03550} {arXiv:2011.03550 [hep-ph]} \BibitemShut {NoStop}%
\bibitem [{\citenamefont {Faroughy}(2021)}]{Faroughy:2020gas}%
  \BibitemOpen
  \bibfield  {author} {\bibinfo {author} {\bibfnamefont {Darius~A.}\ \bibnamefont {Faroughy}},\ }\bibfield  {title} {\enquote {\bibinfo {title} {{Uncovering hidden new physics patterns in collider events using Bayesian probabilistic models}},}\ }\href {\doibase 10.22323/1.390.0238} {\bibfield  {journal} {\bibinfo  {journal} {PoS}\ }\textbf {\bibinfo {volume} {ICHEP2020}},\ \bibinfo {pages} {238} (\bibinfo {year} {2021})},\ \Eprint {http://arxiv.org/abs/2012.08579} {arXiv:2012.08579 [hep-ph]} \BibitemShut {NoStop}%
\bibitem [{\citenamefont {Stein}\ \emph {et~al.}(2020)\citenamefont {Stein}, \citenamefont {Seljak},\ and\ \citenamefont {Dai}}]{Stein:2020rou}%
  \BibitemOpen
  \bibfield  {author} {\bibinfo {author} {\bibfnamefont {George}\ \bibnamefont {Stein}}, \bibinfo {author} {\bibfnamefont {Uros}\ \bibnamefont {Seljak}}, \ and\ \bibinfo {author} {\bibfnamefont {Biwei}\ \bibnamefont {Dai}},\ }\bibfield  {title} {\enquote {\bibinfo {title} {{Unsupervised in-distribution anomaly detection of new physics through conditional density estimation}},}\ }in\ \href@noop {} {\emph {\bibinfo {booktitle} {{34th Conference on Neural Information Processing Systems}}}}\ (\bibinfo {year} {2020})\ \Eprint {http://arxiv.org/abs/2012.11638} {arXiv:2012.11638 [cs.LG]} \BibitemShut {NoStop}%
\bibitem [{\citenamefont {Kasieczka}\ \emph {et~al.}(2021{\natexlab{a}})\citenamefont {Kasieczka} \emph {et~al.}}]{Kasieczka:2021xcg}%
  \BibitemOpen
  \bibfield  {author} {\bibinfo {author} {\bibfnamefont {Gregor}\ \bibnamefont {Kasieczka}} \emph {et~al.},\ }\bibfield  {title} {\enquote {\bibinfo {title} {{The LHC Olympics 2020 a community challenge for anomaly detection in high energy physics}},}\ }\href {\doibase 10.1088/1361-6633/ac36b9} {\bibfield  {journal} {\bibinfo  {journal} {Rept. Prog. Phys.}\ }\textbf {\bibinfo {volume} {84}},\ \bibinfo {pages} {124201} (\bibinfo {year} {2021}{\natexlab{a}})},\ \Eprint {http://arxiv.org/abs/2101.08320} {arXiv:2101.08320 [hep-ph]} \BibitemShut {NoStop}%
\bibitem [{\citenamefont {Chakravarti}\ \emph {et~al.}(2021)\citenamefont {Chakravarti}, \citenamefont {Kuusela}, \citenamefont {Lei},\ and\ \citenamefont {Wasserman}}]{Chakravarti:2021svb}%
  \BibitemOpen
  \bibfield  {author} {\bibinfo {author} {\bibfnamefont {Purvasha}\ \bibnamefont {Chakravarti}}, \bibinfo {author} {\bibfnamefont {Mikael}\ \bibnamefont {Kuusela}}, \bibinfo {author} {\bibfnamefont {Jing}\ \bibnamefont {Lei}}, \ and\ \bibinfo {author} {\bibfnamefont {Larry}\ \bibnamefont {Wasserman}},\ }\bibfield  {title} {\enquote {\bibinfo {title} {{Model-Independent Detection of New Physics Signals Using Interpretable Semi-Supervised Classifier Tests}},}\ }\href@noop {} {\  (\bibinfo {year} {2021})},\ \Eprint {http://arxiv.org/abs/2102.07679} {arXiv:2102.07679 [stat.AP]} \BibitemShut {NoStop}%
\bibitem [{\citenamefont {Batson}\ \emph {et~al.}(2021)\citenamefont {Batson}, \citenamefont {Haaf}, \citenamefont {Kahn},\ and\ \citenamefont {Roberts}}]{Batson:2021agz}%
  \BibitemOpen
  \bibfield  {author} {\bibinfo {author} {\bibfnamefont {Joshua}\ \bibnamefont {Batson}}, \bibinfo {author} {\bibfnamefont {C.~Grace}\ \bibnamefont {Haaf}}, \bibinfo {author} {\bibfnamefont {Yonatan}\ \bibnamefont {Kahn}}, \ and\ \bibinfo {author} {\bibfnamefont {Daniel~A.}\ \bibnamefont {Roberts}},\ }\bibfield  {title} {\enquote {\bibinfo {title} {{Topological Obstructions to Autoencoding}},}\ }\href {\doibase 10.1007/JHEP04(2021)280} {\bibfield  {journal} {\bibinfo  {journal} {JHEP}\ }\textbf {\bibinfo {volume} {04}},\ \bibinfo {pages} {280} (\bibinfo {year} {2021})},\ \Eprint {http://arxiv.org/abs/2102.08380} {arXiv:2102.08380 [hep-ph]} \BibitemShut {NoStop}%
\bibitem [{\citenamefont {Blance}\ and\ \citenamefont {Spannowsky}(2020)}]{Blance:2021gcs}%
  \BibitemOpen
  \bibfield  {author} {\bibinfo {author} {\bibfnamefont {Andrew}\ \bibnamefont {Blance}}\ and\ \bibinfo {author} {\bibfnamefont {Michael}\ \bibnamefont {Spannowsky}},\ }\bibfield  {title} {\enquote {\bibinfo {title} {{Unsupervised event classification with graphs on classical and photonic quantum computers}},}\ }\href {\doibase 10.1007/JHEP08(2021)170} {\bibfield  {journal} {\bibinfo  {journal} {JHEP}\ }\textbf {\bibinfo {volume} {21}},\ \bibinfo {pages} {170} (\bibinfo {year} {2020})},\ \Eprint {http://arxiv.org/abs/2103.03897} {arXiv:2103.03897 [hep-ph]} \BibitemShut {NoStop}%
\bibitem [{\citenamefont {Bortolato}\ \emph {et~al.}(2022)\citenamefont {Bortolato}, \citenamefont {Smolkovi\v{c}}, \citenamefont {Dillon},\ and\ \citenamefont {Kamenik}}]{Bortolato:2021zic}%
  \BibitemOpen
  \bibfield  {author} {\bibinfo {author} {\bibfnamefont {Bla\v{z}}\ \bibnamefont {Bortolato}}, \bibinfo {author} {\bibfnamefont {Aleks}\ \bibnamefont {Smolkovi\v{c}}}, \bibinfo {author} {\bibfnamefont {Barry~M.}\ \bibnamefont {Dillon}}, \ and\ \bibinfo {author} {\bibfnamefont {Jernej~F.}\ \bibnamefont {Kamenik}},\ }\bibfield  {title} {\enquote {\bibinfo {title} {{Bump hunting in latent space}},}\ }\href {\doibase 10.1103/PhysRevD.105.115009} {\bibfield  {journal} {\bibinfo  {journal} {Phys. Rev. D}\ }\textbf {\bibinfo {volume} {105}},\ \bibinfo {pages} {115009} (\bibinfo {year} {2022})},\ \Eprint {http://arxiv.org/abs/2103.06595} {arXiv:2103.06595 [hep-ph]} \BibitemShut {NoStop}%
\bibitem [{\citenamefont {Collins}\ \emph {et~al.}(2021)\citenamefont {Collins}, \citenamefont {Mart\'\i{}n-Ramiro}, \citenamefont {Nachman},\ and\ \citenamefont {Shih}}]{Collins:2021nxn}%
  \BibitemOpen
  \bibfield  {author} {\bibinfo {author} {\bibfnamefont {Jack~H.}\ \bibnamefont {Collins}}, \bibinfo {author} {\bibfnamefont {Pablo}\ \bibnamefont {Mart\'\i{}n-Ramiro}}, \bibinfo {author} {\bibfnamefont {Benjamin}\ \bibnamefont {Nachman}}, \ and\ \bibinfo {author} {\bibfnamefont {David}\ \bibnamefont {Shih}},\ }\bibfield  {title} {\enquote {\bibinfo {title} {{Comparing weak- and unsupervised methods for resonant anomaly detection}},}\ }\href {\doibase 10.1140/epjc/s10052-021-09389-x} {\bibfield  {journal} {\bibinfo  {journal} {Eur. Phys. J. C}\ }\textbf {\bibinfo {volume} {81}},\ \bibinfo {pages} {617} (\bibinfo {year} {2021})},\ \Eprint {http://arxiv.org/abs/2104.02092} {arXiv:2104.02092 [hep-ph]} \BibitemShut {NoStop}%
\bibitem [{\citenamefont {Dillon}\ \emph {et~al.}(2021)\citenamefont {Dillon}, \citenamefont {Plehn}, \citenamefont {Sauer},\ and\ \citenamefont {Sorrenson}}]{Dillon:2021nxw}%
  \BibitemOpen
  \bibfield  {author} {\bibinfo {author} {\bibfnamefont {Barry~M.}\ \bibnamefont {Dillon}}, \bibinfo {author} {\bibfnamefont {Tilman}\ \bibnamefont {Plehn}}, \bibinfo {author} {\bibfnamefont {Christof}\ \bibnamefont {Sauer}}, \ and\ \bibinfo {author} {\bibfnamefont {Peter}\ \bibnamefont {Sorrenson}},\ }\bibfield  {title} {\enquote {\bibinfo {title} {{Better Latent Spaces for Better Autoencoders}},}\ }\href {\doibase 10.21468/SciPostPhys.11.3.061} {\bibfield  {journal} {\bibinfo  {journal} {SciPost Phys.}\ }\textbf {\bibinfo {volume} {11}},\ \bibinfo {pages} {061} (\bibinfo {year} {2021})},\ \Eprint {http://arxiv.org/abs/2104.08291} {arXiv:2104.08291 [hep-ph]} \BibitemShut {NoStop}%
\bibitem [{\citenamefont {Finke}\ \emph {et~al.}(2021)\citenamefont {Finke}, \citenamefont {Kr\"amer}, \citenamefont {Morandini}, \citenamefont {M\"uck},\ and\ \citenamefont {Oleksiyuk}}]{Finke:2021sdf}%
  \BibitemOpen
  \bibfield  {author} {\bibinfo {author} {\bibfnamefont {Thorben}\ \bibnamefont {Finke}}, \bibinfo {author} {\bibfnamefont {Michael}\ \bibnamefont {Kr\"amer}}, \bibinfo {author} {\bibfnamefont {Alessandro}\ \bibnamefont {Morandini}}, \bibinfo {author} {\bibfnamefont {Alexander}\ \bibnamefont {M\"uck}}, \ and\ \bibinfo {author} {\bibfnamefont {Ivan}\ \bibnamefont {Oleksiyuk}},\ }\bibfield  {title} {\enquote {\bibinfo {title} {{Autoencoders for unsupervised anomaly detection in high energy physics}},}\ }\href {\doibase 10.1007/JHEP06(2021)161} {\bibfield  {journal} {\bibinfo  {journal} {JHEP}\ }\textbf {\bibinfo {volume} {06}},\ \bibinfo {pages} {161} (\bibinfo {year} {2021})},\ \Eprint {http://arxiv.org/abs/2104.09051} {arXiv:2104.09051 [hep-ph]} \BibitemShut {NoStop}%
\bibitem [{\citenamefont {Shih}\ \emph {et~al.}(2021)\citenamefont {Shih}, \citenamefont {Buckley}, \citenamefont {Necib},\ and\ \citenamefont {Tamanas}}]{Shih:2021kbt}%
  \BibitemOpen
  \bibfield  {author} {\bibinfo {author} {\bibfnamefont {David}\ \bibnamefont {Shih}}, \bibinfo {author} {\bibfnamefont {Matthew~R.}\ \bibnamefont {Buckley}}, \bibinfo {author} {\bibfnamefont {Lina}\ \bibnamefont {Necib}}, \ and\ \bibinfo {author} {\bibfnamefont {John}\ \bibnamefont {Tamanas}},\ }\bibfield  {title} {\enquote {\bibinfo {title} {{via machinae: Searching for stellar streams using unsupervised machine learning}},}\ }\href {\doibase 10.1093/mnras/stab3372} {\bibfield  {journal} {\bibinfo  {journal} {Mon. Not. Roy. Astron. Soc.}\ }\textbf {\bibinfo {volume} {509}},\ \bibinfo {pages} {5992--6007} (\bibinfo {year} {2021})},\ \Eprint {http://arxiv.org/abs/2104.12789} {arXiv:2104.12789 [astro-ph.GA]} \BibitemShut {NoStop}%
\bibitem [{\citenamefont {Atkinson}\ \emph {et~al.}(2021)\citenamefont {Atkinson}, \citenamefont {Bhardwaj}, \citenamefont {Englert}, \citenamefont {Ngairangbam},\ and\ \citenamefont {Spannowsky}}]{Atkinson:2021nlt}%
  \BibitemOpen
  \bibfield  {author} {\bibinfo {author} {\bibfnamefont {Oliver}\ \bibnamefont {Atkinson}}, \bibinfo {author} {\bibfnamefont {Akanksha}\ \bibnamefont {Bhardwaj}}, \bibinfo {author} {\bibfnamefont {Christoph}\ \bibnamefont {Englert}}, \bibinfo {author} {\bibfnamefont {Vishal~S.}\ \bibnamefont {Ngairangbam}}, \ and\ \bibinfo {author} {\bibfnamefont {Michael}\ \bibnamefont {Spannowsky}},\ }\bibfield  {title} {\enquote {\bibinfo {title} {{Anomaly detection with convolutional Graph Neural Networks}},}\ }\href {\doibase 10.1007/JHEP08(2021)080} {\bibfield  {journal} {\bibinfo  {journal} {JHEP}\ }\textbf {\bibinfo {volume} {08}},\ \bibinfo {pages} {080} (\bibinfo {year} {2021})},\ \Eprint {http://arxiv.org/abs/2105.07988} {arXiv:2105.07988 [hep-ph]} \BibitemShut {NoStop}%
\bibitem [{\citenamefont {Kahn}\ \emph {et~al.}(2021)\citenamefont {Kahn}, \citenamefont {Gonski}, \citenamefont {Ochoa}, \citenamefont {Williams},\ and\ \citenamefont {Brooijmans}}]{Kahn:2021drv}%
  \BibitemOpen
  \bibfield  {author} {\bibinfo {author} {\bibfnamefont {Alan}\ \bibnamefont {Kahn}}, \bibinfo {author} {\bibfnamefont {Julia}\ \bibnamefont {Gonski}}, \bibinfo {author} {\bibfnamefont {In\^es}\ \bibnamefont {Ochoa}}, \bibinfo {author} {\bibfnamefont {Daniel}\ \bibnamefont {Williams}}, \ and\ \bibinfo {author} {\bibfnamefont {Gustaaf}\ \bibnamefont {Brooijmans}},\ }\bibfield  {title} {\enquote {\bibinfo {title} {{Anomalous jet identification via sequence modeling}},}\ }\href {\doibase 10.1088/1748-0221/16/08/P08012} {\bibfield  {journal} {\bibinfo  {journal} {JINST}\ }\textbf {\bibinfo {volume} {16}},\ \bibinfo {pages} {P08012} (\bibinfo {year} {2021})},\ \Eprint {http://arxiv.org/abs/2105.09274} {arXiv:2105.09274 [hep-ph]} \BibitemShut {NoStop}%
\bibitem [{\citenamefont {Aarrestad}\ \emph {et~al.}(2022)\citenamefont {Aarrestad} \emph {et~al.}}]{Aarrestad:2021oeb}%
  \BibitemOpen
  \bibfield  {author} {\bibinfo {author} {\bibfnamefont {Thea}\ \bibnamefont {Aarrestad}} \emph {et~al.},\ }\bibfield  {title} {\enquote {\bibinfo {title} {{The Dark Machines Anomaly Score Challenge: Benchmark Data and Model Independent Event Classification for the Large Hadron Collider}},}\ }\href {\doibase 10.21468/SciPostPhys.12.1.043} {\bibfield  {journal} {\bibinfo  {journal} {SciPost Phys.}\ }\textbf {\bibinfo {volume} {12}},\ \bibinfo {pages} {043} (\bibinfo {year} {2022})},\ \Eprint {http://arxiv.org/abs/2105.14027} {arXiv:2105.14027 [hep-ph]} \BibitemShut {NoStop}%
\bibitem [{\citenamefont {Dorigo}\ \emph {et~al.}(2023)\citenamefont {Dorigo}, \citenamefont {Fumanelli}, \citenamefont {Maccani}, \citenamefont {Mojsovska}, \citenamefont {Strong},\ and\ \citenamefont {Scarpa}}]{Dorigo:2021iyy}%
  \BibitemOpen
  \bibfield  {author} {\bibinfo {author} {\bibfnamefont {Tommaso}\ \bibnamefont {Dorigo}}, \bibinfo {author} {\bibfnamefont {Martina}\ \bibnamefont {Fumanelli}}, \bibinfo {author} {\bibfnamefont {Chiara}\ \bibnamefont {Maccani}}, \bibinfo {author} {\bibfnamefont {Marija}\ \bibnamefont {Mojsovska}}, \bibinfo {author} {\bibfnamefont {Giles~C.}\ \bibnamefont {Strong}}, \ and\ \bibinfo {author} {\bibfnamefont {Bruno}\ \bibnamefont {Scarpa}},\ }\bibfield  {title} {\enquote {\bibinfo {title} {{RanBox: anomaly detection in the copula space}},}\ }\href {\doibase 10.1007/JHEP01(2023)008} {\bibfield  {journal} {\bibinfo  {journal} {JHEP}\ }\textbf {\bibinfo {volume} {01}},\ \bibinfo {pages} {008} (\bibinfo {year} {2023})},\ \Eprint {http://arxiv.org/abs/2106.05747} {arXiv:2106.05747 [physics.data-an]} \BibitemShut {NoStop}%
\bibitem [{\citenamefont {Caron}\ \emph {et~al.}(2022)\citenamefont {Caron}, \citenamefont {Hendriks},\ and\ \citenamefont {Verheyen}}]{Caron:2021wmq}%
  \BibitemOpen
  \bibfield  {author} {\bibinfo {author} {\bibfnamefont {Sascha}\ \bibnamefont {Caron}}, \bibinfo {author} {\bibfnamefont {Luc}\ \bibnamefont {Hendriks}}, \ and\ \bibinfo {author} {\bibfnamefont {Rob}\ \bibnamefont {Verheyen}},\ }\bibfield  {title} {\enquote {\bibinfo {title} {{Rare and Different: Anomaly Scores from a combination of likelihood and out-of-distribution models to detect new physics at the LHC}},}\ }\href {\doibase 10.21468/SciPostPhys.12.2.077} {\bibfield  {journal} {\bibinfo  {journal} {SciPost Phys.}\ }\textbf {\bibinfo {volume} {12}},\ \bibinfo {pages} {077} (\bibinfo {year} {2022})},\ \Eprint {http://arxiv.org/abs/2106.10164} {arXiv:2106.10164 [hep-ph]} \BibitemShut {NoStop}%
\bibitem [{\citenamefont {Govorkova}\ \emph {et~al.}(2022{\natexlab{a}})\citenamefont {Govorkova}, \citenamefont {Puljak}, \citenamefont {Aarrestad}, \citenamefont {Pierini}, \citenamefont {Wo\'zniak},\ and\ \citenamefont {Ngadiuba}}]{Govorkova:2021hqu}%
  \BibitemOpen
  \bibfield  {author} {\bibinfo {author} {\bibfnamefont {Ekaterina}\ \bibnamefont {Govorkova}}, \bibinfo {author} {\bibfnamefont {Ema}\ \bibnamefont {Puljak}}, \bibinfo {author} {\bibfnamefont {Thea}\ \bibnamefont {Aarrestad}}, \bibinfo {author} {\bibfnamefont {Maurizio}\ \bibnamefont {Pierini}}, \bibinfo {author} {\bibfnamefont {Kinga~Anna}\ \bibnamefont {Wo\'zniak}}, \ and\ \bibinfo {author} {\bibfnamefont {Jennifer}\ \bibnamefont {Ngadiuba}},\ }\bibfield  {title} {\enquote {\bibinfo {title} {{LHC physics dataset for unsupervised New Physics detection at 40 MHz}},}\ }\href {\doibase 10.1038/s41597-022-01187-8} {\bibfield  {journal} {\bibinfo  {journal} {Sci. Data}\ }\textbf {\bibinfo {volume} {9}},\ \bibinfo {pages} {118} (\bibinfo {year} {2022}{\natexlab{a}})},\ \Eprint {http://arxiv.org/abs/2107.02157} {arXiv:2107.02157 [physics.data-an]} \BibitemShut {NoStop}%
\bibitem [{\citenamefont {Kasieczka}\ \emph {et~al.}(2021{\natexlab{b}})\citenamefont {Kasieczka}, \citenamefont {Nachman},\ and\ \citenamefont {Shih}}]{Kasieczka:2021tew}%
  \BibitemOpen
  \bibfield  {author} {\bibinfo {author} {\bibfnamefont {Gregor}\ \bibnamefont {Kasieczka}}, \bibinfo {author} {\bibfnamefont {Benjamin}\ \bibnamefont {Nachman}}, \ and\ \bibinfo {author} {\bibfnamefont {David}\ \bibnamefont {Shih}},\ }\bibfield  {title} {\enquote {\bibinfo {title} {{New Methods and Datasets for Group Anomaly Detection From Fundamental Physics}},}\ }in\ \href@noop {} {\emph {\bibinfo {booktitle} {{Conference on Knowledge Discovery and Data Mining}}}}\ (\bibinfo {year} {2021})\ \Eprint {http://arxiv.org/abs/2107.02821} {arXiv:2107.02821 [stat.ML]} \BibitemShut {NoStop}%
\bibitem [{\citenamefont {Volkovich}\ \emph {et~al.}(2022)\citenamefont {Volkovich}, \citenamefont {De~Vito~Halevy},\ and\ \citenamefont {Bressler}}]{Volkovich:2021txe}%
  \BibitemOpen
  \bibfield  {author} {\bibinfo {author} {\bibfnamefont {Sergey}\ \bibnamefont {Volkovich}}, \bibinfo {author} {\bibfnamefont {Federico}\ \bibnamefont {De~Vito~Halevy}}, \ and\ \bibinfo {author} {\bibfnamefont {Shikma}\ \bibnamefont {Bressler}},\ }\bibfield  {title} {\enquote {\bibinfo {title} {{A data-directed paradigm for BSM searches: the bump-hunting example}},}\ }\href {\doibase 10.1140/epjc/s10052-022-10215-1} {\bibfield  {journal} {\bibinfo  {journal} {Eur. Phys. J. C}\ }\textbf {\bibinfo {volume} {82}},\ \bibinfo {pages} {265} (\bibinfo {year} {2022})},\ \Eprint {http://arxiv.org/abs/2107.11573} {arXiv:2107.11573 [hep-ex]} \BibitemShut {NoStop}%
\bibitem [{\citenamefont {Govorkova}\ \emph {et~al.}(2022{\natexlab{b}})\citenamefont {Govorkova} \emph {et~al.}}]{Govorkova:2021utb}%
  \BibitemOpen
  \bibfield  {author} {\bibinfo {author} {\bibfnamefont {Ekaterina}\ \bibnamefont {Govorkova}} \emph {et~al.},\ }\bibfield  {title} {\enquote {\bibinfo {title} {{Autoencoders on field-programmable gate arrays for real-time, unsupervised new physics detection at 40 MHz at the Large Hadron Collider}},}\ }\href {\doibase 10.1038/s42256-022-00441-3} {\bibfield  {journal} {\bibinfo  {journal} {Nature Mach. Intell.}\ }\textbf {\bibinfo {volume} {4}},\ \bibinfo {pages} {154--161} (\bibinfo {year} {2022}{\natexlab{b}})},\ \Eprint {http://arxiv.org/abs/2108.03986} {arXiv:2108.03986 [physics.ins-det]} \BibitemShut {NoStop}%
\bibitem [{\citenamefont {Hallin}\ \emph {et~al.}(2022)\citenamefont {Hallin}, \citenamefont {Isaacson}, \citenamefont {Kasieczka}, \citenamefont {Krause}, \citenamefont {Nachman}, \citenamefont {Quadfasel}, \citenamefont {Schlaffer}, \citenamefont {Shih},\ and\ \citenamefont {Sommerhalder}}]{Hallin:2021wme}%
  \BibitemOpen
  \bibfield  {author} {\bibinfo {author} {\bibfnamefont {Anna}\ \bibnamefont {Hallin}}, \bibinfo {author} {\bibfnamefont {Joshua}\ \bibnamefont {Isaacson}}, \bibinfo {author} {\bibfnamefont {Gregor}\ \bibnamefont {Kasieczka}}, \bibinfo {author} {\bibfnamefont {Claudius}\ \bibnamefont {Krause}}, \bibinfo {author} {\bibfnamefont {Benjamin}\ \bibnamefont {Nachman}}, \bibinfo {author} {\bibfnamefont {Tobias}\ \bibnamefont {Quadfasel}}, \bibinfo {author} {\bibfnamefont {Matthias}\ \bibnamefont {Schlaffer}}, \bibinfo {author} {\bibfnamefont {David}\ \bibnamefont {Shih}}, \ and\ \bibinfo {author} {\bibfnamefont {Manuel}\ \bibnamefont {Sommerhalder}},\ }\bibfield  {title} {\enquote {\bibinfo {title} {{Classifying anomalies through outer density estimation}},}\ }\href {\doibase 10.1103/PhysRevD.106.055006} {\bibfield  {journal} {\bibinfo  {journal} {Phys. Rev. D}\ }\textbf {\bibinfo {volume} {106}},\ \bibinfo {pages} {055006} (\bibinfo {year} {2022})},\ \Eprint {http://arxiv.org/abs/2109.00546} {arXiv:2109.00546
  [hep-ph]} \BibitemShut {NoStop}%
\bibitem [{\citenamefont {Ostdiek}(2022)}]{Ostdiek:2021bem}%
  \BibitemOpen
  \bibfield  {author} {\bibinfo {author} {\bibfnamefont {Bryan}\ \bibnamefont {Ostdiek}},\ }\bibfield  {title} {\enquote {\bibinfo {title} {{Deep Set Auto Encoders for Anomaly Detection in Particle Physics}},}\ }\href {\doibase 10.21468/SciPostPhys.12.1.045} {\bibfield  {journal} {\bibinfo  {journal} {SciPost Phys.}\ }\textbf {\bibinfo {volume} {12}},\ \bibinfo {pages} {045} (\bibinfo {year} {2022})},\ \Eprint {http://arxiv.org/abs/2109.01695} {arXiv:2109.01695 [hep-ph]} \BibitemShut {NoStop}%
\bibitem [{\citenamefont {Fraser}\ \emph {et~al.}(2022)\citenamefont {Fraser}, \citenamefont {Homiller}, \citenamefont {Mishra}, \citenamefont {Ostdiek},\ and\ \citenamefont {Schwartz}}]{Fraser:2021lxm}%
  \BibitemOpen
  \bibfield  {author} {\bibinfo {author} {\bibfnamefont {Katherine}\ \bibnamefont {Fraser}}, \bibinfo {author} {\bibfnamefont {Samuel}\ \bibnamefont {Homiller}}, \bibinfo {author} {\bibfnamefont {Rashmish~K.}\ \bibnamefont {Mishra}}, \bibinfo {author} {\bibfnamefont {Bryan}\ \bibnamefont {Ostdiek}}, \ and\ \bibinfo {author} {\bibfnamefont {Matthew~D.}\ \bibnamefont {Schwartz}},\ }\bibfield  {title} {\enquote {\bibinfo {title} {{Challenges for unsupervised anomaly detection in particle physics}},}\ }\href {\doibase 10.1007/JHEP03(2022)066} {\bibfield  {journal} {\bibinfo  {journal} {JHEP}\ }\textbf {\bibinfo {volume} {03}},\ \bibinfo {pages} {066} (\bibinfo {year} {2022})},\ \Eprint {http://arxiv.org/abs/2110.06948} {arXiv:2110.06948 [hep-ph]} \BibitemShut {NoStop}%
\bibitem [{\citenamefont {Jawahar}\ \emph {et~al.}(2022)\citenamefont {Jawahar}, \citenamefont {Aarrestad}, \citenamefont {Chernyavskaya}, \citenamefont {Pierini}, \citenamefont {Wozniak}, \citenamefont {Ngadiuba}, \citenamefont {Duarte},\ and\ \citenamefont {Tsan}}]{Jawahar:2021vyu}%
  \BibitemOpen
  \bibfield  {author} {\bibinfo {author} {\bibfnamefont {Pratik}\ \bibnamefont {Jawahar}}, \bibinfo {author} {\bibfnamefont {Thea}\ \bibnamefont {Aarrestad}}, \bibinfo {author} {\bibfnamefont {Nadezda}\ \bibnamefont {Chernyavskaya}}, \bibinfo {author} {\bibfnamefont {Maurizio}\ \bibnamefont {Pierini}}, \bibinfo {author} {\bibfnamefont {Kinga~A.}\ \bibnamefont {Wozniak}}, \bibinfo {author} {\bibfnamefont {Jennifer}\ \bibnamefont {Ngadiuba}}, \bibinfo {author} {\bibfnamefont {Javier}\ \bibnamefont {Duarte}}, \ and\ \bibinfo {author} {\bibfnamefont {Steven}\ \bibnamefont {Tsan}},\ }\bibfield  {title} {\enquote {\bibinfo {title} {{Improving Variational Autoencoders for New Physics Detection at the LHC With Normalizing Flows}},}\ }\href {\doibase 10.3389/fdata.2022.803685} {\bibfield  {journal} {\bibinfo  {journal} {Front. Big Data}\ }\textbf {\bibinfo {volume} {5}},\ \bibinfo {pages} {803685} (\bibinfo {year} {2022})},\ \Eprint {http://arxiv.org/abs/2110.08508} {arXiv:2110.08508 [hep-ph]} \BibitemShut {NoStop}%
\bibitem [{\citenamefont {Herrero-Garcia}\ \emph {et~al.}(2022)\citenamefont {Herrero-Garcia}, \citenamefont {Patrick},\ and\ \citenamefont {Scaffidi}}]{Herrero-Garcia:2021goa}%
  \BibitemOpen
  \bibfield  {author} {\bibinfo {author} {\bibfnamefont {Juan}\ \bibnamefont {Herrero-Garcia}}, \bibinfo {author} {\bibfnamefont {Riley}\ \bibnamefont {Patrick}}, \ and\ \bibinfo {author} {\bibfnamefont {Andre}\ \bibnamefont {Scaffidi}},\ }\bibfield  {title} {\enquote {\bibinfo {title} {{A semi-supervised approach to dark matter searches in direct detection data with machine learning}},}\ }\href {\doibase 10.1088/1475-7516/2022/02/039} {\bibfield  {journal} {\bibinfo  {journal} {JCAP}\ }\textbf {\bibinfo {volume} {02}},\ \bibinfo {pages} {039} (\bibinfo {year} {2022})},\ \Eprint {http://arxiv.org/abs/2110.12248} {arXiv:2110.12248 [hep-ph]} \BibitemShut {NoStop}%
\bibitem [{\citenamefont {Lester}\ and\ \citenamefont {Tombs}(2021)}]{Lester:2021aks}%
  \BibitemOpen
  \bibfield  {author} {\bibinfo {author} {\bibfnamefont {Christopher~G.}\ \bibnamefont {Lester}}\ and\ \bibinfo {author} {\bibfnamefont {Rupert}\ \bibnamefont {Tombs}},\ }\bibfield  {title} {\enquote {\bibinfo {title} {{Using unsupervised learning to detect broken symmetries, with relevance to searches for parity violation in nature. (Previously: ''Stressed GANs snag desserts'')}},}\ }\href@noop {} {\  (\bibinfo {year} {2021})},\ \Eprint {http://arxiv.org/abs/2111.00616} {arXiv:2111.00616 [hep-ph]} \BibitemShut {NoStop}%
\bibitem [{\citenamefont {Aguilar-Saavedra}(2022{\natexlab{a}})}]{Aguilar-Saavedra:2021utu}%
  \BibitemOpen
  \bibfield  {author} {\bibinfo {author} {\bibfnamefont {J.~A.}\ \bibnamefont {Aguilar-Saavedra}},\ }\bibfield  {title} {\enquote {\bibinfo {title} {{Anomaly detection from mass unspecific jet tagging}},}\ }\href {\doibase 10.1140/epjc/s10052-022-10058-w} {\bibfield  {journal} {\bibinfo  {journal} {Eur. Phys. J. C}\ }\textbf {\bibinfo {volume} {82}},\ \bibinfo {pages} {130} (\bibinfo {year} {2022}{\natexlab{a}})},\ \Eprint {http://arxiv.org/abs/2111.02647} {arXiv:2111.02647 [hep-ph]} \BibitemShut {NoStop}%
\bibitem [{\citenamefont {Tombs}\ and\ \citenamefont {Lester}(2022)}]{Tombs:2021wae}%
  \BibitemOpen
  \bibfield  {author} {\bibinfo {author} {\bibfnamefont {Rupert}\ \bibnamefont {Tombs}}\ and\ \bibinfo {author} {\bibfnamefont {Christopher~G.}\ \bibnamefont {Lester}},\ }\bibfield  {title} {\enquote {\bibinfo {title} {{A method to challenge symmetries in data with self-supervised learning}},}\ }\href {\doibase 10.1088/1748-0221/17/08/P08024} {\bibfield  {journal} {\bibinfo  {journal} {JINST}\ }\textbf {\bibinfo {volume} {17}},\ \bibinfo {pages} {P08024} (\bibinfo {year} {2022})},\ \Eprint {http://arxiv.org/abs/2111.05442} {arXiv:2111.05442 [hep-ph]} \BibitemShut {NoStop}%
\bibitem [{\citenamefont {Mikuni}\ \emph {et~al.}(2022)\citenamefont {Mikuni}, \citenamefont {Nachman},\ and\ \citenamefont {Shih}}]{Mikuni:2021nwn}%
  \BibitemOpen
  \bibfield  {author} {\bibinfo {author} {\bibfnamefont {Vinicius}\ \bibnamefont {Mikuni}}, \bibinfo {author} {\bibfnamefont {Benjamin}\ \bibnamefont {Nachman}}, \ and\ \bibinfo {author} {\bibfnamefont {David}\ \bibnamefont {Shih}},\ }\bibfield  {title} {\enquote {\bibinfo {title} {{Online-compatible unsupervised nonresonant anomaly detection}},}\ }\href {\doibase 10.1103/PhysRevD.105.055006} {\bibfield  {journal} {\bibinfo  {journal} {Phys. Rev. D}\ }\textbf {\bibinfo {volume} {105}},\ \bibinfo {pages} {055006} (\bibinfo {year} {2022})},\ \Eprint {http://arxiv.org/abs/2111.06417} {arXiv:2111.06417 [cs.LG]} \BibitemShut {NoStop}%
\bibitem [{\citenamefont {Chekanov}\ and\ \citenamefont {Hopkins}(2022)}]{Chekanov:2021pus}%
  \BibitemOpen
  \bibfield  {author} {\bibinfo {author} {\bibfnamefont {Sergei}\ \bibnamefont {Chekanov}}\ and\ \bibinfo {author} {\bibfnamefont {Walter}\ \bibnamefont {Hopkins}},\ }\bibfield  {title} {\enquote {\bibinfo {title} {{Event-Based Anomaly Detection for Searches for New Physics}},}\ }\href {\doibase 10.3390/universe8100494} {\bibfield  {journal} {\bibinfo  {journal} {Universe}\ }\textbf {\bibinfo {volume} {8}},\ \bibinfo {pages} {494} (\bibinfo {year} {2022})},\ \Eprint {http://arxiv.org/abs/2111.12119} {arXiv:2111.12119 [hep-ph]} \BibitemShut {NoStop}%
\bibitem [{\citenamefont {d'Agnolo}\ \emph {et~al.}(2022)\citenamefont {d'Agnolo}, \citenamefont {Grosso}, \citenamefont {Pierini}, \citenamefont {Wulzer},\ and\ \citenamefont {Zanetti}}]{dAgnolo:2021aun}%
  \BibitemOpen
  \bibfield  {author} {\bibinfo {author} {\bibfnamefont {Raffaele~Tito}\ \bibnamefont {d'Agnolo}}, \bibinfo {author} {\bibfnamefont {Gaia}\ \bibnamefont {Grosso}}, \bibinfo {author} {\bibfnamefont {Maurizio}\ \bibnamefont {Pierini}}, \bibinfo {author} {\bibfnamefont {Andrea}\ \bibnamefont {Wulzer}}, \ and\ \bibinfo {author} {\bibfnamefont {Marco}\ \bibnamefont {Zanetti}},\ }\bibfield  {title} {\enquote {\bibinfo {title} {{Learning new physics from an imperfect machine}},}\ }\href {\doibase 10.1140/epjc/s10052-022-10226-y} {\bibfield  {journal} {\bibinfo  {journal} {Eur. Phys. J. C}\ }\textbf {\bibinfo {volume} {82}},\ \bibinfo {pages} {275} (\bibinfo {year} {2022})},\ \Eprint {http://arxiv.org/abs/2111.13633} {arXiv:2111.13633 [hep-ph]} \BibitemShut {NoStop}%
\bibitem [{\citenamefont {Canelli}\ \emph {et~al.}(2022)\citenamefont {Canelli}, \citenamefont {de~Cosa}, \citenamefont {Pottier}, \citenamefont {Niedziela}, \citenamefont {Pedro},\ and\ \citenamefont {Pierini}}]{Canelli:2021aps}%
  \BibitemOpen
  \bibfield  {author} {\bibinfo {author} {\bibfnamefont {Florencia}\ \bibnamefont {Canelli}}, \bibinfo {author} {\bibfnamefont {Annapaola}\ \bibnamefont {de~Cosa}}, \bibinfo {author} {\bibfnamefont {Luc~Le}\ \bibnamefont {Pottier}}, \bibinfo {author} {\bibfnamefont {Jeremi}\ \bibnamefont {Niedziela}}, \bibinfo {author} {\bibfnamefont {Kevin}\ \bibnamefont {Pedro}}, \ and\ \bibinfo {author} {\bibfnamefont {Maurizio}\ \bibnamefont {Pierini}},\ }\bibfield  {title} {\enquote {\bibinfo {title} {{Autoencoders for semivisible jet detection}},}\ }\href {\doibase 10.1007/JHEP02(2022)074} {\bibfield  {journal} {\bibinfo  {journal} {JHEP}\ }\textbf {\bibinfo {volume} {02}},\ \bibinfo {pages} {074} (\bibinfo {year} {2022})},\ \Eprint {http://arxiv.org/abs/2112.02864} {arXiv:2112.02864 [hep-ph]} \BibitemShut {NoStop}%
\bibitem [{\citenamefont {Ngairangbam}\ \emph {et~al.}(2022)\citenamefont {Ngairangbam}, \citenamefont {Spannowsky},\ and\ \citenamefont {Takeuchi}}]{Ngairangbam:2021yma}%
  \BibitemOpen
  \bibfield  {author} {\bibinfo {author} {\bibfnamefont {Vishal~S.}\ \bibnamefont {Ngairangbam}}, \bibinfo {author} {\bibfnamefont {Michael}\ \bibnamefont {Spannowsky}}, \ and\ \bibinfo {author} {\bibfnamefont {Michihisa}\ \bibnamefont {Takeuchi}},\ }\bibfield  {title} {\enquote {\bibinfo {title} {{Anomaly detection in high-energy physics using a quantum autoencoder}},}\ }\href {\doibase 10.1103/PhysRevD.105.095004} {\bibfield  {journal} {\bibinfo  {journal} {Phys. Rev. D}\ }\textbf {\bibinfo {volume} {105}},\ \bibinfo {pages} {095004} (\bibinfo {year} {2022})},\ \Eprint {http://arxiv.org/abs/2112.04958} {arXiv:2112.04958 [hep-ph]} \BibitemShut {NoStop}%
\bibitem [{\citenamefont {Aguilar-Saavedra}(2022{\natexlab{b}})}]{Aguilar-Saavedra:2022ejy}%
  \BibitemOpen
  \bibfield  {author} {\bibinfo {author} {\bibfnamefont {J.~A.}\ \bibnamefont {Aguilar-Saavedra}},\ }\bibfield  {title} {\enquote {\bibinfo {title} {{Taming modeling uncertainties with mass unspecific supervised tagging}},}\ }\href {\doibase 10.1140/epjc/s10052-022-10221-3} {\bibfield  {journal} {\bibinfo  {journal} {Eur. Phys. J. C}\ }\textbf {\bibinfo {volume} {82}},\ \bibinfo {pages} {270} (\bibinfo {year} {2022}{\natexlab{b}})},\ \Eprint {http://arxiv.org/abs/2201.11143} {arXiv:2201.11143 [hep-ph]} \BibitemShut {NoStop}%
\bibitem [{\citenamefont {Buss}\ \emph {et~al.}(2023)\citenamefont {Buss}, \citenamefont {Dillon}, \citenamefont {Finke}, \citenamefont {Kr\"amer}, \citenamefont {Morandini}, \citenamefont {M\"uck}, \citenamefont {Oleksiyuk},\ and\ \citenamefont {Plehn}}]{Buss:2022lxw}%
  \BibitemOpen
  \bibfield  {author} {\bibinfo {author} {\bibfnamefont {Thorsten}\ \bibnamefont {Buss}}, \bibinfo {author} {\bibfnamefont {Barry~M.}\ \bibnamefont {Dillon}}, \bibinfo {author} {\bibfnamefont {Thorben}\ \bibnamefont {Finke}}, \bibinfo {author} {\bibfnamefont {Michael}\ \bibnamefont {Kr\"amer}}, \bibinfo {author} {\bibfnamefont {Alessandro}\ \bibnamefont {Morandini}}, \bibinfo {author} {\bibfnamefont {Alexander}\ \bibnamefont {M\"uck}}, \bibinfo {author} {\bibfnamefont {Ivan}\ \bibnamefont {Oleksiyuk}}, \ and\ \bibinfo {author} {\bibfnamefont {Tilman}\ \bibnamefont {Plehn}},\ }\bibfield  {title} {\enquote {\bibinfo {title} {{What's anomalous in LHC jets?}}}\ }\href {\doibase 10.21468/SciPostPhys.15.4.168} {\bibfield  {journal} {\bibinfo  {journal} {SciPost Phys.}\ }\textbf {\bibinfo {volume} {15}},\ \bibinfo {pages} {168} (\bibinfo {year} {2023})},\ \Eprint {http://arxiv.org/abs/2202.00686} {arXiv:2202.00686 [hep-ph]} \BibitemShut {NoStop}%
\bibitem [{\citenamefont {Bradshaw}\ \emph {et~al.}(2022)\citenamefont {Bradshaw}, \citenamefont {Chang},\ and\ \citenamefont {Ostdiek}}]{Bradshaw:2022qev}%
  \BibitemOpen
  \bibfield  {author} {\bibinfo {author} {\bibfnamefont {Layne}\ \bibnamefont {Bradshaw}}, \bibinfo {author} {\bibfnamefont {Spencer}\ \bibnamefont {Chang}}, \ and\ \bibinfo {author} {\bibfnamefont {Bryan}\ \bibnamefont {Ostdiek}},\ }\bibfield  {title} {\enquote {\bibinfo {title} {{Creating simple, interpretable anomaly detectors for new physics in jet substructure}},}\ }\href {\doibase 10.1103/PhysRevD.106.035014} {\bibfield  {journal} {\bibinfo  {journal} {Phys. Rev. D}\ }\textbf {\bibinfo {volume} {106}},\ \bibinfo {pages} {035014} (\bibinfo {year} {2022})},\ \Eprint {http://arxiv.org/abs/2203.01343} {arXiv:2203.01343 [hep-ph]} \BibitemShut {NoStop}%
\bibitem [{\citenamefont {Birman}\ \emph {et~al.}(2022)\citenamefont {Birman}, \citenamefont {Nachman}, \citenamefont {Sebbah}, \citenamefont {Sela}, \citenamefont {Turetz},\ and\ \citenamefont {Bressler}}]{Birman:2022xzu}%
  \BibitemOpen
  \bibfield  {author} {\bibinfo {author} {\bibfnamefont {Mattias}\ \bibnamefont {Birman}}, \bibinfo {author} {\bibfnamefont {Benjamin}\ \bibnamefont {Nachman}}, \bibinfo {author} {\bibfnamefont {Raphael}\ \bibnamefont {Sebbah}}, \bibinfo {author} {\bibfnamefont {Gal}\ \bibnamefont {Sela}}, \bibinfo {author} {\bibfnamefont {Ophir}\ \bibnamefont {Turetz}}, \ and\ \bibinfo {author} {\bibfnamefont {Shikma}\ \bibnamefont {Bressler}},\ }\bibfield  {title} {\enquote {\bibinfo {title} {{Data-directed search for new physics based on symmetries of the SM}},}\ }\href {\doibase 10.1140/epjc/s10052-022-10454-2} {\bibfield  {journal} {\bibinfo  {journal} {Eur. Phys. J. C}\ }\textbf {\bibinfo {volume} {82}},\ \bibinfo {pages} {508} (\bibinfo {year} {2022})},\ \Eprint {http://arxiv.org/abs/2203.07529} {arXiv:2203.07529 [hep-ph]} \BibitemShut {NoStop}%
\bibitem [{\citenamefont {Raine}\ \emph {et~al.}(2023)\citenamefont {Raine}, \citenamefont {Klein}, \citenamefont {Sengupta},\ and\ \citenamefont {Golling}}]{Raine:2022hht}%
  \BibitemOpen
  \bibfield  {author} {\bibinfo {author} {\bibfnamefont {John~Andrew}\ \bibnamefont {Raine}}, \bibinfo {author} {\bibfnamefont {Samuel}\ \bibnamefont {Klein}}, \bibinfo {author} {\bibfnamefont {Debajyoti}\ \bibnamefont {Sengupta}}, \ and\ \bibinfo {author} {\bibfnamefont {Tobias}\ \bibnamefont {Golling}},\ }\bibfield  {title} {\enquote {\bibinfo {title} {{CURTAINs for your sliding window: Constructing unobserved regions by transforming adjacent intervals}},}\ }\href {\doibase 10.3389/fdata.2023.899345} {\bibfield  {journal} {\bibinfo  {journal} {Front. Big Data}\ }\textbf {\bibinfo {volume} {6}},\ \bibinfo {pages} {899345} (\bibinfo {year} {2023})},\ \Eprint {http://arxiv.org/abs/2203.09470} {arXiv:2203.09470 [hep-ph]} \BibitemShut {NoStop}%
\bibitem [{\citenamefont {Letizia}\ \emph {et~al.}(2022)\citenamefont {Letizia}, \citenamefont {Losapio}, \citenamefont {Rando}, \citenamefont {Grosso}, \citenamefont {Wulzer}, \citenamefont {Pierini}, \citenamefont {Zanetti},\ and\ \citenamefont {Rosasco}}]{Letizia:2022xbe}%
  \BibitemOpen
  \bibfield  {author} {\bibinfo {author} {\bibfnamefont {Marco}\ \bibnamefont {Letizia}}, \bibinfo {author} {\bibfnamefont {Gianvito}\ \bibnamefont {Losapio}}, \bibinfo {author} {\bibfnamefont {Marco}\ \bibnamefont {Rando}}, \bibinfo {author} {\bibfnamefont {Gaia}\ \bibnamefont {Grosso}}, \bibinfo {author} {\bibfnamefont {Andrea}\ \bibnamefont {Wulzer}}, \bibinfo {author} {\bibfnamefont {Maurizio}\ \bibnamefont {Pierini}}, \bibinfo {author} {\bibfnamefont {Marco}\ \bibnamefont {Zanetti}}, \ and\ \bibinfo {author} {\bibfnamefont {Lorenzo}\ \bibnamefont {Rosasco}},\ }\bibfield  {title} {\enquote {\bibinfo {title} {{Learning new physics efficiently with nonparametric methods}},}\ }\href {\doibase 10.1140/epjc/s10052-022-10830-y} {\bibfield  {journal} {\bibinfo  {journal} {Eur. Phys. J. C}\ }\textbf {\bibinfo {volume} {82}},\ \bibinfo {pages} {879} (\bibinfo {year} {2022})},\ \Eprint {http://arxiv.org/abs/2204.02317} {arXiv:2204.02317 [hep-ph]} \BibitemShut {NoStop}%
\bibitem [{\citenamefont {Fanelli}\ \emph {et~al.}(2022)\citenamefont {Fanelli}, \citenamefont {Giroux},\ and\ \citenamefont {Papandreou}}]{Fanelli:2022xwl}%
  \BibitemOpen
  \bibfield  {author} {\bibinfo {author} {\bibfnamefont {C.}~\bibnamefont {Fanelli}}, \bibinfo {author} {\bibfnamefont {J.}~\bibnamefont {Giroux}}, \ and\ \bibinfo {author} {\bibfnamefont {Z.}~\bibnamefont {Papandreou}},\ }\bibfield  {title} {\enquote {\bibinfo {title} {{\textquoteleft{}Flux+Mutability\textquoteright{}: a conditional generative approach to one-class classification and anomaly detection}},}\ }\href {\doibase 10.1088/2632-2153/ac9bcb} {\bibfield  {journal} {\bibinfo  {journal} {Mach. Learn. Sci. Tech.}\ }\textbf {\bibinfo {volume} {3}},\ \bibinfo {pages} {045012} (\bibinfo {year} {2022})},\ \Eprint {http://arxiv.org/abs/2204.08609} {arXiv:2204.08609 [cs.LG]} \BibitemShut {NoStop}%
\bibitem [{\citenamefont {Finke}\ \emph {et~al.}(2022)\citenamefont {Finke}, \citenamefont {Kr\"amer}, \citenamefont {Lipp},\ and\ \citenamefont {M\"uck}}]{Finke:2022lsu}%
  \BibitemOpen
  \bibfield  {author} {\bibinfo {author} {\bibfnamefont {Thorben}\ \bibnamefont {Finke}}, \bibinfo {author} {\bibfnamefont {Michael}\ \bibnamefont {Kr\"amer}}, \bibinfo {author} {\bibfnamefont {Maximilian}\ \bibnamefont {Lipp}}, \ and\ \bibinfo {author} {\bibfnamefont {Alexander}\ \bibnamefont {M\"uck}},\ }\bibfield  {title} {\enquote {\bibinfo {title} {{Boosting mono-jet searches with model-agnostic machine learning}},}\ }\href {\doibase 10.1007/JHEP08(2022)015} {\bibfield  {journal} {\bibinfo  {journal} {JHEP}\ }\textbf {\bibinfo {volume} {08}},\ \bibinfo {pages} {015} (\bibinfo {year} {2022})},\ \Eprint {http://arxiv.org/abs/2204.11889} {arXiv:2204.11889 [hep-ph]} \BibitemShut {NoStop}%
\bibitem [{\citenamefont {Verheyen}(2022)}]{Verheyen:2022tov}%
  \BibitemOpen
  \bibfield  {author} {\bibinfo {author} {\bibfnamefont {Rob}\ \bibnamefont {Verheyen}},\ }\bibfield  {title} {\enquote {\bibinfo {title} {{Event Generation and Density Estimation with Surjective Normalizing Flows}},}\ }\href {\doibase 10.21468/SciPostPhys.13.3.047} {\bibfield  {journal} {\bibinfo  {journal} {SciPost Phys.}\ }\textbf {\bibinfo {volume} {13}},\ \bibinfo {pages} {047} (\bibinfo {year} {2022})},\ \Eprint {http://arxiv.org/abs/2205.01697} {arXiv:2205.01697 [hep-ph]} \BibitemShut {NoStop}%
\bibitem [{\citenamefont {Dillon}\ \emph {et~al.}(2022)\citenamefont {Dillon}, \citenamefont {Mastandrea},\ and\ \citenamefont {Nachman}}]{Dillon:2022tmm}%
  \BibitemOpen
  \bibfield  {author} {\bibinfo {author} {\bibfnamefont {Barry~M.}\ \bibnamefont {Dillon}}, \bibinfo {author} {\bibfnamefont {Radha}\ \bibnamefont {Mastandrea}}, \ and\ \bibinfo {author} {\bibfnamefont {Benjamin}\ \bibnamefont {Nachman}},\ }\bibfield  {title} {\enquote {\bibinfo {title} {{Self-supervised anomaly detection for new physics}},}\ }\href {\doibase 10.1103/PhysRevD.106.056005} {\bibfield  {journal} {\bibinfo  {journal} {Phys. Rev. D}\ }\textbf {\bibinfo {volume} {106}},\ \bibinfo {pages} {056005} (\bibinfo {year} {2022})},\ \Eprint {http://arxiv.org/abs/2205.10380} {arXiv:2205.10380 [hep-ph]} \BibitemShut {NoStop}%
\bibitem [{\citenamefont {Alvi}\ \emph {et~al.}(2023)\citenamefont {Alvi}, \citenamefont {Bauer},\ and\ \citenamefont {Nachman}}]{Alvi:2022fkk}%
  \BibitemOpen
  \bibfield  {author} {\bibinfo {author} {\bibfnamefont {Sulaiman}\ \bibnamefont {Alvi}}, \bibinfo {author} {\bibfnamefont {Christian~W.}\ \bibnamefont {Bauer}}, \ and\ \bibinfo {author} {\bibfnamefont {Benjamin}\ \bibnamefont {Nachman}},\ }\bibfield  {title} {\enquote {\bibinfo {title} {{Quantum anomaly detection for collider physics}},}\ }\href {\doibase 10.1007/JHEP02(2023)220} {\bibfield  {journal} {\bibinfo  {journal} {JHEP}\ }\textbf {\bibinfo {volume} {02}},\ \bibinfo {pages} {220} (\bibinfo {year} {2023})},\ \Eprint {http://arxiv.org/abs/2206.08391} {arXiv:2206.08391 [hep-ph]} \BibitemShut {NoStop}%
\bibitem [{\citenamefont {Dillon}\ \emph {et~al.}(2023)\citenamefont {Dillon}, \citenamefont {Favaro}, \citenamefont {Plehn}, \citenamefont {Sorrenson},\ and\ \citenamefont {Kr\"amer}}]{Dillon:2022mkq}%
  \BibitemOpen
  \bibfield  {author} {\bibinfo {author} {\bibfnamefont {Barry~M.}\ \bibnamefont {Dillon}}, \bibinfo {author} {\bibfnamefont {Luigi}\ \bibnamefont {Favaro}}, \bibinfo {author} {\bibfnamefont {Tilman}\ \bibnamefont {Plehn}}, \bibinfo {author} {\bibfnamefont {Peter}\ \bibnamefont {Sorrenson}}, \ and\ \bibinfo {author} {\bibfnamefont {Michael}\ \bibnamefont {Kr\"amer}},\ }\bibfield  {title} {\enquote {\bibinfo {title} {{A normalized autoencoder for LHC triggers}},}\ }\href {\doibase 10.21468/SciPostPhysCore.6.4.074} {\bibfield  {journal} {\bibinfo  {journal} {SciPost Phys. Core}\ }\textbf {\bibinfo {volume} {6}},\ \bibinfo {pages} {074} (\bibinfo {year} {2023})},\ \Eprint {http://arxiv.org/abs/2206.14225} {arXiv:2206.14225 [hep-ph]} \BibitemShut {NoStop}%
\bibitem [{\citenamefont {Caron}\ \emph {et~al.}(2023)\citenamefont {Caron}, \citenamefont {de~Austri},\ and\ \citenamefont {Zhang}}]{Caron:2022wrw}%
  \BibitemOpen
  \bibfield  {author} {\bibinfo {author} {\bibfnamefont {Sascha}\ \bibnamefont {Caron}}, \bibinfo {author} {\bibfnamefont {Roberto~Ruiz}\ \bibnamefont {de~Austri}}, \ and\ \bibinfo {author} {\bibfnamefont {Zhongyi}\ \bibnamefont {Zhang}},\ }\bibfield  {title} {\enquote {\bibinfo {title} {{Mixture-of-Theories training: can we find new physics and anomalies better by mixing physical theories?}}}\ }\href {\doibase 10.1007/JHEP03(2023)004} {\bibfield  {journal} {\bibinfo  {journal} {JHEP}\ }\textbf {\bibinfo {volume} {03}},\ \bibinfo {pages} {004} (\bibinfo {year} {2023})},\ \Eprint {http://arxiv.org/abs/2207.07631} {arXiv:2207.07631 [hep-ph]} \BibitemShut {NoStop}%
\bibitem [{\citenamefont {Park}\ \emph {et~al.}(2023)\citenamefont {Park}, \citenamefont {Harris},\ and\ \citenamefont {Ostdiek}}]{Park:2022zov}%
  \BibitemOpen
  \bibfield  {author} {\bibinfo {author} {\bibfnamefont {Sang~Eon}\ \bibnamefont {Park}}, \bibinfo {author} {\bibfnamefont {Philip}\ \bibnamefont {Harris}}, \ and\ \bibinfo {author} {\bibfnamefont {Bryan}\ \bibnamefont {Ostdiek}},\ }\bibfield  {title} {\enquote {\bibinfo {title} {{Neural embedding: learning the embedding of the manifold of physics data}},}\ }\href {\doibase 10.1007/JHEP07(2023)108} {\bibfield  {journal} {\bibinfo  {journal} {JHEP}\ }\textbf {\bibinfo {volume} {07}},\ \bibinfo {pages} {108} (\bibinfo {year} {2023})},\ \Eprint {http://arxiv.org/abs/2208.05484} {arXiv:2208.05484 [hep-ph]} \BibitemShut {NoStop}%
\bibitem [{\citenamefont {Kasieczka}\ \emph {et~al.}(2023)\citenamefont {Kasieczka}, \citenamefont {Mastandrea}, \citenamefont {Mikuni}, \citenamefont {Nachman}, \citenamefont {Pettee},\ and\ \citenamefont {Shih}}]{Kasieczka:2022naq}%
  \BibitemOpen
  \bibfield  {author} {\bibinfo {author} {\bibfnamefont {Gregor}\ \bibnamefont {Kasieczka}}, \bibinfo {author} {\bibfnamefont {Radha}\ \bibnamefont {Mastandrea}}, \bibinfo {author} {\bibfnamefont {Vinicius}\ \bibnamefont {Mikuni}}, \bibinfo {author} {\bibfnamefont {Benjamin}\ \bibnamefont {Nachman}}, \bibinfo {author} {\bibfnamefont {Mariel}\ \bibnamefont {Pettee}}, \ and\ \bibinfo {author} {\bibfnamefont {David}\ \bibnamefont {Shih}},\ }\bibfield  {title} {\enquote {\bibinfo {title} {{Anomaly detection under coordinate transformations}},}\ }\href {\doibase 10.1103/PhysRevD.107.015009} {\bibfield  {journal} {\bibinfo  {journal} {Phys. Rev. D}\ }\textbf {\bibinfo {volume} {107}},\ \bibinfo {pages} {015009} (\bibinfo {year} {2023})},\ \Eprint {http://arxiv.org/abs/2209.06225} {arXiv:2209.06225 [hep-ph]} \BibitemShut {NoStop}%
\bibitem [{\citenamefont {Kamenik}\ and\ \citenamefont {Szewc}(2023)}]{Kamenik:2022qxs}%
  \BibitemOpen
  \bibfield  {author} {\bibinfo {author} {\bibfnamefont {Jernej~F.}\ \bibnamefont {Kamenik}}\ and\ \bibinfo {author} {\bibfnamefont {Manuel}\ \bibnamefont {Szewc}},\ }\bibfield  {title} {\enquote {\bibinfo {title} {{Null hypothesis test for anomaly detection}},}\ }\href {\doibase 10.1016/j.physletb.2023.137836} {\bibfield  {journal} {\bibinfo  {journal} {Phys. Lett. B}\ }\textbf {\bibinfo {volume} {840}},\ \bibinfo {pages} {137836} (\bibinfo {year} {2023})},\ \Eprint {http://arxiv.org/abs/2210.02226} {arXiv:2210.02226 [hep-ph]} \BibitemShut {NoStop}%
\bibitem [{\citenamefont {Hallin}\ \emph {et~al.}(2023)\citenamefont {Hallin}, \citenamefont {Kasieczka}, \citenamefont {Quadfasel}, \citenamefont {Shih},\ and\ \citenamefont {Sommerhalder}}]{Hallin:2022eoq}%
  \BibitemOpen
  \bibfield  {author} {\bibinfo {author} {\bibfnamefont {Anna}\ \bibnamefont {Hallin}}, \bibinfo {author} {\bibfnamefont {Gregor}\ \bibnamefont {Kasieczka}}, \bibinfo {author} {\bibfnamefont {Tobias}\ \bibnamefont {Quadfasel}}, \bibinfo {author} {\bibfnamefont {David}\ \bibnamefont {Shih}}, \ and\ \bibinfo {author} {\bibfnamefont {Manuel}\ \bibnamefont {Sommerhalder}},\ }\bibfield  {title} {\enquote {\bibinfo {title} {{Resonant anomaly detection without background sculpting}},}\ }\href {\doibase 10.1103/PhysRevD.107.114012} {\bibfield  {journal} {\bibinfo  {journal} {Phys. Rev. D}\ }\textbf {\bibinfo {volume} {107}},\ \bibinfo {pages} {114012} (\bibinfo {year} {2023})},\ \Eprint {http://arxiv.org/abs/2210.14924} {arXiv:2210.14924 [hep-ph]} \BibitemShut {NoStop}%
\bibitem [{\citenamefont {Araz}\ and\ \citenamefont {Spannowsky}(2023)}]{Araz:2022zxk}%
  \BibitemOpen
  \bibfield  {author} {\bibinfo {author} {\bibfnamefont {Jack~Y.}\ \bibnamefont {Araz}}\ and\ \bibinfo {author} {\bibfnamefont {Michael}\ \bibnamefont {Spannowsky}},\ }\bibfield  {title} {\enquote {\bibinfo {title} {{Quantum-probabilistic Hamiltonian learning for generative modeling and anomaly detection}},}\ }\href {\doibase 10.1103/PhysRevA.108.062422} {\bibfield  {journal} {\bibinfo  {journal} {Phys. Rev. A}\ }\textbf {\bibinfo {volume} {108}},\ \bibinfo {pages} {062422} (\bibinfo {year} {2023})},\ \Eprint {http://arxiv.org/abs/2211.03803} {arXiv:2211.03803 [quant-ph]} \BibitemShut {NoStop}%
\bibitem [{\citenamefont {Mastandrea}\ and\ \citenamefont {Nachman}(2022)}]{Mastandrea:2022vas}%
  \BibitemOpen
  \bibfield  {author} {\bibinfo {author} {\bibfnamefont {Radha}\ \bibnamefont {Mastandrea}}\ and\ \bibinfo {author} {\bibfnamefont {Benjamin}\ \bibnamefont {Nachman}},\ }\bibfield  {title} {\enquote {\bibinfo {title} {{Efficiently Moving Instead of Reweighting Collider Events with Machine Learning}},}\ }in\ \href@noop {} {\emph {\bibinfo {booktitle} {{36th Conference on Neural Information Processing Systems}: {Workshop on Machine Learning and the Physical Sciences}}}}\ (\bibinfo {year} {2022})\ \Eprint {http://arxiv.org/abs/2212.06155} {arXiv:2212.06155 [hep-ph]} \BibitemShut {NoStop}%
\bibitem [{\citenamefont {Schuhmacher}\ \emph {et~al.}(2023)\citenamefont {Schuhmacher}, \citenamefont {Boggia}, \citenamefont {Belis}, \citenamefont {Puljak}, \citenamefont {Grossi}, \citenamefont {Pierini}, \citenamefont {Vallecorsa}, \citenamefont {Tacchino}, \citenamefont {Barkoutsos},\ and\ \citenamefont {Tavernelli}}]{Schuhmacher:2023pro}%
  \BibitemOpen
  \bibfield  {author} {\bibinfo {author} {\bibfnamefont {Julian}\ \bibnamefont {Schuhmacher}}, \bibinfo {author} {\bibfnamefont {Laura}\ \bibnamefont {Boggia}}, \bibinfo {author} {\bibfnamefont {Vasilis}\ \bibnamefont {Belis}}, \bibinfo {author} {\bibfnamefont {Ema}\ \bibnamefont {Puljak}}, \bibinfo {author} {\bibfnamefont {Michele}\ \bibnamefont {Grossi}}, \bibinfo {author} {\bibfnamefont {Maurizio}\ \bibnamefont {Pierini}}, \bibinfo {author} {\bibfnamefont {Sofia}\ \bibnamefont {Vallecorsa}}, \bibinfo {author} {\bibfnamefont {Francesco}\ \bibnamefont {Tacchino}}, \bibinfo {author} {\bibfnamefont {Panagiotis}\ \bibnamefont {Barkoutsos}}, \ and\ \bibinfo {author} {\bibfnamefont {Ivano}\ \bibnamefont {Tavernelli}},\ }\bibfield  {title} {\enquote {\bibinfo {title} {{Unravelling physics beyond the standard model with classical and quantum anomaly detection}},}\ }\href {\doibase 10.1088/2632-2153/ad07f7} {\bibfield  {journal} {\bibinfo  {journal} {Mach. Learn. Sci. Tech.}\ }\textbf {\bibinfo {volume} {4}},\
  \bibinfo {pages} {045031} (\bibinfo {year} {2023})},\ \Eprint {http://arxiv.org/abs/2301.10787} {arXiv:2301.10787 [hep-ex]} \BibitemShut {NoStop}%
\bibitem [{\citenamefont {Golling}\ \emph {et~al.}(2023)\citenamefont {Golling} \emph {et~al.}}]{Golling:2023juz}%
  \BibitemOpen
  \bibfield  {author} {\bibinfo {author} {\bibfnamefont {Tobias}\ \bibnamefont {Golling}} \emph {et~al.},\ }\bibfield  {title} {\enquote {\bibinfo {title} {{The Mass-ive Issue: Anomaly Detection in Jet Physics}},}\ }in\ \href@noop {} {\emph {\bibinfo {booktitle} {{34th Conference on Neural Information Processing Systems}}}}\ (\bibinfo {year} {2023})\ \Eprint {http://arxiv.org/abs/2303.14134} {arXiv:2303.14134 [hep-ph]} \BibitemShut {NoStop}%
\bibitem [{\citenamefont {Roche}\ \emph {et~al.}(2024)\citenamefont {Roche}, \citenamefont {Bayer}, \citenamefont {Carlson}, \citenamefont {Ouligian}, \citenamefont {Serhiayenka}, \citenamefont {Stelzer},\ and\ \citenamefont {Hong}}]{Roche:2023int}%
  \BibitemOpen
  \bibfield  {author} {\bibinfo {author} {\bibfnamefont {Stephen}\ \bibnamefont {Roche}}, \bibinfo {author} {\bibfnamefont {Quincy}\ \bibnamefont {Bayer}}, \bibinfo {author} {\bibfnamefont {Benjamin}\ \bibnamefont {Carlson}}, \bibinfo {author} {\bibfnamefont {William}\ \bibnamefont {Ouligian}}, \bibinfo {author} {\bibfnamefont {Pavel}\ \bibnamefont {Serhiayenka}}, \bibinfo {author} {\bibfnamefont {Joerg}\ \bibnamefont {Stelzer}}, \ and\ \bibinfo {author} {\bibfnamefont {Tae~Min}\ \bibnamefont {Hong}},\ }\bibfield  {title} {\enquote {\bibinfo {title} {{Nanosecond anomaly detection with decision trees and real-time application to exotic Higgs decays}},}\ }\href {\doibase 10.1038/s41467-024-47704-8} {\bibfield  {journal} {\bibinfo  {journal} {Nature Commun.}\ }\textbf {\bibinfo {volume} {15}},\ \bibinfo {pages} {3527} (\bibinfo {year} {2024})},\ \Eprint {http://arxiv.org/abs/2304.03836} {arXiv:2304.03836 [hep-ex]} \BibitemShut {NoStop}%
\bibitem [{\citenamefont {Sengupta}\ \emph {et~al.}(2023)\citenamefont {Sengupta}, \citenamefont {Klein}, \citenamefont {Raine},\ and\ \citenamefont {Golling}}]{Sengupta:2023xqy}%
  \BibitemOpen
  \bibfield  {author} {\bibinfo {author} {\bibfnamefont {Debajyoti}\ \bibnamefont {Sengupta}}, \bibinfo {author} {\bibfnamefont {Samuel}\ \bibnamefont {Klein}}, \bibinfo {author} {\bibfnamefont {John~Andrew}\ \bibnamefont {Raine}}, \ and\ \bibinfo {author} {\bibfnamefont {Tobias}\ \bibnamefont {Golling}},\ }\bibfield  {title} {\enquote {\bibinfo {title} {{CURTAINs Flows For Flows: Constructing Unobserved Regions with Maximum Likelihood Estimation}},}\ }\href@noop {} {\  (\bibinfo {year} {2023})},\ \Eprint {http://arxiv.org/abs/2305.04646} {arXiv:2305.04646 [hep-ph]} \BibitemShut {NoStop}%
\bibitem [{\citenamefont {Aguilar-Saavedra}\ \emph {et~al.}(2023)\citenamefont {Aguilar-Saavedra}, \citenamefont {Arganda}, \citenamefont {Joaquim}, \citenamefont {Sand\'a~Seoane},\ and\ \citenamefont {Seabra}}]{Aguilar-Saavedra:2023pde}%
  \BibitemOpen
  \bibfield  {author} {\bibinfo {author} {\bibfnamefont {J.~A.}\ \bibnamefont {Aguilar-Saavedra}}, \bibinfo {author} {\bibfnamefont {E.}~\bibnamefont {Arganda}}, \bibinfo {author} {\bibfnamefont {F.~R.}\ \bibnamefont {Joaquim}}, \bibinfo {author} {\bibfnamefont {R.~M.}\ \bibnamefont {Sand\'a~Seoane}}, \ and\ \bibinfo {author} {\bibfnamefont {J.~F.}\ \bibnamefont {Seabra}},\ }\bibfield  {title} {\enquote {\bibinfo {title} {{Gradient Boosting MUST taggers for highly-boosted jets}},}\ }\href@noop {} {\  (\bibinfo {year} {2023})},\ \Eprint {http://arxiv.org/abs/2305.04957} {arXiv:2305.04957 [hep-ph]} \BibitemShut {NoStop}%
\bibitem [{\citenamefont {Vaslin}\ \emph {et~al.}(2023)\citenamefont {Vaslin}, \citenamefont {Barra},\ and\ \citenamefont {Donini}}]{Vaslin:2023lig}%
  \BibitemOpen
  \bibfield  {author} {\bibinfo {author} {\bibfnamefont {Louis}\ \bibnamefont {Vaslin}}, \bibinfo {author} {\bibfnamefont {Vincent}\ \bibnamefont {Barra}}, \ and\ \bibinfo {author} {\bibfnamefont {Julien}\ \bibnamefont {Donini}},\ }\bibfield  {title} {\enquote {\bibinfo {title} {{GAN-AE: an anomaly detection algorithm for New Physics search in LHC data}},}\ }\href {\doibase 10.1140/epjc/s10052-023-12169-4} {\bibfield  {journal} {\bibinfo  {journal} {Eur. Phys. J. C}\ }\textbf {\bibinfo {volume} {83}},\ \bibinfo {pages} {1008} (\bibinfo {year} {2023})},\ \Eprint {http://arxiv.org/abs/2305.15179} {arXiv:2305.15179 [hep-ex]} \BibitemShut {NoStop}%
\bibitem [{\citenamefont {Aad}\ \emph {et~al.}(2023)\citenamefont {Aad} \emph {et~al.}}]{ATLAS:2023azi}%
  \BibitemOpen
  \bibfield  {author} {\bibinfo {author} {\bibfnamefont {Georges}\ \bibnamefont {Aad}} \emph {et~al.} (\bibinfo {collaboration} {ATLAS}),\ }\bibfield  {title} {\enquote {\bibinfo {title} {{Anomaly detection search for new resonances decaying into a Higgs boson and a generic new particle $X$ in hadronic final states using $\sqrt{s} = 13$ TeV $pp$ collisions with the ATLAS detector}},}\ }\href {\doibase 10.1103/PhysRevD.108.052009} {\bibfield  {journal} {\bibinfo  {journal} {Phys. Rev. D}\ }\textbf {\bibinfo {volume} {108}},\ \bibinfo {pages} {052009} (\bibinfo {year} {2023})},\ \Eprint {http://arxiv.org/abs/2306.03637} {arXiv:2306.03637 [hep-ex]} \BibitemShut {NoStop}%
\bibitem [{\citenamefont {Mikuni}\ and\ \citenamefont {Nachman}(2024)}]{Mikuni:2023tok}%
  \BibitemOpen
  \bibfield  {author} {\bibinfo {author} {\bibfnamefont {Vinicius}\ \bibnamefont {Mikuni}}\ and\ \bibinfo {author} {\bibfnamefont {Benjamin}\ \bibnamefont {Nachman}},\ }\bibfield  {title} {\enquote {\bibinfo {title} {{High-dimensional and Permutation Invariant Anomaly Detection}},}\ }\href {\doibase 10.21468/SciPostPhys.16.3.062} {\bibfield  {journal} {\bibinfo  {journal} {SciPost Phys.}\ }\textbf {\bibinfo {volume} {16}},\ \bibinfo {pages} {062} (\bibinfo {year} {2024})},\ \Eprint {http://arxiv.org/abs/2306.03933} {arXiv:2306.03933 [hep-ph]} \BibitemShut {NoStop}%
\bibitem [{\citenamefont {Golling}\ \emph {et~al.}(2024)\citenamefont {Golling}, \citenamefont {Kasieczka}, \citenamefont {Krause}, \citenamefont {Mastandrea}, \citenamefont {Nachman}, \citenamefont {Raine}, \citenamefont {Sengupta}, \citenamefont {Shih},\ and\ \citenamefont {Sommerhalder}}]{Golling:2023yjq}%
  \BibitemOpen
  \bibfield  {author} {\bibinfo {author} {\bibfnamefont {Tobias}\ \bibnamefont {Golling}}, \bibinfo {author} {\bibfnamefont {Gregor}\ \bibnamefont {Kasieczka}}, \bibinfo {author} {\bibfnamefont {Claudius}\ \bibnamefont {Krause}}, \bibinfo {author} {\bibfnamefont {Radha}\ \bibnamefont {Mastandrea}}, \bibinfo {author} {\bibfnamefont {Benjamin}\ \bibnamefont {Nachman}}, \bibinfo {author} {\bibfnamefont {John~Andrew}\ \bibnamefont {Raine}}, \bibinfo {author} {\bibfnamefont {Debajyoti}\ \bibnamefont {Sengupta}}, \bibinfo {author} {\bibfnamefont {David}\ \bibnamefont {Shih}}, \ and\ \bibinfo {author} {\bibfnamefont {Manuel}\ \bibnamefont {Sommerhalder}},\ }\bibfield  {title} {\enquote {\bibinfo {title} {{The interplay of machine learning-based resonant anomaly detection methods}},}\ }\href {\doibase 10.1140/epjc/s10052-024-12607-x} {\bibfield  {journal} {\bibinfo  {journal} {Eur. Phys. J. C}\ }\textbf {\bibinfo {volume} {84}},\ \bibinfo {pages} {241} (\bibinfo {year} {2024})},\ \Eprint
  {http://arxiv.org/abs/2307.11157} {arXiv:2307.11157 [hep-ph]} \BibitemShut {NoStop}%
\bibitem [{\citenamefont {Chekanov}\ and\ \citenamefont {Zhang}(2024)}]{Chekanov:2023uot}%
  \BibitemOpen
  \bibfield  {author} {\bibinfo {author} {\bibfnamefont {Sergei~V.}\ \bibnamefont {Chekanov}}\ and\ \bibinfo {author} {\bibfnamefont {Rui}\ \bibnamefont {Zhang}},\ }\bibfield  {title} {\enquote {\bibinfo {title} {{Enhancing the hunt for new phenomena in dijet final states using anomaly detection filters at the high-luminosity large Hadron Collider}},}\ }\href {\doibase 10.1140/epjp/s13360-024-05018-0} {\bibfield  {journal} {\bibinfo  {journal} {Eur. Phys. J. Plus}\ }\textbf {\bibinfo {volume} {139}},\ \bibinfo {pages} {237} (\bibinfo {year} {2024})},\ \Eprint {http://arxiv.org/abs/2308.02671} {arXiv:2308.02671 [hep-ex]} \BibitemShut {NoStop}%
\bibitem [{\citenamefont {Abadjiev}\ \emph {et~al.}(2024)\citenamefont {Abadjiev} \emph {et~al.}}]{CMSECAL:2023fvz}%
  \BibitemOpen
  \bibfield  {author} {\bibinfo {author} {\bibfnamefont {D.}~\bibnamefont {Abadjiev}} \emph {et~al.} (\bibinfo {collaboration} {CMS ECAL}),\ }\bibfield  {title} {\enquote {\bibinfo {title} {{Autoencoder-Based Anomaly Detection System for Online Data Quality Monitoring of the CMS Electromagnetic Calorimeter}},}\ }\href {\doibase 10.1007/s41781-024-00118-z} {\bibfield  {journal} {\bibinfo  {journal} {Comput. Softw. Big Sci.}\ }\textbf {\bibinfo {volume} {8}},\ \bibinfo {pages} {11} (\bibinfo {year} {2024})},\ \Eprint {http://arxiv.org/abs/2309.10157} {arXiv:2309.10157 [physics.ins-det]} \BibitemShut {NoStop}%
\bibitem [{\citenamefont {Bickendorf}\ \emph {et~al.}(2024)\citenamefont {Bickendorf}, \citenamefont {Drees}, \citenamefont {Kasieczka}, \citenamefont {Krause},\ and\ \citenamefont {Shih}}]{Bickendorf:2023nej}%
  \BibitemOpen
  \bibfield  {author} {\bibinfo {author} {\bibfnamefont {Gerrit}\ \bibnamefont {Bickendorf}}, \bibinfo {author} {\bibfnamefont {Manuel}\ \bibnamefont {Drees}}, \bibinfo {author} {\bibfnamefont {Gregor}\ \bibnamefont {Kasieczka}}, \bibinfo {author} {\bibfnamefont {Claudius}\ \bibnamefont {Krause}}, \ and\ \bibinfo {author} {\bibfnamefont {David}\ \bibnamefont {Shih}},\ }\bibfield  {title} {\enquote {\bibinfo {title} {{Combining resonant and tail-based anomaly detection}},}\ }\href {\doibase 10.1103/PhysRevD.109.096031} {\bibfield  {journal} {\bibinfo  {journal} {Phys. Rev. D}\ }\textbf {\bibinfo {volume} {109}},\ \bibinfo {pages} {096031} (\bibinfo {year} {2024})},\ \Eprint {http://arxiv.org/abs/2309.12918} {arXiv:2309.12918 [hep-ph]} \BibitemShut {NoStop}%
\bibitem [{\citenamefont {Finke}\ \emph {et~al.}(2024)\citenamefont {Finke}, \citenamefont {Hein}, \citenamefont {Kasieczka}, \citenamefont {Kr\"amer}, \citenamefont {M\"uck}, \citenamefont {Prangchaikul}, \citenamefont {Quadfasel}, \citenamefont {Shih},\ and\ \citenamefont {Sommerhalder}}]{Finke:2023ltw}%
  \BibitemOpen
  \bibfield  {author} {\bibinfo {author} {\bibfnamefont {Thorben}\ \bibnamefont {Finke}}, \bibinfo {author} {\bibfnamefont {Marie}\ \bibnamefont {Hein}}, \bibinfo {author} {\bibfnamefont {Gregor}\ \bibnamefont {Kasieczka}}, \bibinfo {author} {\bibfnamefont {Michael}\ \bibnamefont {Kr\"amer}}, \bibinfo {author} {\bibfnamefont {Alexander}\ \bibnamefont {M\"uck}}, \bibinfo {author} {\bibfnamefont {Parada}\ \bibnamefont {Prangchaikul}}, \bibinfo {author} {\bibfnamefont {Tobias}\ \bibnamefont {Quadfasel}}, \bibinfo {author} {\bibfnamefont {David}\ \bibnamefont {Shih}}, \ and\ \bibinfo {author} {\bibfnamefont {Manuel}\ \bibnamefont {Sommerhalder}},\ }\bibfield  {title} {\enquote {\bibinfo {title} {{Tree-based algorithms for weakly supervised anomaly detection}},}\ }\href {\doibase 10.1103/PhysRevD.109.034033} {\bibfield  {journal} {\bibinfo  {journal} {Phys. Rev. D}\ }\textbf {\bibinfo {volume} {109}},\ \bibinfo {pages} {034033} (\bibinfo {year} {2024})},\ \Eprint {http://arxiv.org/abs/2309.13111} {arXiv:2309.13111
  [hep-ph]} \BibitemShut {NoStop}%
\bibitem [{\citenamefont {Buhmann}\ \emph {et~al.}(2024)\citenamefont {Buhmann}, \citenamefont {Ewen}, \citenamefont {Kasieczka}, \citenamefont {Mikuni}, \citenamefont {Nachman},\ and\ \citenamefont {Shih}}]{Buhmann:2023acn}%
  \BibitemOpen
  \bibfield  {author} {\bibinfo {author} {\bibfnamefont {Erik}\ \bibnamefont {Buhmann}}, \bibinfo {author} {\bibfnamefont {Cedric}\ \bibnamefont {Ewen}}, \bibinfo {author} {\bibfnamefont {Gregor}\ \bibnamefont {Kasieczka}}, \bibinfo {author} {\bibfnamefont {Vinicius}\ \bibnamefont {Mikuni}}, \bibinfo {author} {\bibfnamefont {Benjamin}\ \bibnamefont {Nachman}}, \ and\ \bibinfo {author} {\bibfnamefont {David}\ \bibnamefont {Shih}},\ }\bibfield  {title} {\enquote {\bibinfo {title} {{Full phase space resonant anomaly detection}},}\ }\href {\doibase 10.1103/PhysRevD.109.055015} {\bibfield  {journal} {\bibinfo  {journal} {Phys. Rev. D}\ }\textbf {\bibinfo {volume} {109}},\ \bibinfo {pages} {055015} (\bibinfo {year} {2024})},\ \Eprint {http://arxiv.org/abs/2310.06897} {arXiv:2310.06897 [hep-ph]} \BibitemShut {NoStop}%
\bibitem [{\citenamefont {Freytsis}\ \emph {et~al.}(2024)\citenamefont {Freytsis}, \citenamefont {Perelstein},\ and\ \citenamefont {San}}]{Freytsis:2023cjr}%
  \BibitemOpen
  \bibfield  {author} {\bibinfo {author} {\bibfnamefont {Marat}\ \bibnamefont {Freytsis}}, \bibinfo {author} {\bibfnamefont {Maxim}\ \bibnamefont {Perelstein}}, \ and\ \bibinfo {author} {\bibfnamefont {Yik~Chuen}\ \bibnamefont {San}},\ }\bibfield  {title} {\enquote {\bibinfo {title} {{Anomaly detection in the presence of irrelevant features}},}\ }\href {\doibase 10.1007/JHEP02(2024)220} {\bibfield  {journal} {\bibinfo  {journal} {JHEP}\ }\textbf {\bibinfo {volume} {02}},\ \bibinfo {pages} {220} (\bibinfo {year} {2024})},\ \Eprint {http://arxiv.org/abs/2310.13057} {arXiv:2310.13057 [hep-ph]} \BibitemShut {NoStop}%
\bibitem [{\citenamefont {Cohen}\ \emph {et~al.}(2014)\citenamefont {Cohen}, \citenamefont {Jankowiak}, \citenamefont {Lisanti}, \citenamefont {Lou},\ and\ \citenamefont {Wacker}}]{Cohen:2014epa}%
  \BibitemOpen
  \bibfield  {author} {\bibinfo {author} {\bibfnamefont {Timothy}\ \bibnamefont {Cohen}}, \bibinfo {author} {\bibfnamefont {Martin}\ \bibnamefont {Jankowiak}}, \bibinfo {author} {\bibfnamefont {Mariangela}\ \bibnamefont {Lisanti}}, \bibinfo {author} {\bibfnamefont {Hou~Keong}\ \bibnamefont {Lou}}, \ and\ \bibinfo {author} {\bibfnamefont {Jay~G.}\ \bibnamefont {Wacker}},\ }\bibfield  {title} {\enquote {\bibinfo {title} {{Jet Substructure Templates: Data-driven QCD Backgrounds for Fat Jet Searches}},}\ }\href {\doibase 10.1007/JHEP05(2014)005} {\bibfield  {journal} {\bibinfo  {journal} {JHEP}\ }\textbf {\bibinfo {volume} {05}},\ \bibinfo {pages} {005} (\bibinfo {year} {2014})},\ \Eprint {http://arxiv.org/abs/1402.0516} {arXiv:1402.0516 [hep-ph]} \BibitemShut {NoStop}%
\bibitem [{\citenamefont {Lin}\ \emph {et~al.}(2019)\citenamefont {Lin}, \citenamefont {Bhimji},\ and\ \citenamefont {Nachman}}]{Lin:2019htn}%
  \BibitemOpen
  \bibfield  {author} {\bibinfo {author} {\bibfnamefont {Joshua}\ \bibnamefont {Lin}}, \bibinfo {author} {\bibfnamefont {Wahid}\ \bibnamefont {Bhimji}}, \ and\ \bibinfo {author} {\bibfnamefont {Benjamin}\ \bibnamefont {Nachman}},\ }\bibfield  {title} {\enquote {\bibinfo {title} {{Machine Learning Templates for QCD Factorization in the Search for Physics Beyond the Standard Model}},}\ }\href {\doibase 10.1007/JHEP05(2019)181} {\bibfield  {journal} {\bibinfo  {journal} {JHEP}\ }\textbf {\bibinfo {volume} {05}},\ \bibinfo {pages} {181} (\bibinfo {year} {2019})},\ \Eprint {http://arxiv.org/abs/1903.02556} {arXiv:1903.02556 [hep-ph]} \BibitemShut {NoStop}%
\bibitem [{\citenamefont {Dery}\ \emph {et~al.}(2017)\citenamefont {Dery}, \citenamefont {Nachman}, \citenamefont {Rubbo},\ and\ \citenamefont {Schwartzman}}]{Dery:2017fap}%
  \BibitemOpen
  \bibfield  {author} {\bibinfo {author} {\bibfnamefont {Lucio~Mwinmaarong}\ \bibnamefont {Dery}}, \bibinfo {author} {\bibfnamefont {Benjamin}\ \bibnamefont {Nachman}}, \bibinfo {author} {\bibfnamefont {Francesco}\ \bibnamefont {Rubbo}}, \ and\ \bibinfo {author} {\bibfnamefont {Ariel}\ \bibnamefont {Schwartzman}},\ }\bibfield  {title} {\enquote {\bibinfo {title} {{Weakly Supervised Classification in High Energy Physics}},}\ }\href {\doibase 10.1007/JHEP05(2017)145} {\bibfield  {journal} {\bibinfo  {journal} {JHEP}\ }\textbf {\bibinfo {volume} {05}},\ \bibinfo {pages} {145} (\bibinfo {year} {2017})},\ \Eprint {http://arxiv.org/abs/1702.00414} {arXiv:1702.00414 [hep-ph]} \BibitemShut {NoStop}%
\bibitem [{\citenamefont {Cohen}\ \emph {et~al.}(2018)\citenamefont {Cohen}, \citenamefont {Freytsis},\ and\ \citenamefont {Ostdiek}}]{Cohen:2017exh}%
  \BibitemOpen
  \bibfield  {author} {\bibinfo {author} {\bibfnamefont {Timothy}\ \bibnamefont {Cohen}}, \bibinfo {author} {\bibfnamefont {Marat}\ \bibnamefont {Freytsis}}, \ and\ \bibinfo {author} {\bibfnamefont {Bryan}\ \bibnamefont {Ostdiek}},\ }\bibfield  {title} {\enquote {\bibinfo {title} {{(Machine) Learning to Do More with Less}},}\ }\href {\doibase 10.1007/JHEP02(2018)034} {\bibfield  {journal} {\bibinfo  {journal} {JHEP}\ }\textbf {\bibinfo {volume} {02}},\ \bibinfo {pages} {034} (\bibinfo {year} {2018})},\ \Eprint {http://arxiv.org/abs/1706.09451} {arXiv:1706.09451 [hep-ph]} \BibitemShut {NoStop}%
\bibitem [{\citenamefont {Metodiev}\ \emph {et~al.}(2017)\citenamefont {Metodiev}, \citenamefont {Nachman},\ and\ \citenamefont {Thaler}}]{Metodiev:2017vrx}%
  \BibitemOpen
  \bibfield  {author} {\bibinfo {author} {\bibfnamefont {Eric~M.}\ \bibnamefont {Metodiev}}, \bibinfo {author} {\bibfnamefont {Benjamin}\ \bibnamefont {Nachman}}, \ and\ \bibinfo {author} {\bibfnamefont {Jesse}\ \bibnamefont {Thaler}},\ }\bibfield  {title} {\enquote {\bibinfo {title} {{Classification without labels: Learning from mixed samples in high energy physics}},}\ }\href {\doibase 10.1007/JHEP10(2017)174} {\bibfield  {journal} {\bibinfo  {journal} {JHEP}\ }\textbf {\bibinfo {volume} {10}},\ \bibinfo {pages} {174} (\bibinfo {year} {2017})},\ \Eprint {http://arxiv.org/abs/1708.02949} {arXiv:1708.02949 [hep-ph]} \BibitemShut {NoStop}%
\bibitem [{\citenamefont {Komiske}\ \emph {et~al.}(2018{\natexlab{a}})\citenamefont {Komiske}, \citenamefont {Metodiev}, \citenamefont {Nachman},\ and\ \citenamefont {Schwartz}}]{Komiske:2018oaa}%
  \BibitemOpen
  \bibfield  {author} {\bibinfo {author} {\bibfnamefont {Patrick~T.}\ \bibnamefont {Komiske}}, \bibinfo {author} {\bibfnamefont {Eric~M.}\ \bibnamefont {Metodiev}}, \bibinfo {author} {\bibfnamefont {Benjamin}\ \bibnamefont {Nachman}}, \ and\ \bibinfo {author} {\bibfnamefont {Matthew~D.}\ \bibnamefont {Schwartz}},\ }\bibfield  {title} {\enquote {\bibinfo {title} {{Learning to classify from impure samples with high-dimensional data}},}\ }\href {\doibase 10.1103/PhysRevD.98.011502} {\bibfield  {journal} {\bibinfo  {journal} {Phys. Rev. D}\ }\textbf {\bibinfo {volume} {98}},\ \bibinfo {pages} {011502} (\bibinfo {year} {2018}{\natexlab{a}})},\ \Eprint {http://arxiv.org/abs/1801.10158} {arXiv:1801.10158 [hep-ph]} \BibitemShut {NoStop}%
\bibitem [{\citenamefont {Sirunyan}\ \emph {et~al.}(2020)\citenamefont {Sirunyan} \emph {et~al.}}]{Sirunyan:2019jud}%
  \BibitemOpen
  \bibfield  {author} {\bibinfo {author} {\bibfnamefont {Albert~M}\ \bibnamefont {Sirunyan}} \emph {et~al.} (\bibinfo {collaboration} {CMS}),\ }\bibfield  {title} {\enquote {\bibinfo {title} {{Measurement of the $\mathrm{t\bar{t}}\mathrm{b\bar{b}}$ production cross section in the all-jet final state in pp collisions at $\sqrt{s} =$ 13 TeV}},}\ }\href {\doibase 10.1016/j.physletb.2020.135285} {\bibfield  {journal} {\bibinfo  {journal} {Phys. Lett. B}\ }\textbf {\bibinfo {volume} {803}},\ \bibinfo {pages} {135285} (\bibinfo {year} {2020})},\ \Eprint {http://arxiv.org/abs/1909.05306} {arXiv:1909.05306 [hep-ex]} \BibitemShut {NoStop}%
\bibitem [{\citenamefont {Metodiev}\ and\ \citenamefont {Thaler}(2018)}]{Metodiev:2018ftz}%
  \BibitemOpen
  \bibfield  {author} {\bibinfo {author} {\bibfnamefont {Eric~M.}\ \bibnamefont {Metodiev}}\ and\ \bibinfo {author} {\bibfnamefont {Jesse}\ \bibnamefont {Thaler}},\ }\bibfield  {title} {\enquote {\bibinfo {title} {{Jet Topics: Disentangling Quarks and Gluons at Colliders}},}\ }\href {\doibase 10.1103/PhysRevLett.120.241602} {\bibfield  {journal} {\bibinfo  {journal} {Phys. Rev. Lett.}\ }\textbf {\bibinfo {volume} {120}},\ \bibinfo {pages} {241602} (\bibinfo {year} {2018})},\ \Eprint {http://arxiv.org/abs/1802.00008} {arXiv:1802.00008 [hep-ph]} \BibitemShut {NoStop}%
\bibitem [{\citenamefont {Komiske}\ \emph {et~al.}(2018{\natexlab{b}})\citenamefont {Komiske}, \citenamefont {Metodiev},\ and\ \citenamefont {Thaler}}]{Komiske:2018vkc}%
  \BibitemOpen
  \bibfield  {author} {\bibinfo {author} {\bibfnamefont {Patrick~T.}\ \bibnamefont {Komiske}}, \bibinfo {author} {\bibfnamefont {Eric~M.}\ \bibnamefont {Metodiev}}, \ and\ \bibinfo {author} {\bibfnamefont {Jesse}\ \bibnamefont {Thaler}},\ }\bibfield  {title} {\enquote {\bibinfo {title} {{An operational definition of quark and gluon jets}},}\ }\href {\doibase 10.1007/JHEP11(2018)059} {\bibfield  {journal} {\bibinfo  {journal} {JHEP}\ }\textbf {\bibinfo {volume} {11}},\ \bibinfo {pages} {059} (\bibinfo {year} {2018}{\natexlab{b}})},\ \Eprint {http://arxiv.org/abs/1809.01140} {arXiv:1809.01140 [hep-ph]} \BibitemShut {NoStop}%
\bibitem [{\citenamefont {Aad}\ \emph {et~al.}(2019)\citenamefont {Aad} \emph {et~al.}}]{Aad:2019onw}%
  \BibitemOpen
  \bibfield  {author} {\bibinfo {author} {\bibfnamefont {Georges}\ \bibnamefont {Aad}} \emph {et~al.} (\bibinfo {collaboration} {ATLAS}),\ }\bibfield  {title} {\enquote {\bibinfo {title} {{Properties of jet fragmentation using charged particles measured with the ATLAS detector in $pp$ collisions at $\sqrt{s}=13$ TeV}},}\ }\href {\doibase 10.1103/PhysRevD.100.052011} {\bibfield  {journal} {\bibinfo  {journal} {Phys. Rev. D}\ }\textbf {\bibinfo {volume} {100}},\ \bibinfo {pages} {052011} (\bibinfo {year} {2019})},\ \Eprint {http://arxiv.org/abs/1906.09254} {arXiv:1906.09254 [hep-ex]} \BibitemShut {NoStop}%
\bibitem [{\citenamefont {Komiske}\ \emph {et~al.}(2022)\citenamefont {Komiske}, \citenamefont {Kryhin},\ and\ \citenamefont {Thaler}}]{Komiske:2022vxg}%
  \BibitemOpen
  \bibfield  {author} {\bibinfo {author} {\bibfnamefont {Patrick~T.}\ \bibnamefont {Komiske}}, \bibinfo {author} {\bibfnamefont {Serhii}\ \bibnamefont {Kryhin}}, \ and\ \bibinfo {author} {\bibfnamefont {Jesse}\ \bibnamefont {Thaler}},\ }\bibfield  {title} {\enquote {\bibinfo {title} {{Disentangling quarks and gluons in CMS open data}},}\ }\href {\doibase 10.1103/PhysRevD.106.094021} {\bibfield  {journal} {\bibinfo  {journal} {Phys. Rev. D}\ }\textbf {\bibinfo {volume} {106}},\ \bibinfo {pages} {094021} (\bibinfo {year} {2022})},\ \Eprint {http://arxiv.org/abs/2205.04459} {arXiv:2205.04459 [hep-ph]} \BibitemShut {NoStop}%
\bibitem [{\citenamefont {Salam}(2010)}]{Salam:2010nqg}%
  \BibitemOpen
  \bibfield  {author} {\bibinfo {author} {\bibfnamefont {Gavin~P.}\ \bibnamefont {Salam}},\ }\bibfield  {title} {\enquote {\bibinfo {title} {{Towards Jetography}},}\ }\href {\doibase 10.1140/epjc/s10052-010-1314-6} {\bibfield  {journal} {\bibinfo  {journal} {Eur. Phys. J. C}\ }\textbf {\bibinfo {volume} {67}},\ \bibinfo {pages} {637--686} (\bibinfo {year} {2010})},\ \Eprint {http://arxiv.org/abs/0906.1833} {arXiv:0906.1833 [hep-ph]} \BibitemShut {NoStop}%
\bibitem [{\citenamefont {Larkoski}\ \emph {et~al.}(2020)\citenamefont {Larkoski}, \citenamefont {Moult},\ and\ \citenamefont {Nachman}}]{Larkoski:2017jix}%
  \BibitemOpen
  \bibfield  {author} {\bibinfo {author} {\bibfnamefont {Andrew~J.}\ \bibnamefont {Larkoski}}, \bibinfo {author} {\bibfnamefont {Ian}\ \bibnamefont {Moult}}, \ and\ \bibinfo {author} {\bibfnamefont {Benjamin}\ \bibnamefont {Nachman}},\ }\bibfield  {title} {\enquote {\bibinfo {title} {{Jet Substructure at the Large Hadron Collider: A Review of Recent Advances in Theory and Machine Learning}},}\ }\href {\doibase 10.1016/j.physrep.2019.11.001} {\bibfield  {journal} {\bibinfo  {journal} {Phys. Rept.}\ }\textbf {\bibinfo {volume} {841}},\ \bibinfo {pages} {1--63} (\bibinfo {year} {2020})},\ \Eprint {http://arxiv.org/abs/1709.04464} {arXiv:1709.04464 [hep-ph]} \BibitemShut {NoStop}%
\bibitem [{\citenamefont {Kogler}\ \emph {et~al.}(2019)\citenamefont {Kogler} \emph {et~al.}}]{Kogler:2018hem}%
  \BibitemOpen
  \bibfield  {author} {\bibinfo {author} {\bibfnamefont {Roman}\ \bibnamefont {Kogler}} \emph {et~al.},\ }\bibfield  {title} {\enquote {\bibinfo {title} {{Jet Substructure at the Large Hadron Collider: Experimental Review}},}\ }\href {\doibase 10.1103/RevModPhys.91.045003} {\bibfield  {journal} {\bibinfo  {journal} {Rev. Mod. Phys.}\ }\textbf {\bibinfo {volume} {91}},\ \bibinfo {pages} {045003} (\bibinfo {year} {2019})},\ \Eprint {http://arxiv.org/abs/1803.06991} {arXiv:1803.06991 [hep-ex]} \BibitemShut {NoStop}%
\bibitem [{\citenamefont {Marzani}\ \emph {et~al.}(2019)\citenamefont {Marzani}, \citenamefont {Soyez},\ and\ \citenamefont {Spannowsky}}]{Marzani:2019hun}%
  \BibitemOpen
  \bibfield  {author} {\bibinfo {author} {\bibfnamefont {Simone}\ \bibnamefont {Marzani}}, \bibinfo {author} {\bibfnamefont {Gregory}\ \bibnamefont {Soyez}}, \ and\ \bibinfo {author} {\bibfnamefont {Michael}\ \bibnamefont {Spannowsky}},\ }\href {\doibase 10.1007/978-3-030-15709-8} {\emph {\bibinfo {title} {{Looking inside jets: an introduction to jet substructure and boosted-object phenomenology}}}},\ Vol.\ \bibinfo {volume} {958}\ (\bibinfo  {publisher} {Springer},\ \bibinfo {year} {2019})\ \Eprint {http://arxiv.org/abs/1901.10342} {arXiv:1901.10342 [hep-ph]} \BibitemShut {NoStop}%
\bibitem [{\citenamefont {Bauer}\ \emph {et~al.}(2001)\citenamefont {Bauer}, \citenamefont {Fleming}, \citenamefont {Pirjol},\ and\ \citenamefont {Stewart}}]{Bauer:2000yr}%
  \BibitemOpen
  \bibfield  {author} {\bibinfo {author} {\bibfnamefont {Christian~W.}\ \bibnamefont {Bauer}}, \bibinfo {author} {\bibfnamefont {Sean}\ \bibnamefont {Fleming}}, \bibinfo {author} {\bibfnamefont {Dan}\ \bibnamefont {Pirjol}}, \ and\ \bibinfo {author} {\bibfnamefont {Iain~W.}\ \bibnamefont {Stewart}},\ }\bibfield  {title} {\enquote {\bibinfo {title} {{An Effective field theory for collinear and soft gluons: Heavy to light decays}},}\ }\href {\doibase 10.1103/PhysRevD.63.114020} {\bibfield  {journal} {\bibinfo  {journal} {Phys. Rev. D}\ }\textbf {\bibinfo {volume} {63}},\ \bibinfo {pages} {114020} (\bibinfo {year} {2001})},\ \Eprint {http://arxiv.org/abs/hep-ph/0011336} {arXiv:hep-ph/0011336} \BibitemShut {NoStop}%
\bibitem [{\citenamefont {Bauer}\ and\ \citenamefont {Stewart}(2001)}]{Bauer:2001ct}%
  \BibitemOpen
  \bibfield  {author} {\bibinfo {author} {\bibfnamefont {Christian~W.}\ \bibnamefont {Bauer}}\ and\ \bibinfo {author} {\bibfnamefont {Iain~W.}\ \bibnamefont {Stewart}},\ }\bibfield  {title} {\enquote {\bibinfo {title} {{Invariant operators in collinear effective theory}},}\ }\href {\doibase 10.1016/S0370-2693(01)00902-9} {\bibfield  {journal} {\bibinfo  {journal} {Phys. Lett. B}\ }\textbf {\bibinfo {volume} {516}},\ \bibinfo {pages} {134--142} (\bibinfo {year} {2001})},\ \Eprint {http://arxiv.org/abs/hep-ph/0107001} {arXiv:hep-ph/0107001} \BibitemShut {NoStop}%
\bibitem [{\citenamefont {Bauer}\ \emph {et~al.}(2002{\natexlab{a}})\citenamefont {Bauer}, \citenamefont {Pirjol},\ and\ \citenamefont {Stewart}}]{Bauer:2001yt}%
  \BibitemOpen
  \bibfield  {author} {\bibinfo {author} {\bibfnamefont {Christian~W.}\ \bibnamefont {Bauer}}, \bibinfo {author} {\bibfnamefont {Dan}\ \bibnamefont {Pirjol}}, \ and\ \bibinfo {author} {\bibfnamefont {Iain~W.}\ \bibnamefont {Stewart}},\ }\bibfield  {title} {\enquote {\bibinfo {title} {{Soft collinear factorization in effective field theory}},}\ }\href {\doibase 10.1103/PhysRevD.65.054022} {\bibfield  {journal} {\bibinfo  {journal} {Phys. Rev. D}\ }\textbf {\bibinfo {volume} {65}},\ \bibinfo {pages} {054022} (\bibinfo {year} {2002}{\natexlab{a}})},\ \Eprint {http://arxiv.org/abs/hep-ph/0109045} {arXiv:hep-ph/0109045} \BibitemShut {NoStop}%
\bibitem [{\citenamefont {Bauer}\ \emph {et~al.}(2002{\natexlab{b}})\citenamefont {Bauer}, \citenamefont {Fleming}, \citenamefont {Pirjol}, \citenamefont {Rothstein},\ and\ \citenamefont {Stewart}}]{Bauer:2002nz}%
  \BibitemOpen
  \bibfield  {author} {\bibinfo {author} {\bibfnamefont {Christian~W.}\ \bibnamefont {Bauer}}, \bibinfo {author} {\bibfnamefont {Sean}\ \bibnamefont {Fleming}}, \bibinfo {author} {\bibfnamefont {Dan}\ \bibnamefont {Pirjol}}, \bibinfo {author} {\bibfnamefont {Ira~Z.}\ \bibnamefont {Rothstein}}, \ and\ \bibinfo {author} {\bibfnamefont {Iain~W.}\ \bibnamefont {Stewart}},\ }\bibfield  {title} {\enquote {\bibinfo {title} {{Hard scattering factorization from effective field theory}},}\ }\href {\doibase 10.1103/PhysRevD.66.014017} {\bibfield  {journal} {\bibinfo  {journal} {Phys. Rev. D}\ }\textbf {\bibinfo {volume} {66}},\ \bibinfo {pages} {014017} (\bibinfo {year} {2002}{\natexlab{b}})},\ \Eprint {http://arxiv.org/abs/hep-ph/0202088} {arXiv:hep-ph/0202088} \BibitemShut {NoStop}%
\bibitem [{\citenamefont {Beneke}\ \emph {et~al.}(2002)\citenamefont {Beneke}, \citenamefont {Chapovsky}, \citenamefont {Diehl},\ and\ \citenamefont {Feldmann}}]{Beneke:2002ph}%
  \BibitemOpen
  \bibfield  {author} {\bibinfo {author} {\bibfnamefont {M.}~\bibnamefont {Beneke}}, \bibinfo {author} {\bibfnamefont {A.~P.}\ \bibnamefont {Chapovsky}}, \bibinfo {author} {\bibfnamefont {M.}~\bibnamefont {Diehl}}, \ and\ \bibinfo {author} {\bibfnamefont {T.}~\bibnamefont {Feldmann}},\ }\bibfield  {title} {\enquote {\bibinfo {title} {{Soft collinear effective theory and heavy to light currents beyond leading power}},}\ }\href {\doibase 10.1016/S0550-3213(02)00687-9} {\bibfield  {journal} {\bibinfo  {journal} {Nucl. Phys. B}\ }\textbf {\bibinfo {volume} {643}},\ \bibinfo {pages} {431--476} (\bibinfo {year} {2002})},\ \Eprint {http://arxiv.org/abs/hep-ph/0206152} {arXiv:hep-ph/0206152} \BibitemShut {NoStop}%
\bibitem [{\citenamefont {Collins}\ \emph {et~al.}(1989)\citenamefont {Collins}, \citenamefont {Soper},\ and\ \citenamefont {Sterman}}]{Collins:1989gx}%
  \BibitemOpen
  \bibfield  {author} {\bibinfo {author} {\bibfnamefont {John~C.}\ \bibnamefont {Collins}}, \bibinfo {author} {\bibfnamefont {Davison~E.}\ \bibnamefont {Soper}}, \ and\ \bibinfo {author} {\bibfnamefont {George~F.}\ \bibnamefont {Sterman}},\ }\bibfield  {title} {\enquote {\bibinfo {title} {{Factorization of Hard Processes in QCD}},}\ }\href {\doibase 10.1142/9789814503266_0001} {\bibfield  {journal} {\bibinfo  {journal} {Adv. Ser. Direct. High Energy Phys.}\ }\textbf {\bibinfo {volume} {5}},\ \bibinfo {pages} {1--91} (\bibinfo {year} {1989})},\ \Eprint {http://arxiv.org/abs/hep-ph/0409313} {arXiv:hep-ph/0409313} \BibitemShut {NoStop}%
\bibitem [{\citenamefont {Bauer}\ \emph {et~al.}(2009)\citenamefont {Bauer}, \citenamefont {Hornig},\ and\ \citenamefont {Tackmann}}]{Bauer:2008jx}%
  \BibitemOpen
  \bibfield  {author} {\bibinfo {author} {\bibfnamefont {Christian~W.}\ \bibnamefont {Bauer}}, \bibinfo {author} {\bibfnamefont {Andrew}\ \bibnamefont {Hornig}}, \ and\ \bibinfo {author} {\bibfnamefont {Frank~J.}\ \bibnamefont {Tackmann}},\ }\bibfield  {title} {\enquote {\bibinfo {title} {{Factorization for generic jet production}},}\ }\href {\doibase 10.1103/PhysRevD.79.114013} {\bibfield  {journal} {\bibinfo  {journal} {Phys. Rev. D}\ }\textbf {\bibinfo {volume} {79}},\ \bibinfo {pages} {114013} (\bibinfo {year} {2009})},\ \Eprint {http://arxiv.org/abs/0808.2191} {arXiv:0808.2191 [hep-ph]} \BibitemShut {NoStop}%
\bibitem [{\citenamefont {Feige}\ and\ \citenamefont {Schwartz}(2014)}]{Feige:2014wja}%
  \BibitemOpen
  \bibfield  {author} {\bibinfo {author} {\bibfnamefont {Ilya}\ \bibnamefont {Feige}}\ and\ \bibinfo {author} {\bibfnamefont {Matthew~D.}\ \bibnamefont {Schwartz}},\ }\bibfield  {title} {\enquote {\bibinfo {title} {{Hard-Soft-Collinear Factorization to All Orders}},}\ }\href {\doibase 10.1103/PhysRevD.90.105020} {\bibfield  {journal} {\bibinfo  {journal} {Phys. Rev. D}\ }\textbf {\bibinfo {volume} {90}},\ \bibinfo {pages} {105020} (\bibinfo {year} {2014})},\ \Eprint {http://arxiv.org/abs/1403.6472} {arXiv:1403.6472 [hep-ph]} \BibitemShut {NoStop}%
\bibitem [{\citenamefont {Sterman}(2022)}]{Sterman:2022gyf}%
  \BibitemOpen
  \bibfield  {author} {\bibinfo {author} {\bibfnamefont {George}\ \bibnamefont {Sterman}},\ }\bibfield  {title} {\enquote {\bibinfo {title} {{Comments on collinear factorization}},}\ }in\ \href@noop {} {\emph {\bibinfo {booktitle} {{Snowmass 2021}}}}\ (\bibinfo {year} {2022})\ \Eprint {http://arxiv.org/abs/2207.06507} {arXiv:2207.06507 [hep-ph]} \BibitemShut {NoStop}%
\bibitem [{\citenamefont {Berdine}\ \emph {et~al.}(2007)\citenamefont {Berdine}, \citenamefont {Kauer},\ and\ \citenamefont {Rainwater}}]{Berdine:2007uv}%
  \BibitemOpen
  \bibfield  {author} {\bibinfo {author} {\bibfnamefont {D.}~\bibnamefont {Berdine}}, \bibinfo {author} {\bibfnamefont {N.}~\bibnamefont {Kauer}}, \ and\ \bibinfo {author} {\bibfnamefont {D.}~\bibnamefont {Rainwater}},\ }\bibfield  {title} {\enquote {\bibinfo {title} {{Breakdown of the Narrow Width Approximation for New Physics}},}\ }\href {\doibase 10.1103/PhysRevLett.99.111601} {\bibfield  {journal} {\bibinfo  {journal} {Phys. Rev. Lett.}\ }\textbf {\bibinfo {volume} {99}},\ \bibinfo {pages} {111601} (\bibinfo {year} {2007})},\ \Eprint {http://arxiv.org/abs/hep-ph/0703058} {arXiv:hep-ph/0703058} \BibitemShut {NoStop}%
\bibitem [{\citenamefont {Uhlemann}\ and\ \citenamefont {Kauer}(2009)}]{Uhlemann:2008pm}%
  \BibitemOpen
  \bibfield  {author} {\bibinfo {author} {\bibfnamefont {C.~F.}\ \bibnamefont {Uhlemann}}\ and\ \bibinfo {author} {\bibfnamefont {N.}~\bibnamefont {Kauer}},\ }\bibfield  {title} {\enquote {\bibinfo {title} {{Narrow-width approximation accuracy}},}\ }\href {\doibase 10.1016/j.nuclphysb.2009.01.022} {\bibfield  {journal} {\bibinfo  {journal} {Nucl. Phys. B}\ }\textbf {\bibinfo {volume} {814}},\ \bibinfo {pages} {195--211} (\bibinfo {year} {2009})},\ \Eprint {http://arxiv.org/abs/0807.4112} {arXiv:0807.4112 [hep-ph]} \BibitemShut {NoStop}%
\bibitem [{\citenamefont {Thaler}\ and\ \citenamefont {Van~Tilburg}(2011)}]{Thaler:2010tr}%
  \BibitemOpen
  \bibfield  {author} {\bibinfo {author} {\bibfnamefont {Jesse}\ \bibnamefont {Thaler}}\ and\ \bibinfo {author} {\bibfnamefont {Ken}\ \bibnamefont {Van~Tilburg}},\ }\bibfield  {title} {\enquote {\bibinfo {title} {{Identifying Boosted Objects with N-subjettiness}},}\ }\href {\doibase 10.1007/JHEP03(2011)015} {\bibfield  {journal} {\bibinfo  {journal} {JHEP}\ }\textbf {\bibinfo {volume} {03}},\ \bibinfo {pages} {015} (\bibinfo {year} {2011})},\ \Eprint {http://arxiv.org/abs/1011.2268} {arXiv:1011.2268 [hep-ph]} \BibitemShut {NoStop}%
\bibitem [{\citenamefont {Thaler}\ and\ \citenamefont {Van~Tilburg}(2012)}]{Thaler:2011gf}%
  \BibitemOpen
  \bibfield  {author} {\bibinfo {author} {\bibfnamefont {Jesse}\ \bibnamefont {Thaler}}\ and\ \bibinfo {author} {\bibfnamefont {Ken}\ \bibnamefont {Van~Tilburg}},\ }\bibfield  {title} {\enquote {\bibinfo {title} {{Maximizing Boosted Top Identification by Minimizing N-subjettiness}},}\ }\href {\doibase 10.1007/JHEP02(2012)093} {\bibfield  {journal} {\bibinfo  {journal} {JHEP}\ }\textbf {\bibinfo {volume} {02}},\ \bibinfo {pages} {093} (\bibinfo {year} {2012})},\ \Eprint {http://arxiv.org/abs/1108.2701} {arXiv:1108.2701 [hep-ph]} \BibitemShut {NoStop}%
\bibitem [{\citenamefont {Larkoski}\ \emph {et~al.}(2014{\natexlab{a}})\citenamefont {Larkoski}, \citenamefont {Moult},\ and\ \citenamefont {Neill}}]{Larkoski:2014gra}%
  \BibitemOpen
  \bibfield  {author} {\bibinfo {author} {\bibfnamefont {Andrew~J.}\ \bibnamefont {Larkoski}}, \bibinfo {author} {\bibfnamefont {Ian}\ \bibnamefont {Moult}}, \ and\ \bibinfo {author} {\bibfnamefont {Duff}\ \bibnamefont {Neill}},\ }\bibfield  {title} {\enquote {\bibinfo {title} {{Power Counting to Better Jet Observables}},}\ }\href {\doibase 10.1007/JHEP12(2014)009} {\bibfield  {journal} {\bibinfo  {journal} {JHEP}\ }\textbf {\bibinfo {volume} {12}},\ \bibinfo {pages} {009} (\bibinfo {year} {2014}{\natexlab{a}})},\ \Eprint {http://arxiv.org/abs/1409.6298} {arXiv:1409.6298 [hep-ph]} \BibitemShut {NoStop}%
\bibitem [{\citenamefont {Larkoski}\ \emph {et~al.}(2015)\citenamefont {Larkoski}, \citenamefont {Moult},\ and\ \citenamefont {Neill}}]{Larkoski:2014zma}%
  \BibitemOpen
  \bibfield  {author} {\bibinfo {author} {\bibfnamefont {Andrew~J.}\ \bibnamefont {Larkoski}}, \bibinfo {author} {\bibfnamefont {Ian}\ \bibnamefont {Moult}}, \ and\ \bibinfo {author} {\bibfnamefont {Duff}\ \bibnamefont {Neill}},\ }\bibfield  {title} {\enquote {\bibinfo {title} {{Building a Better Boosted Top Tagger}},}\ }\href {\doibase 10.1103/PhysRevD.91.034035} {\bibfield  {journal} {\bibinfo  {journal} {Phys. Rev. D}\ }\textbf {\bibinfo {volume} {91}},\ \bibinfo {pages} {034035} (\bibinfo {year} {2015})},\ \Eprint {http://arxiv.org/abs/1411.0665} {arXiv:1411.0665 [hep-ph]} \BibitemShut {NoStop}%
\bibitem [{\citenamefont {Moult}\ \emph {et~al.}(2016)\citenamefont {Moult}, \citenamefont {Necib},\ and\ \citenamefont {Thaler}}]{Moult:2016cvt}%
  \BibitemOpen
  \bibfield  {author} {\bibinfo {author} {\bibfnamefont {Ian}\ \bibnamefont {Moult}}, \bibinfo {author} {\bibfnamefont {Lina}\ \bibnamefont {Necib}}, \ and\ \bibinfo {author} {\bibfnamefont {Jesse}\ \bibnamefont {Thaler}},\ }\bibfield  {title} {\enquote {\bibinfo {title} {{New Angles on Energy Correlation Functions}},}\ }\href {\doibase 10.1007/JHEP12(2016)153} {\bibfield  {journal} {\bibinfo  {journal} {JHEP}\ }\textbf {\bibinfo {volume} {12}},\ \bibinfo {pages} {153} (\bibinfo {year} {2016})},\ \Eprint {http://arxiv.org/abs/1609.07483} {arXiv:1609.07483 [hep-ph]} \BibitemShut {NoStop}%
\bibitem [{\citenamefont {Neyman}\ and\ \citenamefont {Pearson}(1992)}]{nplemma}%
  \BibitemOpen
  \bibfield  {author} {\bibinfo {author} {\bibfnamefont {Jerzy}\ \bibnamefont {Neyman}}\ and\ \bibinfo {author} {\bibfnamefont {Egon~S}\ \bibnamefont {Pearson}},\ }\bibfield  {title} {\enquote {\bibinfo {title} {On the problem of the most efficient tests of statistical hypotheses},}\ }in\ \href@noop {} {\emph {\bibinfo {booktitle} {Breakthroughs in statistics}}}\ (\bibinfo  {publisher} {Springer},\ \bibinfo {year} {1992})\ pp.\ \bibinfo {pages} {73--108}\BibitemShut {NoStop}%
\bibitem [{\citenamefont {Kasieczka}\ \emph {et~al.}(2019)\citenamefont {Kasieczka}, \citenamefont {Nachman},\ and\ \citenamefont {Shih}}]{gregor_kasieczka_2019_2629073}%
  \BibitemOpen
  \bibfield  {author} {\bibinfo {author} {\bibfnamefont {Gregor}\ \bibnamefont {Kasieczka}}, \bibinfo {author} {\bibfnamefont {Ben}\ \bibnamefont {Nachman}}, \ and\ \bibinfo {author} {\bibfnamefont {David}\ \bibnamefont {Shih}},\ }\href {\doibase 10.5281/zenodo.2629073} {\enquote {\bibinfo {title} {{R\&D Dataset for LHC Olympics 2020 Anomaly Detection Challenge}},}\ } (\bibinfo {year} {2019})\BibitemShut {NoStop}%
\bibitem [{\citenamefont {Sjostrand}\ \emph {et~al.}(2008)\citenamefont {Sjostrand}, \citenamefont {Mrenna},\ and\ \citenamefont {Skands}}]{Sjostrand:2007gs}%
  \BibitemOpen
  \bibfield  {author} {\bibinfo {author} {\bibfnamefont {Torbjorn}\ \bibnamefont {Sjostrand}}, \bibinfo {author} {\bibfnamefont {Stephen}\ \bibnamefont {Mrenna}}, \ and\ \bibinfo {author} {\bibfnamefont {Peter~Z.}\ \bibnamefont {Skands}},\ }\bibfield  {title} {\enquote {\bibinfo {title} {{A Brief Introduction to PYTHIA 8.1}},}\ }\href {\doibase 10.1016/j.cpc.2008.01.036} {\bibfield  {journal} {\bibinfo  {journal} {Comput. Phys. Commun.}\ }\textbf {\bibinfo {volume} {178}},\ \bibinfo {pages} {852--867} (\bibinfo {year} {2008})},\ \Eprint {http://arxiv.org/abs/0710.3820} {arXiv:0710.3820 [hep-ph]} \BibitemShut {NoStop}%
\bibitem [{\citenamefont {Sj\"ostrand}\ \emph {et~al.}(2015)\citenamefont {Sj\"ostrand}, \citenamefont {Ask}, \citenamefont {Christiansen}, \citenamefont {Corke}, \citenamefont {Desai}, \citenamefont {Ilten}, \citenamefont {Mrenna}, \citenamefont {Prestel}, \citenamefont {Rasmussen},\ and\ \citenamefont {Skands}}]{Sjostrand:2014zea}%
  \BibitemOpen
  \bibfield  {author} {\bibinfo {author} {\bibfnamefont {Torbj\"orn}\ \bibnamefont {Sj\"ostrand}}, \bibinfo {author} {\bibfnamefont {Stefan}\ \bibnamefont {Ask}}, \bibinfo {author} {\bibfnamefont {Jesper~R.}\ \bibnamefont {Christiansen}}, \bibinfo {author} {\bibfnamefont {Richard}\ \bibnamefont {Corke}}, \bibinfo {author} {\bibfnamefont {Nishita}\ \bibnamefont {Desai}}, \bibinfo {author} {\bibfnamefont {Philip}\ \bibnamefont {Ilten}}, \bibinfo {author} {\bibfnamefont {Stephen}\ \bibnamefont {Mrenna}}, \bibinfo {author} {\bibfnamefont {Stefan}\ \bibnamefont {Prestel}}, \bibinfo {author} {\bibfnamefont {Christine~O.}\ \bibnamefont {Rasmussen}}, \ and\ \bibinfo {author} {\bibfnamefont {Peter~Z.}\ \bibnamefont {Skands}},\ }\bibfield  {title} {\enquote {\bibinfo {title} {{An introduction to PYTHIA 8.2}},}\ }\href {\doibase 10.1016/j.cpc.2015.01.024} {\bibfield  {journal} {\bibinfo  {journal} {Comput. Phys. Commun.}\ }\textbf {\bibinfo {volume} {191}},\ \bibinfo {pages} {159--177} (\bibinfo {year} {2015})},\
  \Eprint {http://arxiv.org/abs/1410.3012} {arXiv:1410.3012 [hep-ph]} \BibitemShut {NoStop}%
\bibitem [{\citenamefont {de~Favereau}\ \emph {et~al.}(2014)\citenamefont {de~Favereau}, \citenamefont {Delaere}, \citenamefont {Demin}, \citenamefont {Giammanco}, \citenamefont {Lema\^\i{}tre}, \citenamefont {Mertens},\ and\ \citenamefont {Selvaggi}}]{deFavereau:2013fsa}%
  \BibitemOpen
  \bibfield  {author} {\bibinfo {author} {\bibfnamefont {J.}~\bibnamefont {de~Favereau}}, \bibinfo {author} {\bibfnamefont {C.}~\bibnamefont {Delaere}}, \bibinfo {author} {\bibfnamefont {P.}~\bibnamefont {Demin}}, \bibinfo {author} {\bibfnamefont {A.}~\bibnamefont {Giammanco}}, \bibinfo {author} {\bibfnamefont {V.}~\bibnamefont {Lema\^\i{}tre}}, \bibinfo {author} {\bibfnamefont {A.}~\bibnamefont {Mertens}}, \ and\ \bibinfo {author} {\bibfnamefont {M.}~\bibnamefont {Selvaggi}} (\bibinfo {collaboration} {DELPHES 3}),\ }\bibfield  {title} {\enquote {\bibinfo {title} {{DELPHES 3, A modular framework for fast simulation of a generic collider experiment}},}\ }\href {\doibase 10.1007/JHEP02(2014)057} {\bibfield  {journal} {\bibinfo  {journal} {JHEP}\ }\textbf {\bibinfo {volume} {02}},\ \bibinfo {pages} {057} (\bibinfo {year} {2014})},\ \Eprint {http://arxiv.org/abs/1307.6346} {arXiv:1307.6346 [hep-ex]} \BibitemShut {NoStop}%
\bibitem [{\citenamefont {Cacciari}\ \emph {et~al.}(2008)\citenamefont {Cacciari}, \citenamefont {Salam},\ and\ \citenamefont {Soyez}}]{Cacciari:2008gp}%
  \BibitemOpen
  \bibfield  {author} {\bibinfo {author} {\bibfnamefont {Matteo}\ \bibnamefont {Cacciari}}, \bibinfo {author} {\bibfnamefont {Gavin~P.}\ \bibnamefont {Salam}}, \ and\ \bibinfo {author} {\bibfnamefont {Gregory}\ \bibnamefont {Soyez}},\ }\bibfield  {title} {\enquote {\bibinfo {title} {{The anti-$k_t$ jet clustering algorithm}},}\ }\href {\doibase 10.1088/1126-6708/2008/04/063} {\bibfield  {journal} {\bibinfo  {journal} {JHEP}\ }\textbf {\bibinfo {volume} {04}},\ \bibinfo {pages} {063} (\bibinfo {year} {2008})},\ \Eprint {http://arxiv.org/abs/0802.1189} {arXiv:0802.1189 [hep-ph]} \BibitemShut {NoStop}%
\bibitem [{\citenamefont {Cacciari}\ \emph {et~al.}(2012)\citenamefont {Cacciari}, \citenamefont {Salam},\ and\ \citenamefont {Soyez}}]{Cacciari:2011ma}%
  \BibitemOpen
  \bibfield  {author} {\bibinfo {author} {\bibfnamefont {Matteo}\ \bibnamefont {Cacciari}}, \bibinfo {author} {\bibfnamefont {Gavin~P.}\ \bibnamefont {Salam}}, \ and\ \bibinfo {author} {\bibfnamefont {Gregory}\ \bibnamefont {Soyez}},\ }\bibfield  {title} {\enquote {\bibinfo {title} {{FastJet User Manual}},}\ }\href {\doibase 10.1140/epjc/s10052-012-1896-2} {\bibfield  {journal} {\bibinfo  {journal} {Eur. Phys. J. C}\ }\textbf {\bibinfo {volume} {72}},\ \bibinfo {pages} {1896} (\bibinfo {year} {2012})},\ \Eprint {http://arxiv.org/abs/1111.6097} {arXiv:1111.6097 [hep-ph]} \BibitemShut {NoStop}%
\bibitem [{\citenamefont {Komiske}\ \emph {et~al.}(2018{\natexlab{c}})\citenamefont {Komiske}, \citenamefont {Metodiev},\ and\ \citenamefont {Thaler}}]{Komiske:2017aww}%
  \BibitemOpen
  \bibfield  {author} {\bibinfo {author} {\bibfnamefont {Patrick~T.}\ \bibnamefont {Komiske}}, \bibinfo {author} {\bibfnamefont {Eric~M.}\ \bibnamefont {Metodiev}}, \ and\ \bibinfo {author} {\bibfnamefont {Jesse}\ \bibnamefont {Thaler}},\ }\bibfield  {title} {\enquote {\bibinfo {title} {{Energy flow polynomials: A complete linear basis for jet substructure}},}\ }\href {\doibase 10.1007/JHEP04(2018)013} {\bibfield  {journal} {\bibinfo  {journal} {JHEP}\ }\textbf {\bibinfo {volume} {04}},\ \bibinfo {pages} {013} (\bibinfo {year} {2018}{\natexlab{c}})},\ \Eprint {http://arxiv.org/abs/1712.07124} {arXiv:1712.07124 [hep-ph]} \BibitemShut {NoStop}%
\bibitem [{\citenamefont {Komiske}\ \emph {et~al.}(2020{\natexlab{a}})\citenamefont {Komiske}, \citenamefont {Metodiev},\ and\ \citenamefont {Thaler}}]{Komiske:2019asc}%
  \BibitemOpen
  \bibfield  {author} {\bibinfo {author} {\bibfnamefont {Patrick~T.}\ \bibnamefont {Komiske}}, \bibinfo {author} {\bibfnamefont {Eric~M.}\ \bibnamefont {Metodiev}}, \ and\ \bibinfo {author} {\bibfnamefont {Jesse}\ \bibnamefont {Thaler}},\ }\bibfield  {title} {\enquote {\bibinfo {title} {{Cutting Multiparticle Correlators Down to Size}},}\ }\href {\doibase 10.1103/PhysRevD.101.036019} {\bibfield  {journal} {\bibinfo  {journal} {Phys. Rev. D}\ }\textbf {\bibinfo {volume} {101}},\ \bibinfo {pages} {036019} (\bibinfo {year} {2020}{\natexlab{a}})},\ \Eprint {http://arxiv.org/abs/1911.04491} {arXiv:1911.04491 [hep-ph]} \BibitemShut {NoStop}%
\bibitem [{\citenamefont {Komiske}\ and\ \citenamefont {Metodiev}(2019)}]{EnergyFlow}%
  \BibitemOpen
  \bibfield  {author} {\bibinfo {author} {\bibfnamefont {Patrick~T.}\ \bibnamefont {Komiske}}\ and\ \bibinfo {author} {\bibfnamefont {Eric~M.}\ \bibnamefont {Metodiev}},\ }\href {https://energyflow.network/} {\enquote {\bibinfo {title} {{EnergyFlow}},}\ } (\bibinfo {year} {2019}),\ \bibinfo {note} {\url{https://energyflow.network/}}\BibitemShut {NoStop}%
\bibitem [{\citenamefont {Srivastava}\ \emph {et~al.}(2014)\citenamefont {Srivastava}, \citenamefont {Hinton}, \citenamefont {Krizhevsky}, \citenamefont {Sutskever},\ and\ \citenamefont {Salakhutdinov}}]{dropout}%
  \BibitemOpen
  \bibfield  {author} {\bibinfo {author} {\bibfnamefont {Nitish}\ \bibnamefont {Srivastava}}, \bibinfo {author} {\bibfnamefont {Geoffrey}\ \bibnamefont {Hinton}}, \bibinfo {author} {\bibfnamefont {Alex}\ \bibnamefont {Krizhevsky}}, \bibinfo {author} {\bibfnamefont {Ilya}\ \bibnamefont {Sutskever}}, \ and\ \bibinfo {author} {\bibfnamefont {Ruslan}\ \bibnamefont {Salakhutdinov}},\ }\bibfield  {title} {\enquote {\bibinfo {title} {Dropout: A simple way to prevent neural networks from overfitting},}\ }\href@noop {} {\bibfield  {journal} {\bibinfo  {journal} {J. Mach. Learn. Res.}\ }\textbf {\bibinfo {volume} {15}},\ \bibinfo {pages} {1929–1958} (\bibinfo {year} {2014})}\BibitemShut {NoStop}%
\bibitem [{\citenamefont {Chollet}(2017)}]{keras}%
  \BibitemOpen
  \bibfield  {author} {\bibinfo {author} {\bibfnamefont {Francois}\ \bibnamefont {Chollet}},\ }\href@noop {} {\enquote {\bibinfo {title} {Keras},}\ }\bibinfo {howpublished} {\url{https://github.com/fchollet/keras}} (\bibinfo {year} {2017})\BibitemShut {NoStop}%
\bibitem [{\citenamefont {Abadi}\ \emph {et~al.}(2016)\citenamefont {Abadi}, \citenamefont {Barham}, \citenamefont {Chen}, \citenamefont {Chen}, \citenamefont {Davis}, \citenamefont {Dean}, \citenamefont {Devin}, \citenamefont {Ghemawat}, \citenamefont {Irving}, \citenamefont {Isard} \emph {et~al.}}]{tensorflow}%
  \BibitemOpen
  \bibfield  {author} {\bibinfo {author} {\bibfnamefont {Mart{\'\i}n}\ \bibnamefont {Abadi}}, \bibinfo {author} {\bibfnamefont {Paul}\ \bibnamefont {Barham}}, \bibinfo {author} {\bibfnamefont {Jianmin}\ \bibnamefont {Chen}}, \bibinfo {author} {\bibfnamefont {Zhifeng}\ \bibnamefont {Chen}}, \bibinfo {author} {\bibfnamefont {Andy}\ \bibnamefont {Davis}}, \bibinfo {author} {\bibfnamefont {Jeffrey}\ \bibnamefont {Dean}}, \bibinfo {author} {\bibfnamefont {Matthieu}\ \bibnamefont {Devin}}, \bibinfo {author} {\bibfnamefont {Sanjay}\ \bibnamefont {Ghemawat}}, \bibinfo {author} {\bibfnamefont {Geoffrey}\ \bibnamefont {Irving}}, \bibinfo {author} {\bibfnamefont {Michael}\ \bibnamefont {Isard}},  \emph {et~al.},\ }\bibfield  {title} {\enquote {\bibinfo {title} {Tensorflow: A system for large-scale machine learning.}}\ }in\ \href@noop {} {\emph {\bibinfo {booktitle} {OSDI}}},\ Vol.~\bibinfo {volume} {16}\ (\bibinfo {year} {2016})\ pp.\ \bibinfo {pages} {265--283}\BibitemShut {NoStop}%
\bibitem [{\citenamefont {Kingma}\ and\ \citenamefont {Ba}(2017)}]{kingma2017adam}%
  \BibitemOpen
  \bibfield  {author} {\bibinfo {author} {\bibfnamefont {Diederik~P.}\ \bibnamefont {Kingma}}\ and\ \bibinfo {author} {\bibfnamefont {Jimmy}\ \bibnamefont {Ba}},\ }\href@noop {} {\enquote {\bibinfo {title} {Adam: A method for stochastic optimization},}\ } (\bibinfo {year} {2017}),\ \Eprint {http://arxiv.org/abs/1412.6980} {arXiv:1412.6980 [cs.LG]} \BibitemShut {NoStop}%
\bibitem [{\citenamefont {Wynne}(2023)}]{repocode}%
  \BibitemOpen
  \bibfield  {author} {\bibinfo {author} {\bibfnamefont {Raymond}\ \bibnamefont {Wynne}},\ }\href@noop {} {\enquote {\bibinfo {title} {{FORCE}},}\ }\bibinfo {howpublished} {\url{https://github.com/r-wynne/FORCE}} (\bibinfo {year} {2023})\BibitemShut {NoStop}%
\bibitem [{\citenamefont {Dolen}\ \emph {et~al.}(2016)\citenamefont {Dolen}, \citenamefont {Harris}, \citenamefont {Marzani}, \citenamefont {Rappoccio},\ and\ \citenamefont {Tran}}]{Dolen:2016kst}%
  \BibitemOpen
  \bibfield  {author} {\bibinfo {author} {\bibfnamefont {James}\ \bibnamefont {Dolen}}, \bibinfo {author} {\bibfnamefont {Philip}\ \bibnamefont {Harris}}, \bibinfo {author} {\bibfnamefont {Simone}\ \bibnamefont {Marzani}}, \bibinfo {author} {\bibfnamefont {Salvatore}\ \bibnamefont {Rappoccio}}, \ and\ \bibinfo {author} {\bibfnamefont {Nhan}\ \bibnamefont {Tran}},\ }\bibfield  {title} {\enquote {\bibinfo {title} {{Thinking outside the ROCs: Designing Decorrelated Taggers (DDT) for jet substructure}},}\ }\href {\doibase 10.1007/JHEP05(2016)156} {\bibfield  {journal} {\bibinfo  {journal} {JHEP}\ }\textbf {\bibinfo {volume} {05}},\ \bibinfo {pages} {156} (\bibinfo {year} {2016})},\ \Eprint {http://arxiv.org/abs/1603.00027} {arXiv:1603.00027 [hep-ph]} \BibitemShut {NoStop}%
\bibitem [{\citenamefont {Shimmin}\ \emph {et~al.}(2017)\citenamefont {Shimmin}, \citenamefont {Sadowski}, \citenamefont {Baldi}, \citenamefont {Weik}, \citenamefont {Whiteson}, \citenamefont {Goul},\ and\ \citenamefont {S\o{}gaard}}]{Shimmin:2017mfk}%
  \BibitemOpen
  \bibfield  {author} {\bibinfo {author} {\bibfnamefont {Chase}\ \bibnamefont {Shimmin}}, \bibinfo {author} {\bibfnamefont {Peter}\ \bibnamefont {Sadowski}}, \bibinfo {author} {\bibfnamefont {Pierre}\ \bibnamefont {Baldi}}, \bibinfo {author} {\bibfnamefont {Edison}\ \bibnamefont {Weik}}, \bibinfo {author} {\bibfnamefont {Daniel}\ \bibnamefont {Whiteson}}, \bibinfo {author} {\bibfnamefont {Edward}\ \bibnamefont {Goul}}, \ and\ \bibinfo {author} {\bibfnamefont {Andreas}\ \bibnamefont {S\o{}gaard}},\ }\bibfield  {title} {\enquote {\bibinfo {title} {{Decorrelated Jet Substructure Tagging using Adversarial Neural Networks}},}\ }\href {\doibase 10.1103/PhysRevD.96.074034} {\bibfield  {journal} {\bibinfo  {journal} {Phys. Rev. D}\ }\textbf {\bibinfo {volume} {96}},\ \bibinfo {pages} {074034} (\bibinfo {year} {2017})},\ \Eprint {http://arxiv.org/abs/1703.03507} {arXiv:1703.03507 [hep-ex]} \BibitemShut {NoStop}%
\bibitem [{\citenamefont {Moult}\ \emph {et~al.}(2018)\citenamefont {Moult}, \citenamefont {Nachman},\ and\ \citenamefont {Neill}}]{Moult:2017okx}%
  \BibitemOpen
  \bibfield  {author} {\bibinfo {author} {\bibfnamefont {Ian}\ \bibnamefont {Moult}}, \bibinfo {author} {\bibfnamefont {Benjamin}\ \bibnamefont {Nachman}}, \ and\ \bibinfo {author} {\bibfnamefont {Duff}\ \bibnamefont {Neill}},\ }\bibfield  {title} {\enquote {\bibinfo {title} {{Convolved Substructure: Analytically Decorrelating Jet Substructure Observables}},}\ }\href {\doibase 10.1007/JHEP05(2018)002} {\bibfield  {journal} {\bibinfo  {journal} {JHEP}\ }\textbf {\bibinfo {volume} {05}},\ \bibinfo {pages} {002} (\bibinfo {year} {2018})},\ \Eprint {http://arxiv.org/abs/1710.06859} {arXiv:1710.06859 [hep-ph]} \BibitemShut {NoStop}%
\bibitem [{\citenamefont {Ganaie}\ \emph {et~al.}(2021)\citenamefont {Ganaie}, \citenamefont {Hu}, \citenamefont {Tanveer},\ and\ \citenamefont {Suganthan}}]{DBLP:journals/corr/abs-2104-02395}%
  \BibitemOpen
  \bibfield  {author} {\bibinfo {author} {\bibfnamefont {Mudasir~Ahmad}\ \bibnamefont {Ganaie}}, \bibinfo {author} {\bibfnamefont {Minghui}\ \bibnamefont {Hu}}, \bibinfo {author} {\bibfnamefont {Mohammad}\ \bibnamefont {Tanveer}}, \ and\ \bibinfo {author} {\bibfnamefont {Ponnuthurai~N.}\ \bibnamefont {Suganthan}},\ }\bibfield  {title} {\enquote {\bibinfo {title} {Ensemble deep learning: {A} review},}\ }\href {https://arxiv.org/abs/2104.02395} {\bibfield  {journal} {\bibinfo  {journal} {CoRR}\ }\textbf {\bibinfo {volume} {abs/2104.02395}} (\bibinfo {year} {2021})},\ \Eprint {http://arxiv.org/abs/2104.02395} {2104.02395} \BibitemShut {NoStop}%
\bibitem [{\citenamefont {Andreassen}\ \emph {et~al.}(2020{\natexlab{b}})\citenamefont {Andreassen}, \citenamefont {Komiske}, \citenamefont {Metodiev}, \citenamefont {Nachman},\ and\ \citenamefont {Thaler}}]{Andreassen:2019cjw}%
  \BibitemOpen
  \bibfield  {author} {\bibinfo {author} {\bibfnamefont {Anders}\ \bibnamefont {Andreassen}}, \bibinfo {author} {\bibfnamefont {Patrick~T.}\ \bibnamefont {Komiske}}, \bibinfo {author} {\bibfnamefont {Eric~M.}\ \bibnamefont {Metodiev}}, \bibinfo {author} {\bibfnamefont {Benjamin}\ \bibnamefont {Nachman}}, \ and\ \bibinfo {author} {\bibfnamefont {Jesse}\ \bibnamefont {Thaler}},\ }\bibfield  {title} {\enquote {\bibinfo {title} {{OmniFold: A Method to Simultaneously Unfold All Observables}},}\ }\href {\doibase 10.1103/PhysRevLett.124.182001} {\bibfield  {journal} {\bibinfo  {journal} {Phys. Rev. Lett.}\ }\textbf {\bibinfo {volume} {124}},\ \bibinfo {pages} {182001} (\bibinfo {year} {2020}{\natexlab{b}})},\ \Eprint {http://arxiv.org/abs/1911.09107} {arXiv:1911.09107 [hep-ph]} \BibitemShut {NoStop}%
\bibitem [{\citenamefont {Bellagente}\ \emph {et~al.}(2020)\citenamefont {Bellagente}, \citenamefont {Butter}, \citenamefont {Kasieczka}, \citenamefont {Plehn}, \citenamefont {Rousselot}, \citenamefont {Winterhalder}, \citenamefont {Ardizzone},\ and\ \citenamefont {K\"othe}}]{Bellagente:2020piv}%
  \BibitemOpen
  \bibfield  {author} {\bibinfo {author} {\bibfnamefont {Marco}\ \bibnamefont {Bellagente}}, \bibinfo {author} {\bibfnamefont {Anja}\ \bibnamefont {Butter}}, \bibinfo {author} {\bibfnamefont {Gregor}\ \bibnamefont {Kasieczka}}, \bibinfo {author} {\bibfnamefont {Tilman}\ \bibnamefont {Plehn}}, \bibinfo {author} {\bibfnamefont {Armand}\ \bibnamefont {Rousselot}}, \bibinfo {author} {\bibfnamefont {Ramon}\ \bibnamefont {Winterhalder}}, \bibinfo {author} {\bibfnamefont {Lynton}\ \bibnamefont {Ardizzone}}, \ and\ \bibinfo {author} {\bibfnamefont {Ullrich}\ \bibnamefont {K\"othe}},\ }\bibfield  {title} {\enquote {\bibinfo {title} {{Invertible Networks or Partons to Detector and Back Again}},}\ }\href {\doibase 10.21468/SciPostPhys.9.5.074} {\bibfield  {journal} {\bibinfo  {journal} {SciPost Phys.}\ }\textbf {\bibinfo {volume} {9}},\ \bibinfo {pages} {074} (\bibinfo {year} {2020})},\ \Eprint {http://arxiv.org/abs/2006.06685} {arXiv:2006.06685 [hep-ph]} \BibitemShut {NoStop}%
\bibitem [{\citenamefont {Andreassen}\ \emph {et~al.}(2021)\citenamefont {Andreassen}, \citenamefont {Komiske}, \citenamefont {Metodiev}, \citenamefont {Nachman}, \citenamefont {Suresh},\ and\ \citenamefont {Thaler}}]{Andreassen:2021zzk}%
  \BibitemOpen
  \bibfield  {author} {\bibinfo {author} {\bibfnamefont {Anders}\ \bibnamefont {Andreassen}}, \bibinfo {author} {\bibfnamefont {Patrick~T.}\ \bibnamefont {Komiske}}, \bibinfo {author} {\bibfnamefont {Eric~M.}\ \bibnamefont {Metodiev}}, \bibinfo {author} {\bibfnamefont {Benjamin}\ \bibnamefont {Nachman}}, \bibinfo {author} {\bibfnamefont {Adi}\ \bibnamefont {Suresh}}, \ and\ \bibinfo {author} {\bibfnamefont {Jesse}\ \bibnamefont {Thaler}},\ }\bibfield  {title} {\enquote {\bibinfo {title} {{Scaffolding Simulations with Deep Learning for High-dimensional Deconvolution}},}\ }in\ \href@noop {} {\emph {\bibinfo {booktitle} {{9th International Conference on Learning Representations}}}}\ (\bibinfo {year} {2021})\ \Eprint {http://arxiv.org/abs/2105.04448} {arXiv:2105.04448 [stat.ML]} \BibitemShut {NoStop}%
\bibitem [{\citenamefont {Arratia}\ \emph {et~al.}(2022)\citenamefont {Arratia} \emph {et~al.}}]{Arratia:2021otl}%
  \BibitemOpen
  \bibfield  {author} {\bibinfo {author} {\bibfnamefont {Miguel}\ \bibnamefont {Arratia}} \emph {et~al.},\ }\bibfield  {title} {\enquote {\bibinfo {title} {{Publishing unbinned differential cross section results}},}\ }\href {\doibase 10.1088/1748-0221/17/01/P01024} {\bibfield  {journal} {\bibinfo  {journal} {JINST}\ }\textbf {\bibinfo {volume} {17}},\ \bibinfo {pages} {P01024} (\bibinfo {year} {2022})},\ \Eprint {http://arxiv.org/abs/2109.13243} {arXiv:2109.13243 [hep-ph]} \BibitemShut {NoStop}%
\bibitem [{\citenamefont {Dasgupta}\ \emph {et~al.}(2013)\citenamefont {Dasgupta}, \citenamefont {Fregoso}, \citenamefont {Marzani},\ and\ \citenamefont {Salam}}]{Dasgupta:2013ihk}%
  \BibitemOpen
  \bibfield  {author} {\bibinfo {author} {\bibfnamefont {Mrinal}\ \bibnamefont {Dasgupta}}, \bibinfo {author} {\bibfnamefont {Alessandro}\ \bibnamefont {Fregoso}}, \bibinfo {author} {\bibfnamefont {Simone}\ \bibnamefont {Marzani}}, \ and\ \bibinfo {author} {\bibfnamefont {Gavin~P.}\ \bibnamefont {Salam}},\ }\bibfield  {title} {\enquote {\bibinfo {title} {{Towards an understanding of jet substructure}},}\ }\href {\doibase 10.1007/JHEP09(2013)029} {\bibfield  {journal} {\bibinfo  {journal} {JHEP}\ }\textbf {\bibinfo {volume} {09}},\ \bibinfo {pages} {029} (\bibinfo {year} {2013})},\ \Eprint {http://arxiv.org/abs/1307.0007} {arXiv:1307.0007 [hep-ph]} \BibitemShut {NoStop}%
\bibitem [{\citenamefont {Larkoski}\ \emph {et~al.}(2014{\natexlab{b}})\citenamefont {Larkoski}, \citenamefont {Marzani}, \citenamefont {Soyez},\ and\ \citenamefont {Thaler}}]{Larkoski:2014wba}%
  \BibitemOpen
  \bibfield  {author} {\bibinfo {author} {\bibfnamefont {Andrew~J.}\ \bibnamefont {Larkoski}}, \bibinfo {author} {\bibfnamefont {Simone}\ \bibnamefont {Marzani}}, \bibinfo {author} {\bibfnamefont {Gregory}\ \bibnamefont {Soyez}}, \ and\ \bibinfo {author} {\bibfnamefont {Jesse}\ \bibnamefont {Thaler}},\ }\bibfield  {title} {\enquote {\bibinfo {title} {{Soft Drop}},}\ }\href {\doibase 10.1007/JHEP05(2014)146} {\bibfield  {journal} {\bibinfo  {journal} {JHEP}\ }\textbf {\bibinfo {volume} {05}},\ \bibinfo {pages} {146} (\bibinfo {year} {2014}{\natexlab{b}})},\ \Eprint {http://arxiv.org/abs/1402.2657} {arXiv:1402.2657 [hep-ph]} \BibitemShut {NoStop}%
\bibitem [{\citenamefont {Frye}\ \emph {et~al.}(2016)\citenamefont {Frye}, \citenamefont {Larkoski}, \citenamefont {Schwartz},\ and\ \citenamefont {Yan}}]{Frye:2016aiz}%
  \BibitemOpen
  \bibfield  {author} {\bibinfo {author} {\bibfnamefont {Christopher}\ \bibnamefont {Frye}}, \bibinfo {author} {\bibfnamefont {Andrew~J.}\ \bibnamefont {Larkoski}}, \bibinfo {author} {\bibfnamefont {Matthew~D.}\ \bibnamefont {Schwartz}}, \ and\ \bibinfo {author} {\bibfnamefont {Kai}\ \bibnamefont {Yan}},\ }\bibfield  {title} {\enquote {\bibinfo {title} {{Factorization for groomed jet substructure beyond the next-to-leading logarithm}},}\ }\href {\doibase 10.1007/JHEP07(2016)064} {\bibfield  {journal} {\bibinfo  {journal} {JHEP}\ }\textbf {\bibinfo {volume} {07}},\ \bibinfo {pages} {064} (\bibinfo {year} {2016})},\ \Eprint {http://arxiv.org/abs/1603.09338} {arXiv:1603.09338 [hep-ph]} \BibitemShut {NoStop}%
\bibitem [{\citenamefont {Komiske}\ \emph {et~al.}(2019)\citenamefont {Komiske}, \citenamefont {Metodiev},\ and\ \citenamefont {Thaler}}]{Komiske:2019fks}%
  \BibitemOpen
  \bibfield  {author} {\bibinfo {author} {\bibfnamefont {Patrick~T.}\ \bibnamefont {Komiske}}, \bibinfo {author} {\bibfnamefont {Eric~M.}\ \bibnamefont {Metodiev}}, \ and\ \bibinfo {author} {\bibfnamefont {Jesse}\ \bibnamefont {Thaler}},\ }\bibfield  {title} {\enquote {\bibinfo {title} {{Metric Space of Collider Events}},}\ }\href {\doibase 10.1103/PhysRevLett.123.041801} {\bibfield  {journal} {\bibinfo  {journal} {Phys. Rev. Lett.}\ }\textbf {\bibinfo {volume} {123}},\ \bibinfo {pages} {041801} (\bibinfo {year} {2019})},\ \Eprint {http://arxiv.org/abs/1902.02346} {arXiv:1902.02346 [hep-ph]} \BibitemShut {NoStop}%
\bibitem [{\citenamefont {Komiske}\ \emph {et~al.}(2020{\natexlab{b}})\citenamefont {Komiske}, \citenamefont {Metodiev},\ and\ \citenamefont {Thaler}}]{Komiske:2020qhg}%
  \BibitemOpen
  \bibfield  {author} {\bibinfo {author} {\bibfnamefont {Patrick~T.}\ \bibnamefont {Komiske}}, \bibinfo {author} {\bibfnamefont {Eric~M.}\ \bibnamefont {Metodiev}}, \ and\ \bibinfo {author} {\bibfnamefont {Jesse}\ \bibnamefont {Thaler}},\ }\bibfield  {title} {\enquote {\bibinfo {title} {{The Hidden Geometry of Particle Collisions}},}\ }\href {\doibase 10.1007/JHEP07(2020)006} {\bibfield  {journal} {\bibinfo  {journal} {JHEP}\ }\textbf {\bibinfo {volume} {07}},\ \bibinfo {pages} {006} (\bibinfo {year} {2020}{\natexlab{b}})},\ \Eprint {http://arxiv.org/abs/2004.04159} {arXiv:2004.04159 [hep-ph]} \BibitemShut {NoStop}%
\end{thebibliography}%

\clearpage

\onecolumngrid

\begin{center}
    \textbf{\large
Anomaly Detection in Collider Physics via Factorized Observables: \\ Supplemental Material}
\end{center}

\twocolumngrid

In this Supplemental Material, we provide additional derivations and results related to the FORCE method.
First, we demonstrate how the conditional expectation minimizes the mean squared error loss.
Then, we analytically validate the effectiveness of FORCE using a straightforward Gaussian example.
We review the energy flow polynomials (EFPs) and highlight the degradation of FORCE performance if they are not normalized.
Finally, we discuss the impact of deviation from factorization.

\section{Conditional Expectation Minimizes Mean Squared Error}

The mean squared error (MSE) loss is the most ubiquitous loss function in machine learning.
Here we provide a short proof of the statement from the main text that the conditional expectation minimizes the MSE.

Consider estimating a random variable $Y$ given observations of another random variables $X$.
In the context of the main body, $Y$ corresponds to $p_T$ and $X$ corresponds to $\mathcal{O}$.
The MSE loss functional $L[f]$ between an estimator $f(X)$ and the true value $Y$ is:
\begin{equation}
L[f]= \mathbb{E}\big[\big(Y - f(X)\big)^2\big],
\end{equation}
where the expectation is taken over the joint distribution of $(X,Y)$.
More explicitly, we can express the MSE in terms of the joint density $P(x,y)$ as:
\begin{equation}
L[f] = \int dx \, dy \, P(x, y) \, \big(y-f(x)\big)^2.
\end{equation}

Through functional variation (i.e.~the Euler-Lagrange method from physics), we can find the estimator $f(x)$ that extremizes $L[f]$.
Since $L[f]$ depends only on $f$ and not its derivatives, the extrema satisfy:
\begin{equation}
\frac{\delta L}{\delta f} = 0.
\end{equation}
Varying the integrand in $L[f]$ with respect to $f(x)$ yields the following constraint:
\begin{equation}
\int dy \, P(x, y) \, 2 \big(y - f(x) \big) = 0.
\end{equation}
Expressing the joint density as $P(x,y) = P(x) \, P(y|x)$ and assuming that $P(x)$ has support everywhere, this constraint is solved by
\begin{equation}
f(x) = \int dy \, P(y | x) \, y  \equiv \mathbb{E}[Y|X=x].
\end{equation}
This shows that the MSE is extremized by the conditional expectation.
From the convexity properties of the MSE, one can further conclude that this is extremum is in fact a minimum.

With sufficient statistics, an alternative way to obtain $\mathbb{E}[Y|X=x]$ would be to directly compute the mean value of $Y$ in a small histogram bin around $x$.
When $x$ is high dimensional, though, as is the case for our substructure analysis, machine learning regression provides better generalization performance than a binned analysis.

\section{A Gaussian Example}

\begin{figure}[t]
    \centering
    \includegraphics[width=0.85\columnwidth]{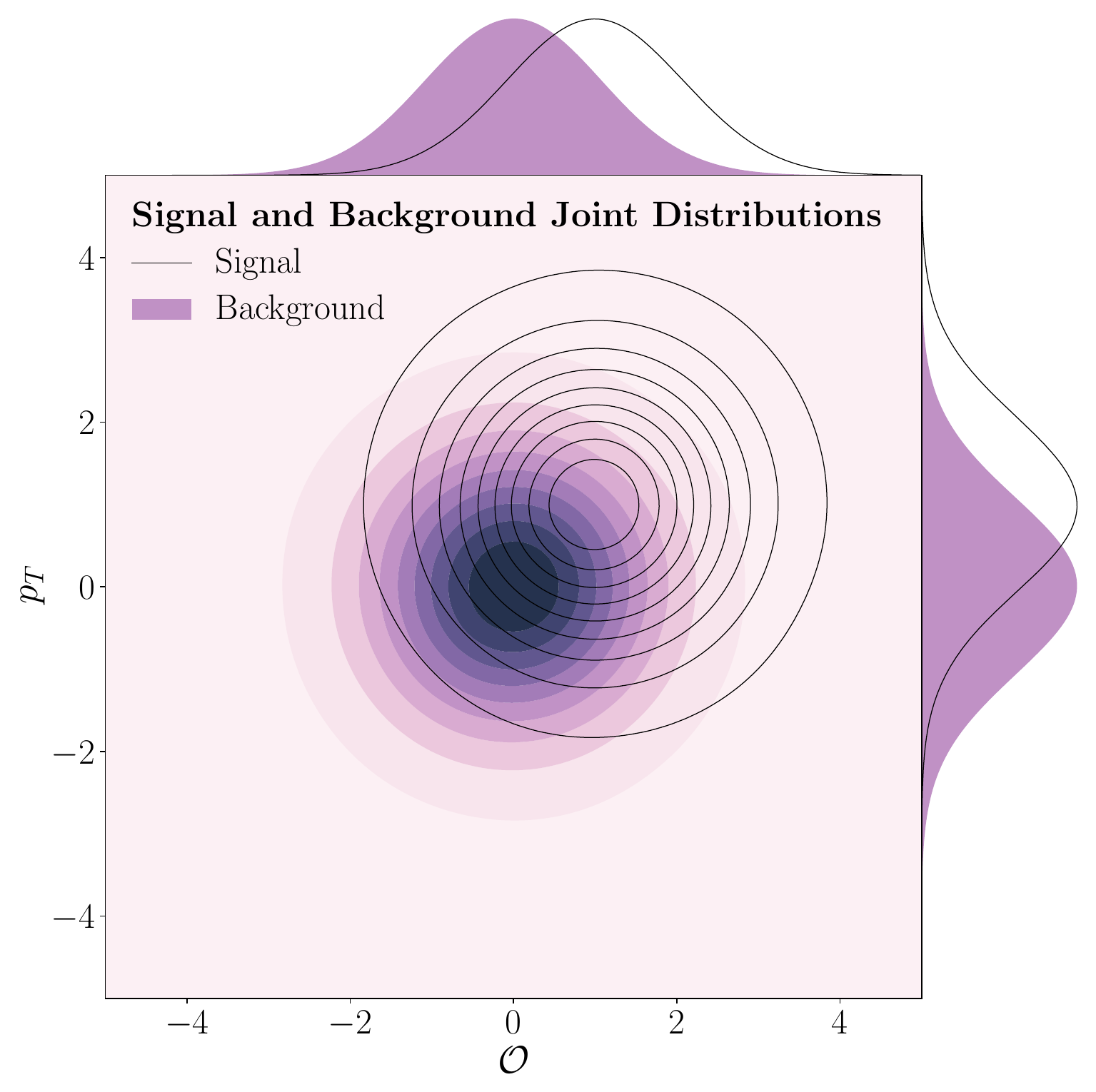}
    \caption{Signal and background joint distributions for an illustrative Gaussian example.}
    \label{fig:sig_back_dist}
\end{figure}

To analytically demonstrate the claims of \CleverName, we provide an illustrative Gaussian example.
Let $\mathcal{N}(\mu, \sigma^2)$ denote the  normal distribution with mean $\mu$ and variance $\sigma^2$.
Consider two variables $p_{T}$ and $\mathcal{O}$ with different distributions for the signal ($S$) and background ($B$):
\begin{align}
\nonumber    p_{T, S} &\sim \mathcal{N}\big(\braket{p_T}_S, (\Delta p_T)^2_S\big),& 
    \mathcal{O}_{S} &\sim \mathcal{N}\big(\braket{\mathcal{O}}_S, (\Delta \mathcal{O})^2_S\big),\\
    p_{T, B} &\sim \mathcal{N}\big(\braket{p_T}_B,  (\Delta p_T)^2_B)\big),&
    \mathcal{O}_{B} &\sim \mathcal{N}\big(\braket{\mathcal{O}}_B, (\Delta \mathcal{O})^2_B\big).
\end{align}
For our numerical analysis, we take $\braket{p_T}_S = \braket{\mathcal{O}}_S = 1$, $\braket{p_T}_B = \braket{\mathcal{O}}_B = 0$, and all variances to be 1.
We fix the background size at 1 million samples and draw the number of signal events to have a signal fraction of $f_S$.
These distributions are shown in \Fig{fig:sig_back_dist} as contour plots.
Since the kinematic variable ($p_T$) is drawn independently from the substructure variable ($\mathcal{O}$), concatenating these signal and background samples with signal fraction $f_S$ yields a data set with distribution defined by Eq. (3) in the main text.

\begin{figure}[t]
    \centering
    \includegraphics[width=0.85\columnwidth]{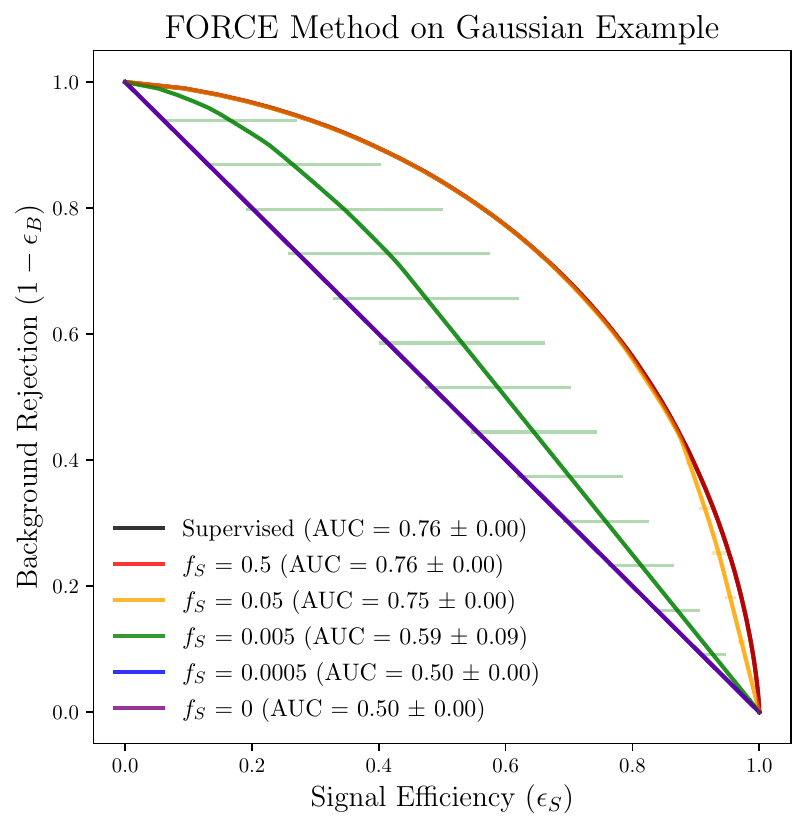}
    \caption{
    Results of applying \CleverName to a simulated Gaussian example. We see that the model is the optimal classifier in the high signal fraction limit (the Supervised black and $f_S=0.5$ red curves coincide), with diminishing discrimination power as the signal fraction is decreased (the blue $f_S=0.0005$ and magenta $f_S=0$ curves coincide).
    The large uncertainties at $f_S = 0.005$ arise from the difficulty of extracting a reliable discriminant from so few signal events.
    }
    \label{fig:gauss_auc}
\end{figure}

To apply \CleverName to this Gaussian example, we use fully connected neural networks with three layers of 100 nodes per layer, training for 10 epochs and ensembling over 10 different models.
We show the decay in statistical power as a function of signal fraction in \Fig{fig:gauss_auc}.
As expected, the network behaves optimally in the large signal fraction limit, and converges to a constant random classifier in the no-signal limit.

\begin{figure}[t]
    \centering
    \includegraphics[width=\columnwidth]{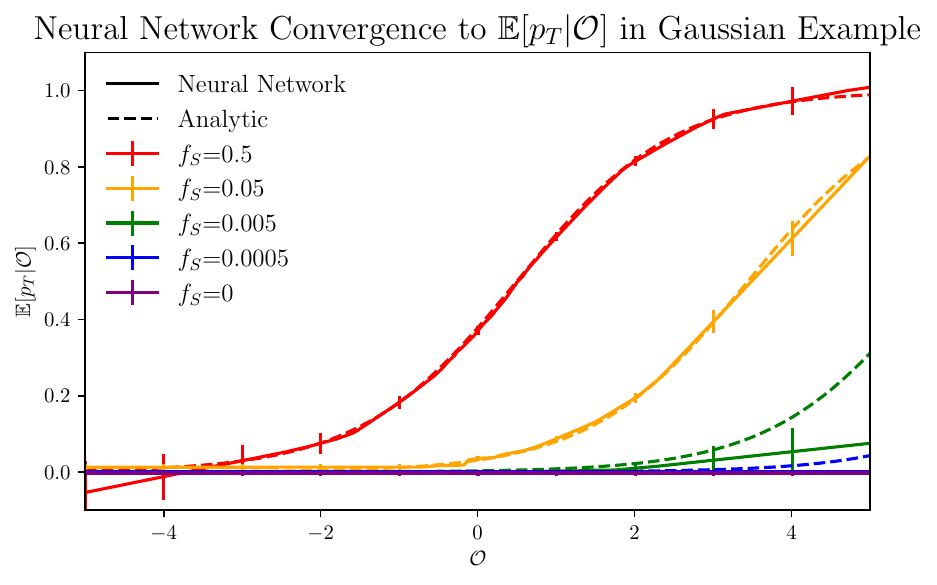}
    \caption{
    The neural network (solid) convergence to the conditional expectation (dashed) for the Gaussian example. We see good convergence in the high signal limit, with decay as the signal fraction is decreased.}
    \label{fig:gauss_conv}
\end{figure}

Furthermore, since we know the analytic form for our signal and background distributions, we can construct the conditional expectation formula explicitly:
\begin{equation}
    \mathbb{E}[p_T|\mathcal{O}] = \braket{p_T}_B +  f_S \frac{ \, (\langle p_T\rangle_S-\langle p_T\rangle_B) L_{S / B}(\mathcal O) }{1 - f_S + f_S \, L_{S / B}(\mathcal O)},
\end{equation}
where the likelihood ratio is:
\begin{equation}    
    L_{S / B}(\mathcal O) = \tfrac{\Delta \mathcal{O}_B}{\Delta \mathcal{O}_S} \exp\left[- \tfrac{{(\mathcal{O} - \braket{O}_S)^2}}{{2 (\Delta \mathcal{O})^2_S}} + \tfrac{{(\mathcal{O} - \braket{\mathcal{O}}_B)^2}}{{2 (\Delta \mathcal{O})^2_B}} \right].
\end{equation}
This allows us to visualize the convergence of the model for various signal fractions, shown in \Fig{fig:gauss_conv}, with optimal performance in the high signal fraction limit and convergence to a constant in the no signal limit.

\section{Energy Flow Polynomials}

Energy flow polynomials (EFPs) are $N$-point correlators on jets that form a discrete linear basis for all infrared and collinear safe (IRC) substructure observables. 
Diagramatically, they can be represented as loopless multigraphs.
For a multigraph $G$ with $N$ vertices and edges $(k, \ell) \in G$, the EFP takes the following form:
\begin{equation}
    \text{EFP}_G = \sum_{i_1=1}^M \ldots \sum_{i_N=1}^M z_{i_1} \ldots z_{i_N}
    \prod_{(k, \ell) \in G} \theta_{i_k, i_\ell},
\end{equation}
where $M$ is the number of constituents in the jet.
Each graph node corresponds to a sum over the energy fraction of the constituents of the jet, while edges correspond to an angular weighting factor between connected particles.
A list of the ``prime'' (i.e.~connected graph) EFPs up to degree 3 is shown \Tab{tab:graphs}.

The definitions for $z_i$ and $\theta_{jk}$ are context dependent and typically chosen to respect a subgroup of Lorentz symmetry to match the collider detector geometry. 
For unpolarized hadron colliders, observables are desired to be invariant under beam-direction boosts as well as rotations about the beam axis. 
Therefore, the particle transverse momentum $p_T$ and rapidity-azimuth ($y$, $\phi$) coordinates are used.
The energy weighting factor is simply:
\begin{equation}
    z_i = p_{T, i}.
\end{equation}
For massless particles, as used in our study, a convenient angular weighting factor is:
\begin{align}
\nonumber \theta_{ij} &= (2 p_i^\mu p_{j\mu}/p_{T,i} p_{T,j})^{\beta/2}\\ &=  (2\cos(\Delta \phi_{ij}) - 2\cosh(\Delta y_{ij}))^{\beta/2}.
\end{align}
In the analysis in the main text, we used $\beta=1$.

\begin{table}[t]
    \centering
    \begin{tabular}{| >{\centering}m{.5in} | >{\centering}m{2.5in} |}\hline
    \textbf{Degree} & \textbf{Connected Multigraphs} \tabularnewline\hline \hline
    $d=0$ & \includegraphics[scale=0.02]{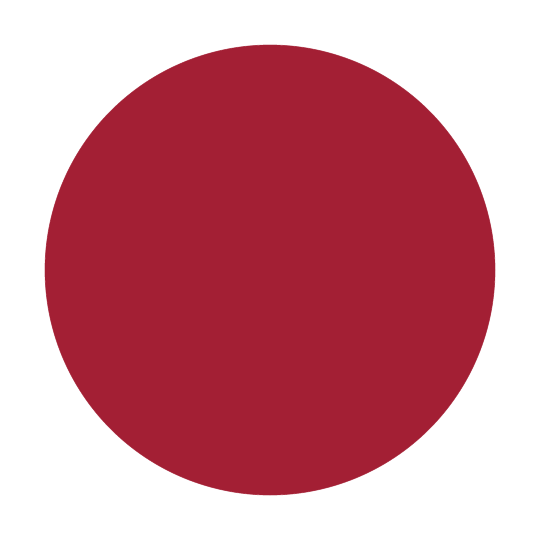}\tabularnewline\hline
    $d=1$ & \vspace{.05in}\includegraphics[scale=0.15]{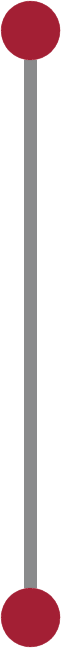}
    \tabularnewline\hline
    $d=2$ & \vspace{.05in}\includegraphics[scale=0.15]{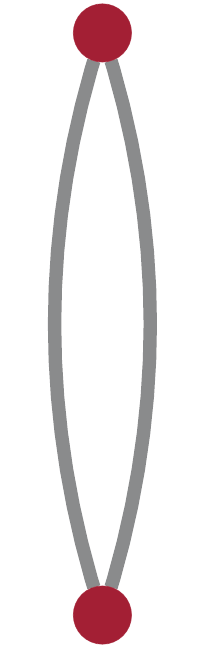}
                                         \includegraphics[scale=0.15]{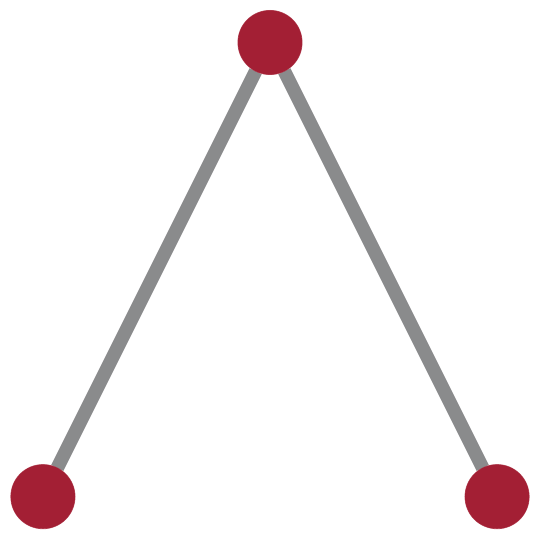} 
    \tabularnewline\hline
    $d=3$ & \vspace{.08in}\includegraphics[scale=0.15]{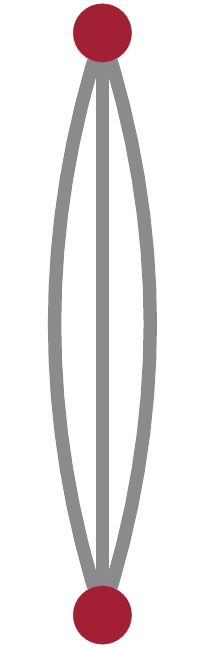}
                                         \includegraphics[scale=0.15]{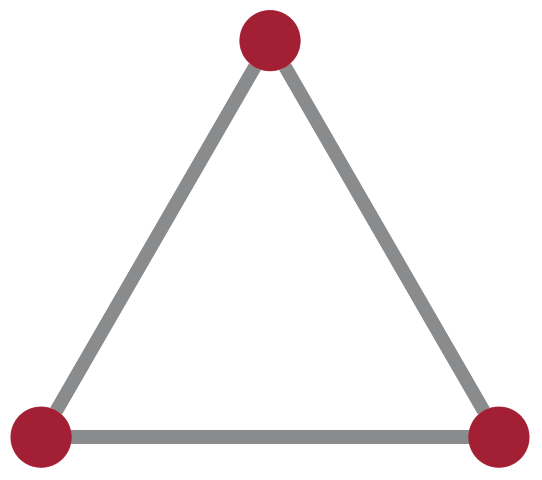}
                                         \includegraphics[scale=0.15]{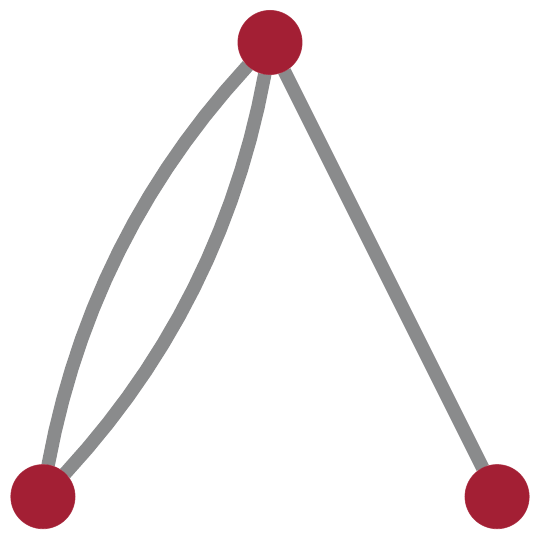}
                                         \includegraphics[scale=0.15]{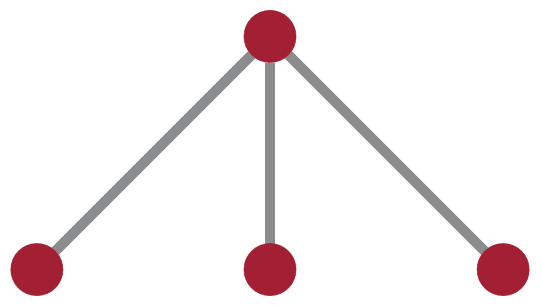}
                                         \includegraphics[scale=0.15]{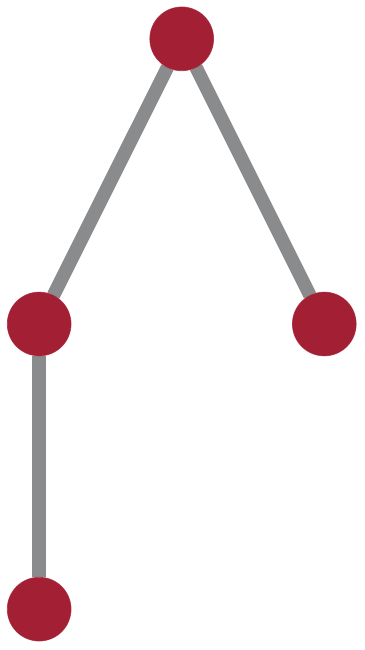}
    \tabularnewline\hline
    \end{tabular}
    \caption{All isomorphic, connected, loopless multigraphs up to degree 3.  Images  from~\url{http://energyflow.network/}.}
    \label{tab:graphs}
\end{table}

\section{Normalization for the Energy Flow Polynomials}

As discussed in the main text, to satisfy the factorization structure of the FORCE method, we need to make our EFP observables quasi-scale- and boost-invariant.
We can do this by normalizing each polynomial by a factor of
\begin{equation}
\label{eq:normalization}
\Bigg(\sum_{i=1}^M p_{T,i}\Bigg)^{N-2d} \left(\displaystyle\sum_{i=1}^M \sum_{j=1}^M p_{T,i} p_{T,j} \theta_{ij}\right)^{d},
\end{equation}
where $N$ and $d$ are the energy and angular degrees of the polynomial.
The reason that yields only quasi-scale and boost invariant observables (and not fully invariant ones) is that the jet radius defines a fixed boundary that does not transform under these transformations.
Thus, scaling $p_T$ and/or $\theta_{ij}$ could either add or subtract particles from the jet, changing the number of terms in the sums.

\begin{figure*}[p]
\centering
\subfloat[][]{\includegraphics[width=0.9\columnwidth]{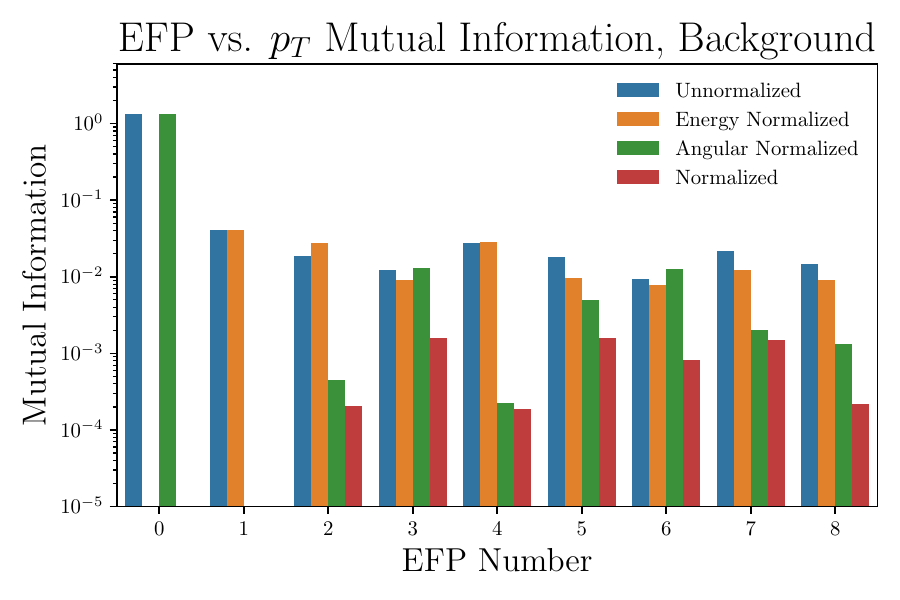}}
$\quad$
\subfloat[][]{\includegraphics[width=0.9\columnwidth]{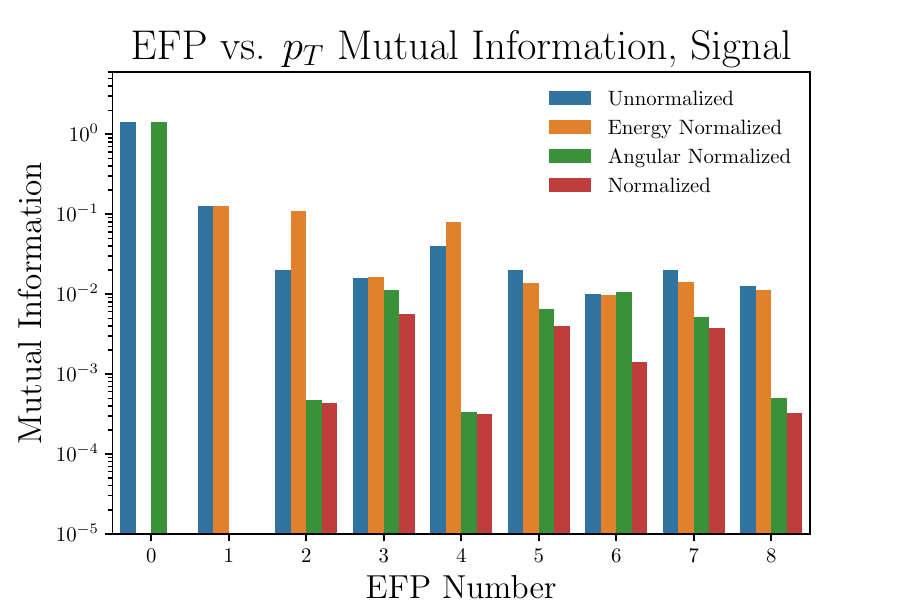}}
\newline
\subfloat[][]{\includegraphics[width=0.9\columnwidth]{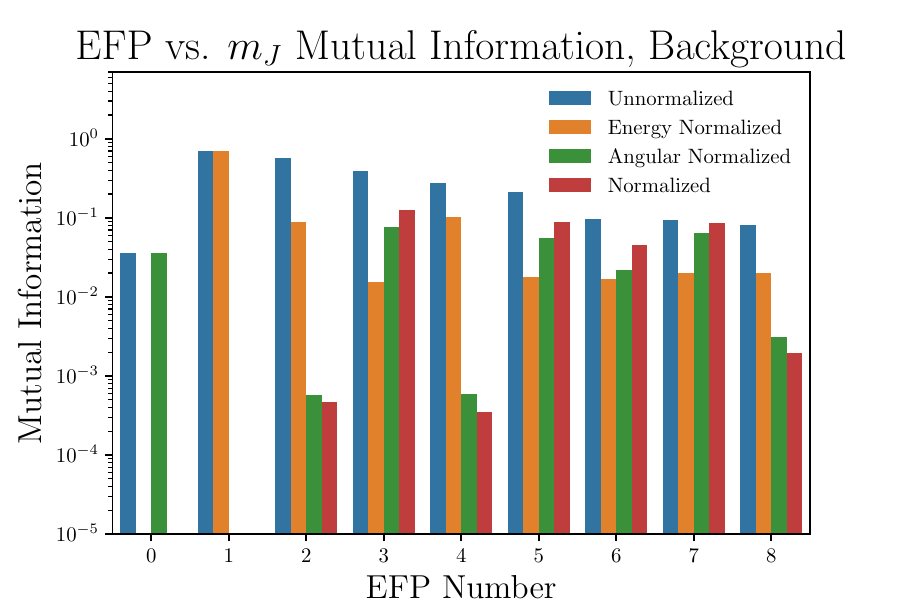}}
$\quad$
\subfloat[][]{\includegraphics[width=0.9\columnwidth]{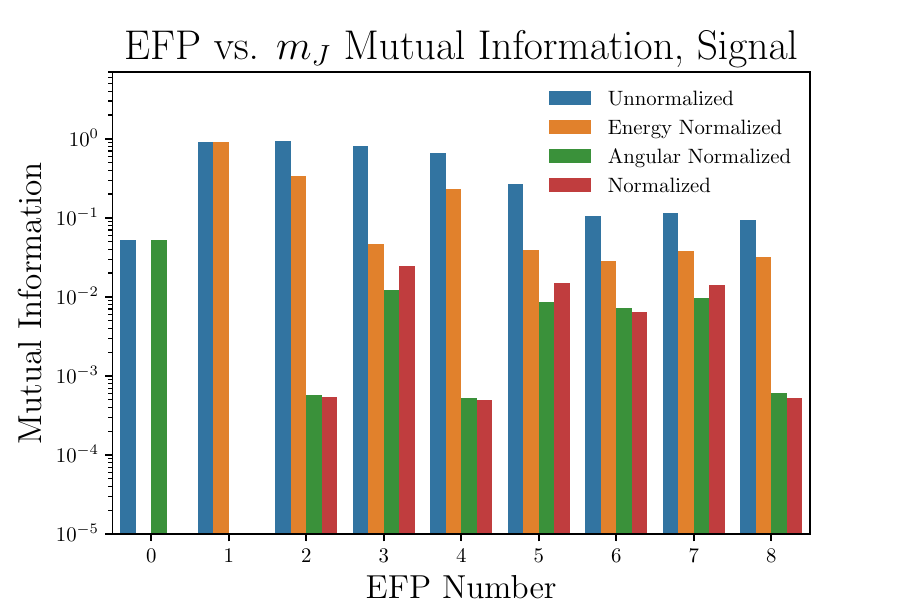}}
\caption{\label{fig:mutual-info}
    Mutual information between the EFPs and (top row) the jet $p_T$ and (bottom row) the jet mass, for (left column) background events and (right column) signal events.
    We consider four different normalization schemes, where EFP 0 is trivial for energy normalization and EFP 1 is trivial for angular normalization.
    Full normalization tends to have the lowest mutual information among the different normalization schemes, motivating its use in \CleverName as a feature that factorizes from $p_T$.
    Strictly speaking, factorization from jet mass is not needed for the \CleverName method to work, but it helps minimize sculpting effects for bump hunting.
}
\end{figure*}

The first term in \Eq{eq:normalization} can be viewed as an energy normalization while the second can be viewed as the angular normalization. 
To check that these normalization factors yield the desired the factorization structure, we calculate the mutual information between the jet's $p_T$ and the EFPs with 4 normalization schemes:
\begin{enumerate}
    \item unnormalized,
    \item only energy normalized,
    \item only angular normalized, and
    \item fully normalized.
\end{enumerate}
The results are shown in the top row of \Fig{fig:mutual-info}, for both signal and background events.
As expected, full normalization tends to yield the smallest mutual information between the EFPs and jet $p_T$ among all the normalization schemes, motivating its use as a factorized observable for the \CleverName method.

One additional benefit of full normalization is that the \CleverName model becomes less correlated from jet mass, as shown in the bottom row of \Fig{fig:mutual-info}.
Jet mass itself corresponds to the unnormalized EFP 2, and after angular normalization, this EFP has minimal correlation with jet mass as expected.
While the other EFPs retain some correlation with jet mass even after full normalization, it is a small effect that does not result in excessive mass sculpting in our case study.

\section{Shuffled Features for Factorized Objects in the LHC Olympics}

As noted in the main text, the $f_S = 0$ learned model did not converge to a constant function, unlike the prediction from the conditional expectation in Eq. (6) in the main text. 
This implies that there is some non-factorized dependence in our realistic case study that is being learned and exploited by the model.

To demonstrate an idealized use case, we construct a dataset that follows Eq. (3) in the main text exactly to check that it has the expected $f_S \to 0$ behavior.
For the signal and background separately, we create joint distributions as the product of its marginals:
\begin{align}
P_{S, \text{shuffle}}(\mathcal{O}, p_T) &= P_{S}(p_T) \, P_{S}(\mathcal{O}), \\
P_{B, \text{shuffle}}(\mathcal{O}, p_T) &= P_{B}(p_T) \, P_{B}(\mathcal{O}).
\end{align}
We call these ``shuffled'' datasets, since they correspond to independently drawing $p_T$ and substructure variables without replacement. 
To create the full dataset with $N$ data points drawn from the factorized joint distribution,
\begin{equation}
    P_{\text{shuffle}}(\mathcal{O}, p_T) = f_S \, P_{S}(p_T) \, P_{S}(\mathcal{O}) + 
    \\f_B \,  P_{B}(p_T) \, P_{B}(\mathcal{O}),
\end{equation}
we draw $f_{S} \, N$ samples from $P_{S, \text{shuffle}}$ and $f_{B} \, N$ samples from $P_{B, \text{shuffle}}$, with $f_S + f_B = 1$, and concatenate these draws.

\begin{figure*}[p]
    \centering
    \includegraphics[width=1.0\columnwidth]{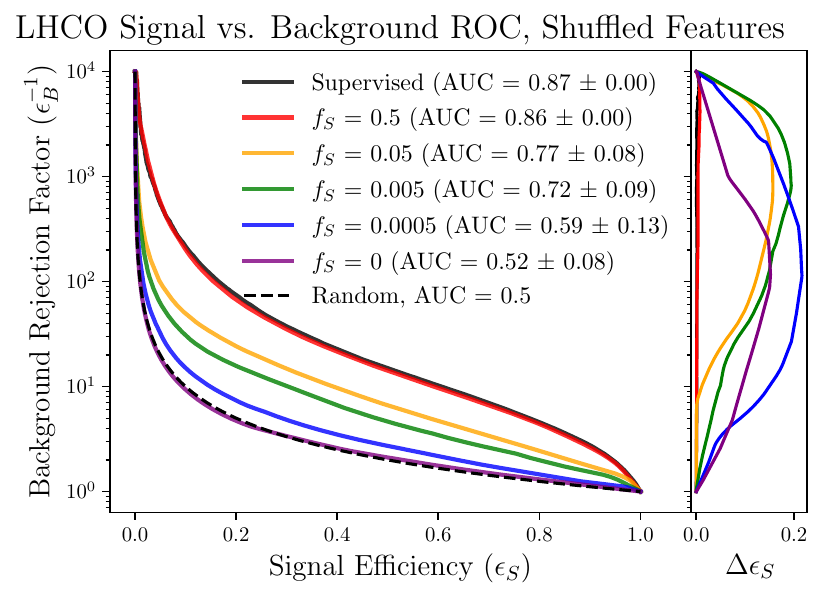}
    \caption{\label{fig:aucs_shuffle}
    The new physics per-jet classification performance of the \CleverName method for different signal fractions $f_S$ with shuffled normalized EFPs. We see that the model behaves optimally in the high-signal limit, approaching the supervised classification performance. In the low-signal limit, the model behaves as a random classifier, as predicted by the main text. Furthermore, there is a smooth decay in statistical power as the signal fraction is swept to zero.}
\end{figure*}

The signal versus background discrimination results for the shuffled dataset are shown in \Fig{fig:aucs_shuffle}. 
We see the expected behavior in both the no-signal and high-signal case, with no-signal functioning as a random classifier and the high-signal limit approaching optimal discrimination.

\begin{figure*}[t]
    \centering
    \subfloat[]{\includegraphics[width=.45\linewidth, height=.3\linewidth]{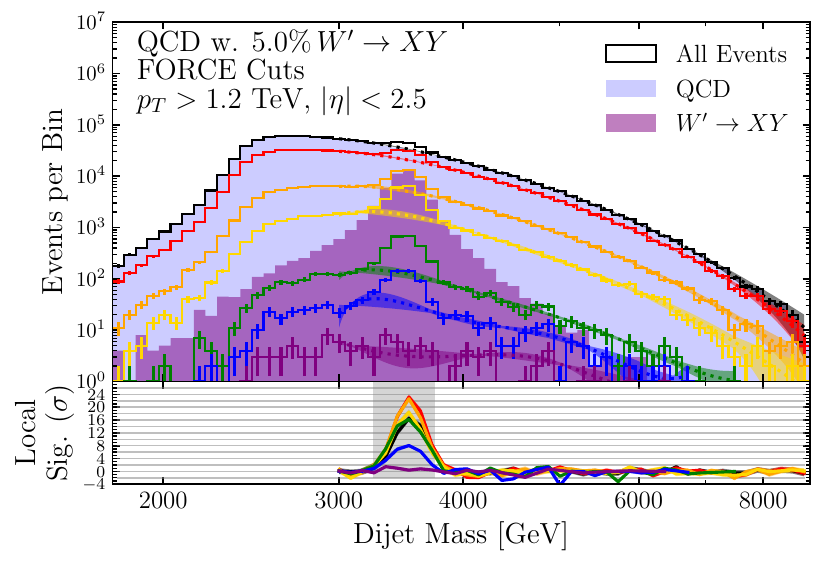}\label{fig:a}}%
    $\qquad$
    \subfloat[]{\includegraphics[width=.45\linewidth, height=.3\linewidth]{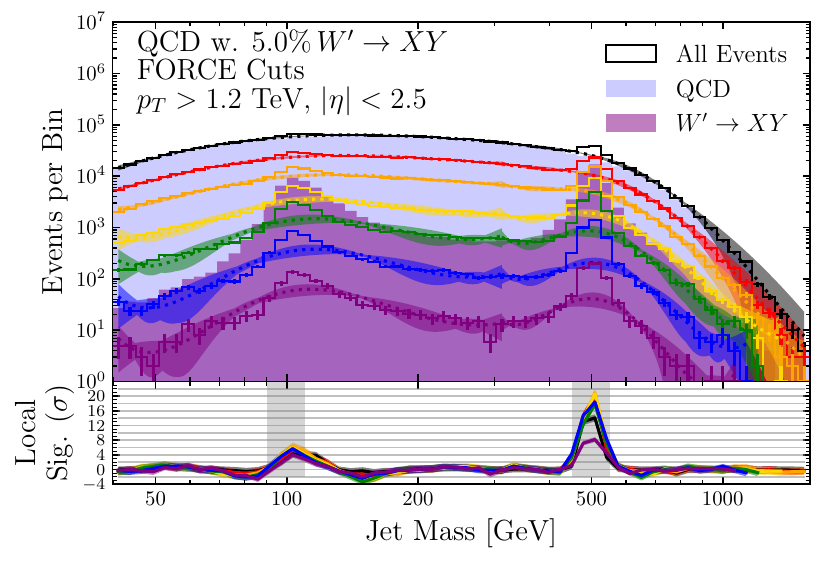}\label{fig:b}}%
    \\
    \subfloat[]{\includegraphics[width=.45\linewidth, height=.3\linewidth]{figures/bump_hunt/dijetmass/bump_hunt_0.005.pdf}\label{fig:c}}%
    $\qquad$
    \subfloat[]{\includegraphics[width=.45\linewidth, height=.3\linewidth]{figures/bump_hunt/jetmass/jetmass_bump_hunt_0.005.pdf}\label{fig:d}}%
    \\
    \subfloat[]{\includegraphics[width=.45\linewidth, height=.3\linewidth]{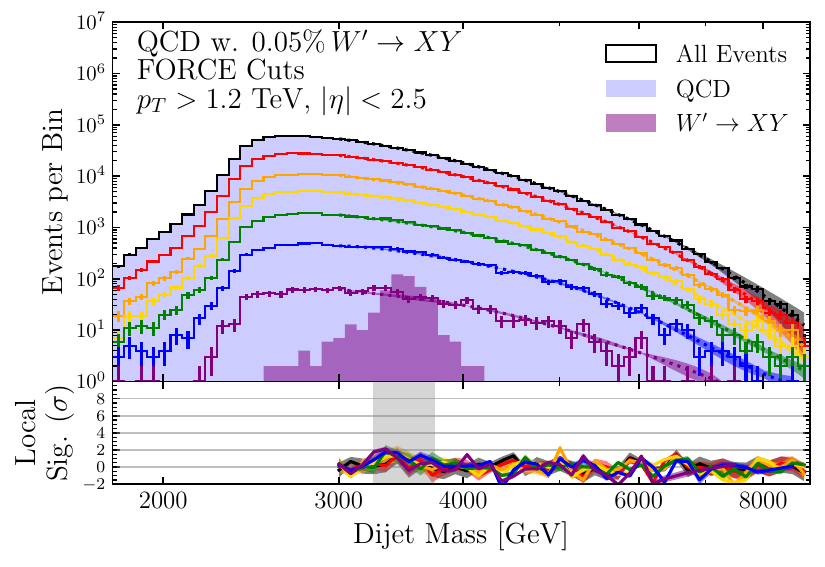}\label{fig:e}}%
    $\qquad$
    \subfloat[]{\includegraphics[width=.45\linewidth, height=.3\linewidth]{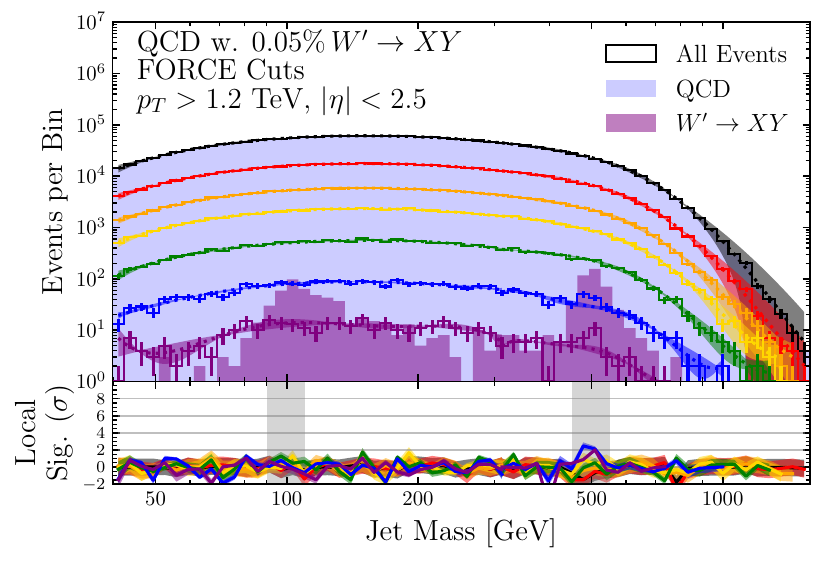}\label{fig:f}}%
    \caption{
    The same as Fig.~2 in the main text, but for (top row) $f_s = 0.05$, (middle row) $f_S = 0.005$ as in the main text, and (bottom row) $f_S = 0.0005$.
    Shown are (left column) dijet and (right column) jet mass distributions in the top panels and local significance values in the lower panels.
    }
    \label{fig:bump}
\end{figure*}

\begin{figure*}[t]
    \centering
    \subfloat[]{\includegraphics[width=.45\linewidth, height=.3\linewidth]{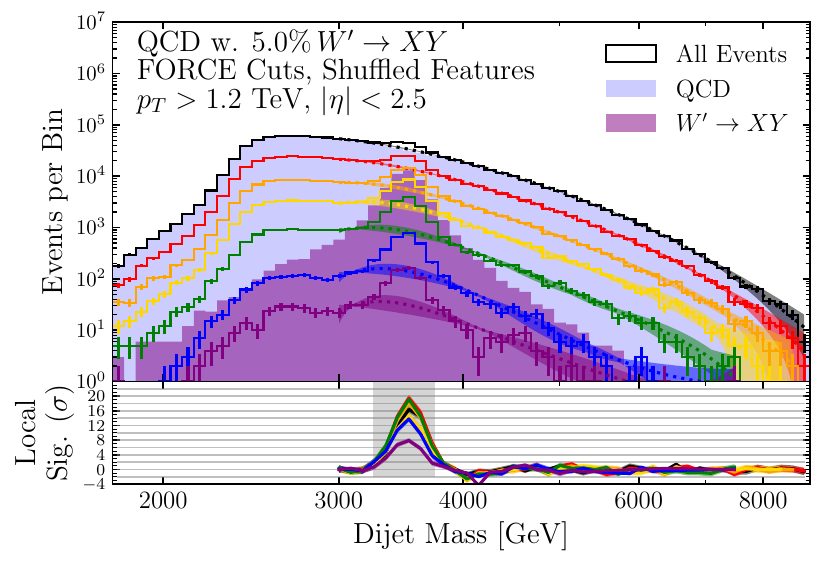}\label{fig:shuffle_a}}%
    $\qquad$
    \subfloat[]{\includegraphics[width=.45\linewidth, height=.3\linewidth]{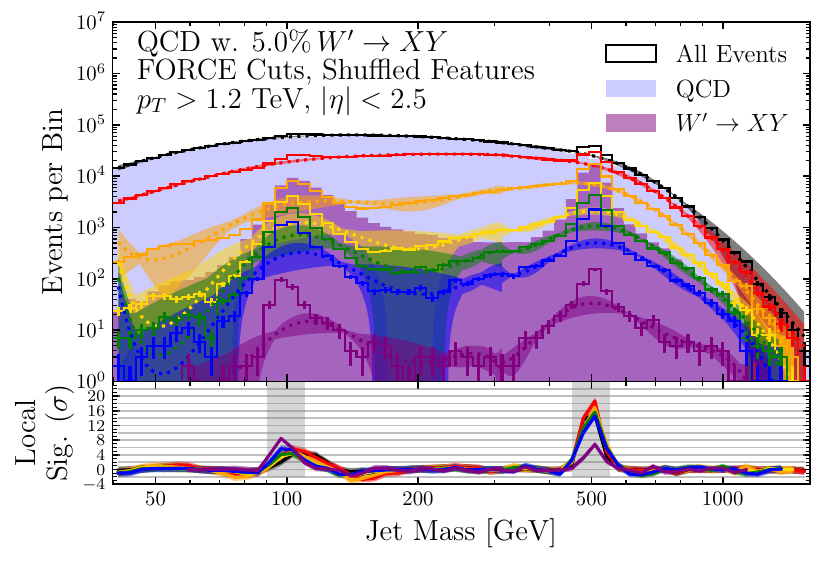}\label{fig:shuffle_b}}
    \\
    \subfloat[]{\includegraphics[width=.45\linewidth, height=.3\linewidth]{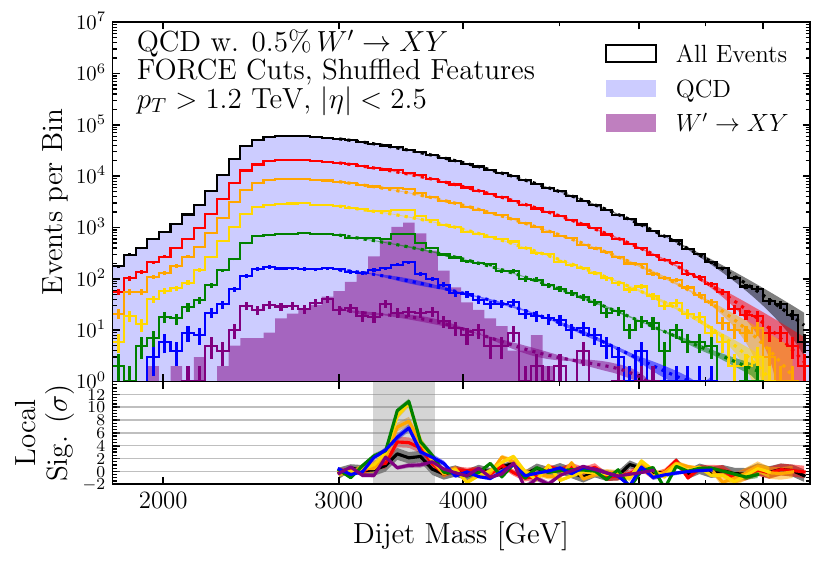}\label{fig:shuffle_c}}%
    $\qquad$
\subfloat[]{\includegraphics[width=.45\linewidth, height=.3\linewidth]{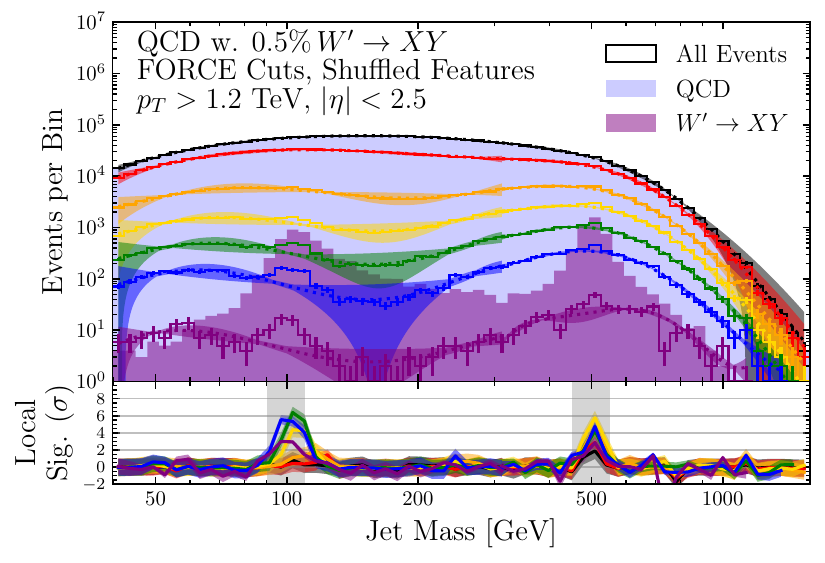}\label{fig:shuffle_d}}%
    \\
    \subfloat[]{\includegraphics[width=.45\linewidth, height=.3\linewidth]{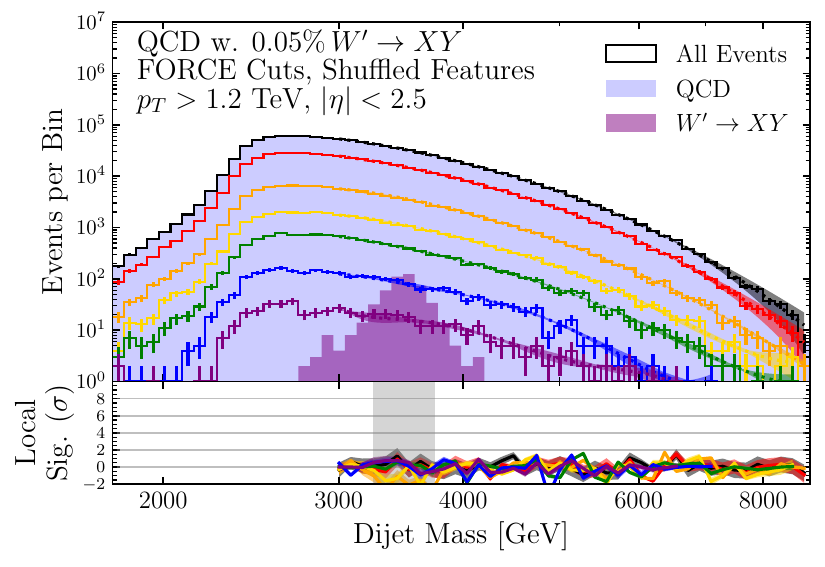}\label{fig:shuffle_e}}%
    $\qquad$
    \subfloat[]{\includegraphics[width=.45\linewidth, height=.3\linewidth]{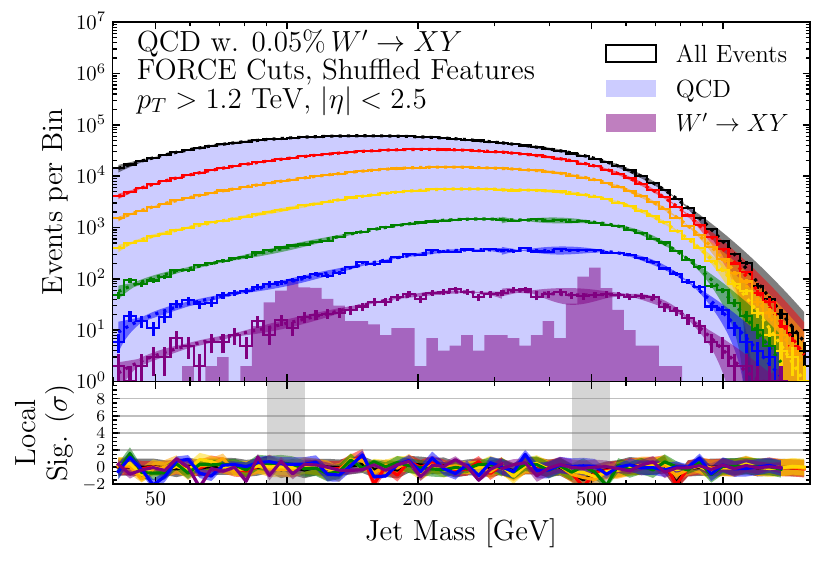}\label{fig:shuffle_f}}%
    \caption{
    The same as \Fig{fig:bump}, but using EFPs trained on shuffled targets where factorization holds exactly.}
    \label{fig:bump_shuffle}
\end{figure*}

For completeness, we repeat the analysis from Fig. 2 in the main text with the jet and dijet mass bump hunt results, but for three different values of the signal fraction.
The results for the original samples are shown in \Fig{fig:bump}, while the results for the shuffled sample are shown in \Fig{fig:bump_shuffle}.
We see that shuffling does not change the qualitative features of these plots, implying that the original samples obeyed factorization sufficiently well for \CleverName to yield trustable results.

\clearpage

\end{document}